\documentclass[a4paper,fleqn,usenatbib]{mnras}
\usepackage{Times}
\usepackage[T1]{fontenc}
\usepackage{ae,aecompl}

\usepackage{graphicx}	% Including figure files
\usepackage{amsmath}	% Advanced maths commands
\usepackage{amssymb}	% Extra maths symbols

\title[CALYMHA Survey: Ly$\alpha$ emitters at $z=2.23$]{The CALYMHA survey: Ly$\alpha$ luminosity function and global escape fraction of Ly$\alpha$ photons at $\bf z=2.23$\thanks{Based on observations obtained on the Isaac Newton Telescope (INT), programs: I13AN002, I14AN002, 088-INT7/14A, I14BN006, 118-INT13/14B \& I15AN008.}}

\author[D. Sobral et al.]{David Sobral$^{1,2}$\thanks{E-mail: d.sobral@lancaster.ac.uk}, Jorryt Matthee$^{2}$, Philip Best$^{3}$, Andra Stroe$^{4}$\thanks{ESO Fellow}, Huub R\"{o}ttgering$^{2}$, \newauthor Iv\'{a}n Oteo$^{3,4}$, Ian Smail$^{5}$, Leah Morabito$^{2}$, Ana Paulino-Afonso$^{6,7}$ \\
$^{1}$ Department of Physics, Lancaster University, Lancaster, LA1 4YB, UK \\
$^{2}$ Leiden Observatory, Leiden University, P.O.\ Box 9513, NL-2300 RA Leiden, The Netherlands \\
$^{3}$ Institute for Astronomy, University of Edinburgh, Royal Observatory, Blackford Hill, Edinburgh EH9 3HJ, UK \\
$^{4}$ European Southern Observatory, Karl-Schwarzschild-Str. 2, 85748 Garching, Germany \\
$^{5}$ Centre for Extragalactic Astronomy, Department of Physics, Durham University, South Road, Durham DH1 3LE UK \\
$^{6}$ Instituto de Astrof\'{\i}sica e Ci\^{e}ncias do Espa\c{c}o, Universidade de Lisboa, OAL, Tapada da Ajuda, PT1349-018 Lisbon, Portugal \\
$^{7}$ Departamento de F\'{i}sica, Faculdade de Ci\^{e}ncias, Universidade de Lisboa, Campo Grande, PT1749-016 Lisbon, Portugal
}

\date{Accepted 2016 November 25. Received 2016 November 25; in original form 2016 July 25.}

\pubyear{2016}

\begin{document}
\label{firstpage}
\pagerange{\pageref{firstpage}--\pageref{lastpage}}
\maketitle

% Abstract of the paper
\begin{abstract}
We present the CAlibrating LYMan-$\alpha$ with H$\alpha$ (CALYMHA) pilot survey and new results on Lyman-$\alpha$ (Ly$\alpha$) selected galaxies at $z\sim2$. We use a custom-built Ly$\alpha$ narrow-band filter at the Isaac Newton Telescope, designed to provide a matched volume coverage to the $z=2.23$ H$\alpha$ HiZELS survey. Here we present the first results for the COSMOS and UDS fields. Our survey currently reaches a 3$\sigma$ line flux limit of $\sim4\times10^{-17}$\,erg\,s$^{-1}$\,cm$^{-2}$, and a Ly$\alpha$ luminosity limit of $\sim10^{42.3}$\,erg\,s$^{-1}$. We find 188 Ly$\alpha$ emitters over $7.3\times10^5$\,Mpc$^{3}$, but also find significant numbers of other line emitting sources corresponding to He{\sc ii}, C{\sc iii}] and C{\sc iv} emission lines. These sources are important contaminants, and we carefully remove them, unlike most previous studies. We find that the Ly$\alpha$ luminosity function at $z=2.23$ is very well described by a Schechter function up to $L_{\rm Ly\alpha}\approx10^{43}$\,erg\,s$^{-1}$ with $L^*=10^{42.59^{+0.16}_{-0.08}}$\,erg\,s$^{-1}$, $\phi^*=10^{-3.09^{+0.14}_{-0.34}}$\,Mpc$^{-3}$ and $\alpha=-1.75\pm0.25$. Above $L_{\rm Ly\alpha}\approx10^{43}$\,erg\,s$^{-1}$ the Ly$\alpha$ luminosity function becomes power-law like, driven by X-ray AGN. We find that Ly$\alpha$-selected emitters have a high escape fraction of $37\pm7$\%, anti-correlated with Ly$\alpha$ luminosity and correlated with Ly$\alpha$ equivalent width. Ly$\alpha$ emitters have ubiquitous large ($\approx40$\,kpc) Ly$\alpha$ haloes, $\sim2\times$ larger than their H$\alpha$ extents. By directly comparing our Ly$\alpha$ and H$\alpha$ luminosity functions we find that the global/overall escape fraction of Ly$\alpha$ photons (within a 13\,kpc radius) from the full population of star-forming galaxies is $5.1\pm0.2$\% at the peak of the star formation history. An extra $3.3\pm0.3$\% of Ly$\alpha$ photons likely still escape, but at larger radii.
\end{abstract}

%%%%%%%%%%%%%%%%%%%%%%%%%%%%%%%%%%%%%%%%%%%%%%%%%%

\section{Introduction}

Understanding galaxy formation and evolution requires significant efforts on both theoretical and observational sides. Observations show that the star formation activity in the Universe was over 10 times higher in the past, reaching a peak at $z\sim2-3$ \citep[e.g.][]{Lilly96, Karim11, Sobral2013}. Most of this increase is explained by typical star formation rates (SFRs) of galaxies at $z\sim2$ being a factor $\sim10\times$ higher than at $z=0$ \citep[e.g.][]{Smit2012,Sobral2014,StroeSobral2015}, likely driven, to first order, by relatively high gas fractions \citep[e.g.][]{Tacconi2010,Saintonge2011,Stott2016}. Beyond $z\sim2-3$, UV and rest-frame optical emission line studies suggest a decline of the star-formation history of the Universe with increasing redshift \citep[e.g.][]{Bouwens15,Khostovan2015}.

While the UV is the main way of photometrically selecting $z>3$ star-forming galaxies, by taking advantage of the Lyman break technique \citep[e.g.][]{Steidel96,Giavalisco2002}, the Lyman-$\alpha$ (Ly$\alpha$) emission line is by far the most used for spectroscopically confirming and studying very distant galaxies \citep[e.g.][]{Ono2012,Oesch2015,Sobral2015CR7,Zitrin2015}. Ly$\alpha$ has also been widely used to obtain large samples of galaxies through the narrow-band selection \citep[e.g.][]{Ouchi2008,Ouchi2010,Matthee2015,Santos2016} and to find distant galaxies with extremely young and likely metal-poor stellar populations \citep[e.g.][]{Kashikawa2012,Sobral2015CR7}. The Ly$\alpha$ line is also used to study the interstellar \citep[e.g.][]{Swinbank2015}, circumgalactic and/or intergalactic medium \citep[e.g.][]{Sargent1980,Hernquist1996}. This is facilitated by the fact that Ly$\alpha$ emission line is intrinsically the brightest emission line in {\sc Hii} regions \citep[e.g.][]{PartridgePeebles1967,Pritchet1994}, and due to the fact that it is redshifted into easily-observed optical wavelengths beyond $z\sim2$ \citep[see also][]{DijkstraReview}.

The Ly$\alpha$ luminosity function has been found to evolve very strongly from $z\sim0$ to $z\sim3$ for relatively faint Ly$\alpha$ emitters \citep[e.g.][]{Ouchi2008, Cowie2010,Barger2012,Drake2016}. At $z\sim2$ the Ly$\alpha$ luminosity function has been studied by e.g. \cite{Hayes2010} and \cite{Konno2016}, with significant disagreements probably explained by the expected strong cosmic variance \citep[see][]{Sobral2015}. \cite{Konno2016} also finds a significant deviation from a Schechter function for $L_{\rm Ly\alpha}>L^*$, consistent with results seen for H$\alpha$ selected samples from \cite{Sobral2016}. However, an important issue that needs to be addressed is the contamination by other lines. Most Ly$\alpha$ surveys assume that contaminants are negligible \citep[e.g.][]{Konno2016}, but that is not necessarily the case \citep[e.g.][]{Matthee2015,Santos2016,Nakajima2016}.

Despite much progress in selecting Ly$\alpha$ emitters through large surveys, the nature and evolution of Ly$\alpha$ sources are still a matter of debate. For example, recent advances with IFU surveys using the MUSE instrument on the VLT \citep[e.g.][]{Bacon15,Karman15} confirm a population of Ly$\alpha$ emitters at $z\sim3-6$ which are completely undetected in the deepest broad-band photometric surveys, due to their very high equivalent widths (EW). Hundreds of similar candidate Ly$\alpha$ emitters were previously discovered by e.g. the Subaru telescope \citep[][]{MalhotraRhoads2004,Murayama2007,Kashikawa2006,Ouchi2008,Ouchi2010}. This is consistent with many Ly$\alpha$ emitters at $z>3$ being typically low mass, blue and likely low metallicity \citep[e.g.][]{Gawiser2007,Gronwall2007,Ono2010,Sobral2015CR7,Nakajima2016}. However, studies closer to the peak of star formation history at $z\sim2$ reveal Ly$\alpha$ sources which differ from those typical characteristics \citep[e.g.][]{Stiavelli2001,Bongiovanni2010,Oteo2015,Hathi2016}. Some are found to be relatively massive, dusty \citep[e.g.][]{Chapman2005,Matthee2016} and red \citep[e.g.][]{Stiavelli2001,Oteo2012,Oteo2015,Sandberg2015}. Below $z\sim3$, studies find that luminous Ly$\alpha$ emitters are progressively AGN dominated and more evolved \citep[][]{Nilsson2009,Cowie2010,Barger2012,Wold2014}, although others can easily be considered analogues of $z>3$ emitters \citep[e.g.][]{Oteo2012b,Barger2012,Erb2016,Trainor2016}.

Many of the key limitations/questions about Ly$\alpha$ emitters result directly from Ly$\alpha$'s complex radiative transfer \citep[e.g.][]{Verhamme2006,Dijkstra2007,Verhamme2008,Gronke2015b,Gronke2016}. The resonant nature of the Ly$\alpha$ line results in Ly$\alpha$ photons scattering in neutral hydrogen, substantially increasing the likelihood of absorption by interstellar dust \citep[e.g.][]{Atek2008,Hayes2015}. Thus, Ly$\alpha$ luminosity can be significantly reduced, or even completely suppressed \citep[e.g.][]{Verhamme2008,Atek09,Hayes2011,Atek2014}. Theoretical galaxy formation models predict f$_{esc}= 2-10$\,\% \citep[e.g.][]{LeDelliou06,Nagamine08,Garel2015} at $z = 2-3$, but are limited by a large number of assumptions which only direct observations can verify. Furthermore, a major limitation for models is the need for a compromise between the resolution required for radiative transfer, and the need to simulate large enough volumes to be representative. For Ly$\alpha$-selected samples (biased towards high Ly$\alpha$ escape fractions)  at $z\sim2-3$ \citep[e.g.][]{Nilsson2009}, the comparison of Ly$\alpha$ with the UV suggests Ly$\alpha$ escape fractions, f$_{\rm esc}$, of $30-60$\,\% \citep[e.g.][]{Wardlow2014,Trainor2015}. 

One way to improve our understanding of Ly$\alpha$ selected sources and its escape fraction is the comparison with a well understood, non-resonant recombination emission line, such as H$\alpha$. \cite{Hayes2010} provided such a study for a relatively small volume at $z=2.2$, finding a global $\sim5$\% escape fraction. More recently, \cite{Matthee2016} studied a sample of $\sim1000$ H$\alpha$-selected galaxies, to find that the Ly$\alpha$ escape fraction strongly depends on the aperture used and on star formation rate (SFR). \cite{Konno2016} have also presented a statistical global escape fraction measurement by comparing their Ly$\alpha$ luminosity function with the UV or with the H$\alpha$ luminosity function from \cite{Sobral2013}. \cite{Sandberg2015} presented an H$\alpha$-Ly$\alpha$ study over the GOODS N field at $z\sim2$, but the small sample size and the typical low luminosity of the sources greatly limits their conclusions. A significant advance can only be obtained with a panoramic survey, covering the full range of environments, and having access to both Ly$\alpha$ and H$\alpha$.

In order to address current shortcomings, we are carrying out the CALYMHA survey: CAlibrating LYMan-$\alpha$ with H$\alpha$. Our survey combines the $z=2.23$ H$\alpha$ emitters from HiZELS \citep{Sobral2013} with Ly$\alpha$ measurements using a custom-made NB filter (see Figure \ref{Profiles}). Here we describe the first CALYMHA observations from our pilot survey. \S\ref{Observations} describes the observations, data reduction and photometry. In \S\ref{Selection_emitters_LAEs} we select emission line candidates, explore their nature and diversity, and select our sample of Ly$\alpha$ emitters at $z=2.23$. \S\ref{methods} presents the methods and corrections used in this paper. \S\ref{results} presents the Ly$\alpha$ luminosity function, its evolution and the Ly$\alpha$ EW distribution. In \S\ref{global_escape_section} we present the results on the Ly$\alpha$ escape fraction and discuss them. Finally, \S\ref{conclusions} presents the conclusions. We use a $\Lambda$CDM cosmology with $H_0$ = 70 km s$^{-1}$\,Mpc$^{-1}$, $\Omega_{\rm M}=0.3$ and $\Omega_{\Lambda} = 0.7$. Magnitudes are measured in $3''$ diameter apertures in the AB system, unless noted otherwise.

%%%%%%%%%%%%%%%%%%%%%%%%%%%%%%%%%%%%%%%%%%%%%%%%%%%%
%
% Table 1 - CALYMHA observations in COSMOS and UDS
%
%%%%%%%%%%%%%%%%%%%%%%%%%%%%%%%%%%%%%%%%%%%%%%%%%%%%
\begin{table*}
\caption{Observation log of all NB392 observations for our CALYMHA survey, including observations undertaken under bad seeing conditions which were not used. t$_{\rm exp}$ is the total exposure time, while the value between brackets is the exposure time effectively used after rejecting all bad frames. We also show the full range of FWHM in all images for each pointing, while in brackets we show the FWHM within the frames that were effectively used (corresponding to the total exposure times also presented in brackets).}
\begin{tabular}{cccccc}
\hline
{\bf Field} & {\bf R.A.} & {\bf Dec.} & {\bf t$_{\rm exp}$ (used)} & {\bf FWHM (used)} & {\bf Dates of observations} \\ 
 & (J2000) & (J2000) & (ks) & ($''$) & (All conditions) \\ \hline
COSMOS 1 & 10 01 59.4 & +02 27 06.5 & 28.4 (8.9) & 2.1 $\pm$ 0.4 (1.8 $\pm$ 0.2) & 2014 Feb 28, Mar 1-4  \\
COSMOS 2 & 10 01 59.4 & +01 53 48.5 & 41.1 (12.9) & 3.4 $\pm$ 1.2 (1.7 $\pm$ 0.1) & 2014 Mar 6,8; 2015 Jan 19-21,24  \\
COSMOS 3 & 10 01 15.0 & +02 49 18.5 & 40.0 (21.5) & 3.4 $\pm$ 1.3 (1.7 $\pm$ 0.1) & 2014 Mar 5,7; 2015 Jan 21-24  \\
COSMOS 4 & 10 00 30.6 & +02 16 00.5 & 105.6 (55) & 1.9 $\pm$ 0.5 (1.6 $\pm$ 0.1) & 2014 Mar 1,7-9,26, Dec 23-26; 2015 Jan 20-22,28 \\
COSMOS 5 & 09 59 46.3 & +01 53 48.5 & 68.7 (11.9) & 3.3 $\pm$ 1.3 (1.8 $\pm$ 0.2) & 2014 Mar 4-7,24-28; 2015 Jan 20,24,25  \\
COSMOS 6 & 09 58 55.7 & +02 38 12.5 & 104.3 (12.2) & 2.7 $\pm$ 0.9 (1.8 $\pm$ 0.1) & 2014 Dec 21,23-25; 2015 Jan 19,23-28  \\
COSMOS 7 & 09 58 17.5 & +02 04 54.5 & 49.8 (12.1) & 2.2 $\pm$ 1.4 (1.9 $\pm$ 0.1) & 2014 Feb 26-28; Mar 1; 2015 Jan 27-28\\ 
UDS 1 & 02 16 43.0 & $-$04 51 48.0 & 81.0 (36.0) & 2.0 $\pm$ 0.9 (1.5 $\pm$ 0.2) & 2014 Feb 28, Mar 1,3, Dec 20,22-25; 2015 Jan 20-27  \\ \hline
\end{tabular}\label{table_obs}
\end{table*}

\section{Observations and Data Reduction}\label{Observations}

\subsection{Observations with INT/WFC}

\begin{figure}
\begin{tabular}{cccc}
\includegraphics[width=8.3cm]{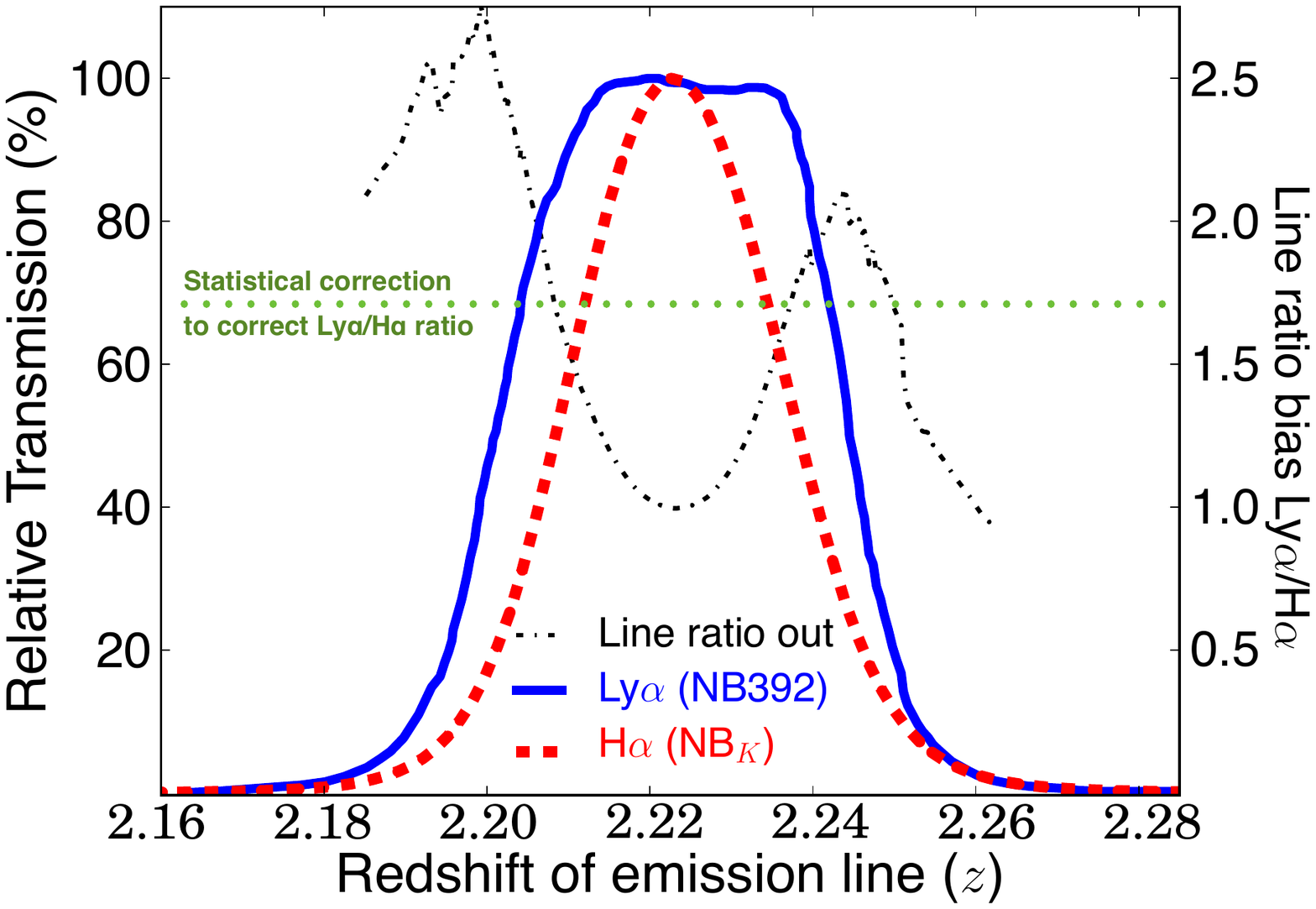}
\end{tabular} 
\caption{The transmission curves of our NB392 filter, primarily targeting the Ly$\alpha$ emission line at $z=2.23$, and our NB$_K$ filter (\citealt{Sobral2013}), which targets H$\alpha$ at the same redshift. We also show how observed line ratios vary as a function of redshift on a source by source basis, while we show the global correction for statistical samples that are randomly distributed in redshift. Note that the most significant biases are found in the wings, but the probability of finding a source, within a statistical sample, in the wings, is extremely low.}
\label{Profiles}
\end{figure}

Observations were obtained with a custom-built narrow-band filter (NB392) for the Isaac Newton Telescope's Wide Field Camera. The NB392 filter ($\lambda_c = 3918${\AA}, $\Delta\lambda = 52${\AA}) was designed by us such that the transmission of the redshifted Ly$\alpha$ line matches that of the redshifted H$\alpha$ line in the NB$_K$ filter (see Figure \ref{Profiles}). The filter was designed to have an H$\alpha$ selected sample as the primary science driver, and thus one requirement was that the filter profile was slightly wider in redshift, so that H$\alpha$ emitters would have close to 100\% transmission in the Ly$\alpha$ filter and also to allow for velocity offsets between Ly$\alpha$ and H$\alpha$ (see Figure \ref{Profiles} and \citealt{Matthee2016}). First light was obtained on May 6 2013, and the last observations presented in this paper were taken on January 27 2015. In total, we have observed for roughly 50 nights (programs: 2013AN002, 2013BN008, 2014AC88, 2014AN002, 2014BN006, 2014BC118) over a wide range of observing conditions. A significant amount of time was lost due to clouds, high humidity, rain, snow, ice, Sahara dust (`calima') and technical failures. With a typical seeing at La Palma/INT of about 1.3-1.5$''$ over our observing runs, and with the filter being at short wavelengths ($u$ band), the median seeing is 1.8$''$ overall in our NB392 filter. Table \ref{table_obs} presents the observations.

%%%%%%%%%%%%%%%%%%%%%%%%%%%%%%%%%%%%%%%
%
% Figure 1 -  Filter profile - also showing line fractions as a function of redshift
%
%%%%%%%%%%%%%%%%%%%%%%%%%%%%%%%%%%%%%%%
%%

\begin{figure*}
\begin{tabular}{cccc}
\includegraphics[width=16.0cm]{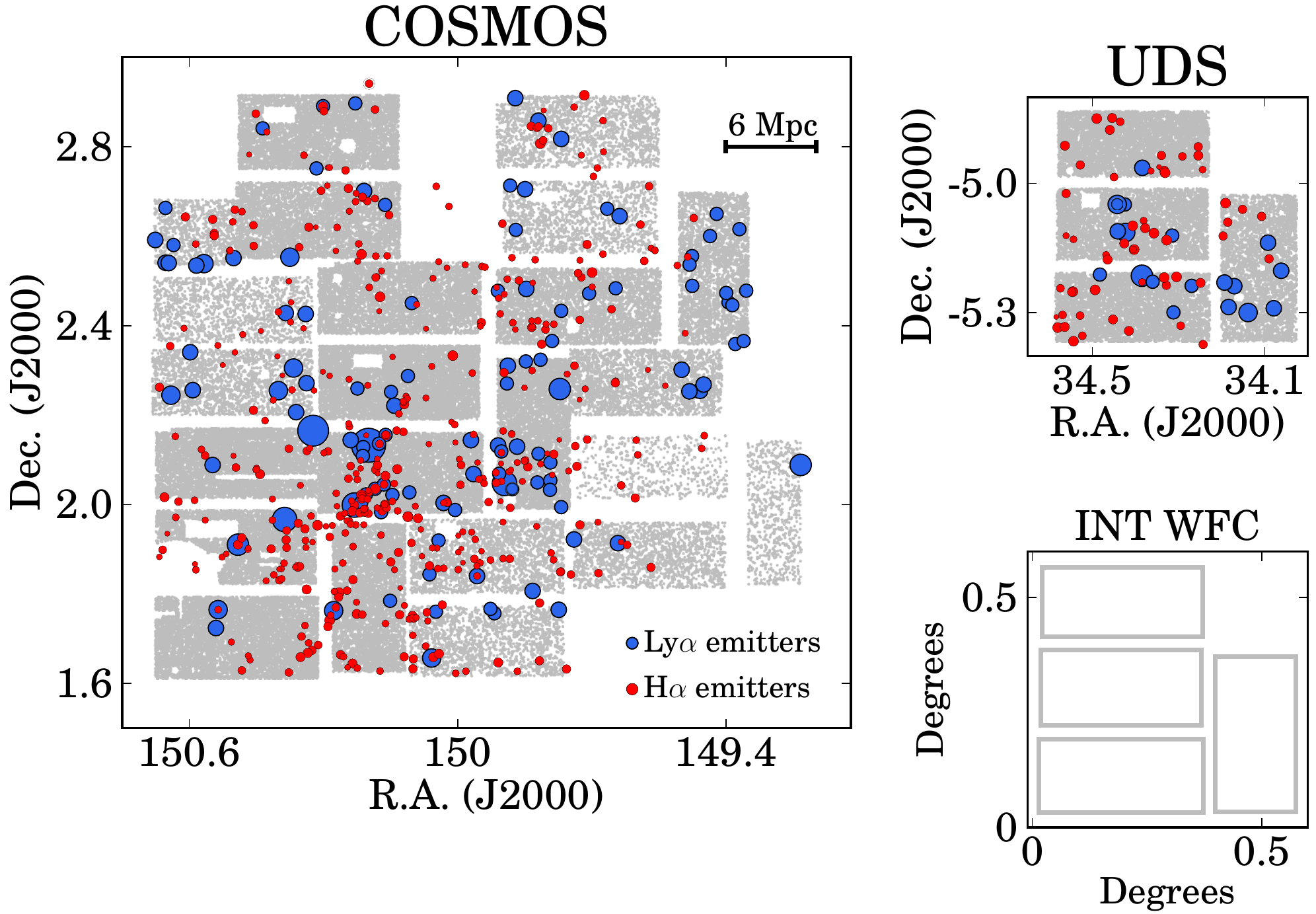}
\end{tabular} 
\caption{On sky distribution of all NB392 detections in COSMOS and UDS, showing the masked regions and highlighting the differences in depth of some of the pointings. Grey points show NB392 sources. On top we show the H$\alpha$ emitters from \citet{Sobral2013} and our Ly$\alpha$ emitters at $z=2.23$, after selecting them out of all NB392 emitters (see \S\ref{selecting_lymanalphaf}). Symbol sizes are scaled with luminosity for Ly$\alpha$ emitters. We also show the field of view of WFC/INT. Note that we only cover a fraction of the full UDS field.}
\label{RADEC}
\end{figure*}

Observations were conducted following a cross dither pattern, each consisting of 5 exposures with typical offsets of $30''$ to fill in the chip gaps (see Figure \ref{RADEC}) and sample the location of bad/hot pixels in an optimal way. The exposure times for individual frames were either 0.2 or 1.0\,ks, depending on whether there was a suitable guide-star available. Auto-guiding was relatively challenging because the guide window also goes through our particularly narrow filter, such that a star needs to be about 5-6 magnitudes brighter than usual to provide high enough signal to noise.

%%%%%%%%%%%%%%%%%%%%%%%%%%%%%%%%%%%%%%%%%%%%%%%%%%%%%%%%%%%%%%
%
% Figure 2 - RA DEC distribution and comparison HAEs vs LAEs and coverage in COSMOS, UDS and FoV of INT/WFC
%
%%%%%%%%%%%%%%%%%%%%%%%%%%%%%%%%%%%%%%%%%%%%%%%%%%%%%%%%%%%%%%

\subsection{Data reduction: NB392}

We reduced our NB392 data with a dedicated pipeline based on \textsc{python}, presented in \cite{Stroe2014} and \cite{StroeSobral2015}. Briefly, the data for each CCD were processed independently. The flats for each night were median-combined, after masking sources, to obtain a ``master-flat". A ``master-bias" for each night of observing was obtained by median-combining biases. The individual exposures were bias-subtracted and sky-flattened to remove electronic camera noise, shadowing effect and normalised for the pixel quantum efficiency. Science exposure pixels that deviated by more than $3\sigma$ from the local median were masked. These are either bad pixels (non-responsive) or hot pixels (typically stable over time) or cosmic rays (varying from frame to frame).

We have removed all frames with insufficient quality for our analysis. This included automatic removal of images which had failed astrometry due to the low number of sources in the image, mostly due to high extinction by clouds. We also rejected images for which any problems may have happened, including focussing and read-out issues. We visually checked all frames and removed a total of 20 frames due to read-out errors, guiding losses and satellite trails. These account for the removal of 2\,\% of data. 

Our observations were conducted in a wide variety of observing conditions. Before combining the data, we study the effect of different rejection criteria in terms of seeing, such that the depth is maximised. We use {\sc SExtractor} \citep{Bertin1996} to measure the median seeing and then stack frames in ranked subsets up to a certain FWHM seeing. We find that the depth (measured in apertures of 3$''$) improves rapidly up to seeing 1.8\,$''$ for our deepest pointing, COSMOS P4 \citep[see][]{Matthee2016}. Other fields reach a greater depth by including frames up to a maximum seeing of 2\,$''$. We therefore use these and reject individual frames with seeing greater than 2$''$ (see Table \ref{table_obs}).

Before stacking we normalise images to the same zero-point (using SDSS $u$ photometry) and match them to the same point spread function (PSF); see \cite{Matthee2016}. We then mask regions in the final stacks which are too noisy, are contaminated by bright stars, or where the S/N is significantly below the average (e.g. gaps between detectors). Figure \ref{RADEC} presents all the NB392 sources detected after masking, with the density of sources scaling with depth achieved in each sub-region. The total area after masking is 1.43\,deg$^2$.

\subsection{Photometric Calibration and survey depth}

\begin{figure}
\includegraphics[width=8.3cm]{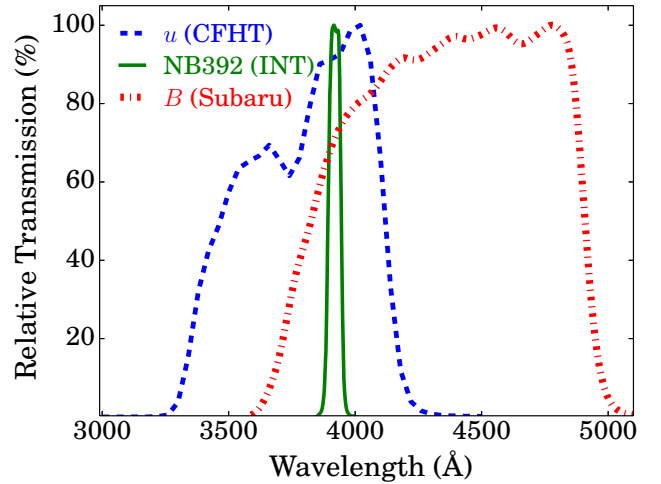}
\caption{The transmission curves of the $u$ (CFHT), NB392 (INT) and $B$ (Subaru) filters used to identify NB392 emitters. We use these 3 filters for the selection of emitters and to measure emission line fluxes and equivalent widths.}
\label{Profiles_u_B}
\end{figure}

The central wavelength of the NB392 filter lies between the $u$ and $B$ bands in the bluest part of the optical (see Figure \ref{Profiles_u_B}), and thus we use both bands to estimate the continuum. We start by PSF matching $u$ and $B$ to NB392 \citep[data from CFHT and Subaru; for full details see][]{Matthee2016}. We use bright unsaturated stars convolved with a Gaussian kernel to the same FWHM \citep[for full details, see][]{Matthee2016}.

%%%%%%%%%%%%%%%%%%%%%%%%%%%%%%%%%%%%%%%
%
% Figure 3 -  Filter profiles: u, NB, B
%
%%%%%%%%%%%%%%%%%%%%%%%%%%%%%%%%%%%%%%%
%%

In principle one could simply use a combination of $u$ and $B$ photometry of several stars in order to calibrate the NB392 data. However, the wavelength range covered by our filter probes the strong stellar CaHK absorption feature, which can vary significantly depending on stellar type and metallicity. Thus, the blind use of stars would introduce significant problems and scatter. In order to solve this potential problem, we use galaxies with photometric redshifts between $z=0.01-1.5$ without any features in our region of interest, which provide flat, robust calibrators \citep[see][]{Matthee2016}. We assure this is the case by selecting only galaxies with a flat continuum, i.e., $u-B\approx0$ color. We then calibrate the zero-point magnitude for the NB392 data using $u$ with those flat sources in the blue as a first order calibration.

%%%%%%%%%%%%%%%%%%%%%%%%%%%%%%%%%%%
%
% Figure 4 - double panel - Showing COSMOS and UDS 
% EW cut and Sigma cuts
% Excess - showing full sample, Emitters, Lyman-alpha selected
%
%%%%%%%%%%%%%%%%%%%%%%%%%%%%%%%%%%%
\begin{figure*}
\centering
\begin{tabular}{cc}
\includegraphics[width=8cm]{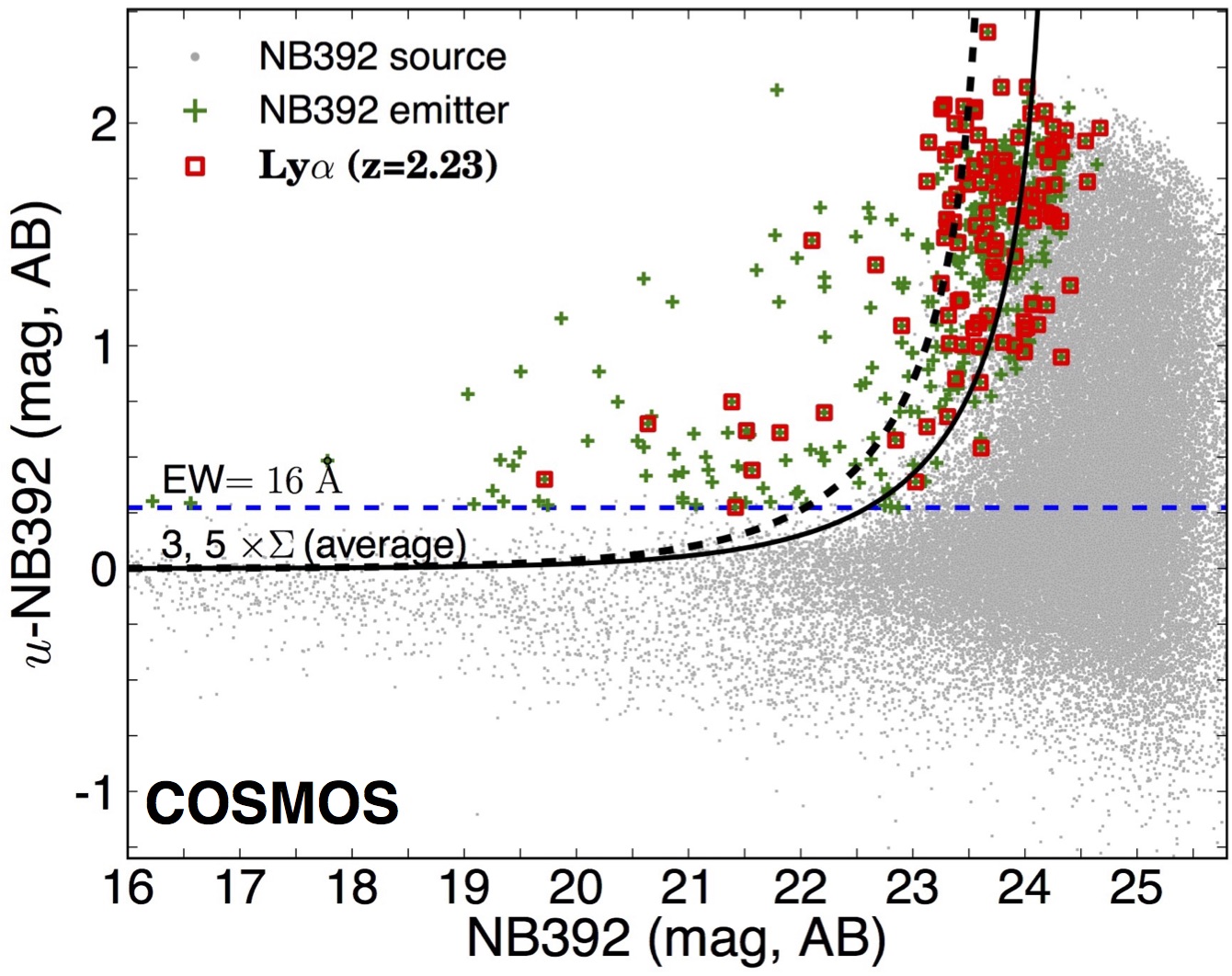}&
\includegraphics[width=8cm]{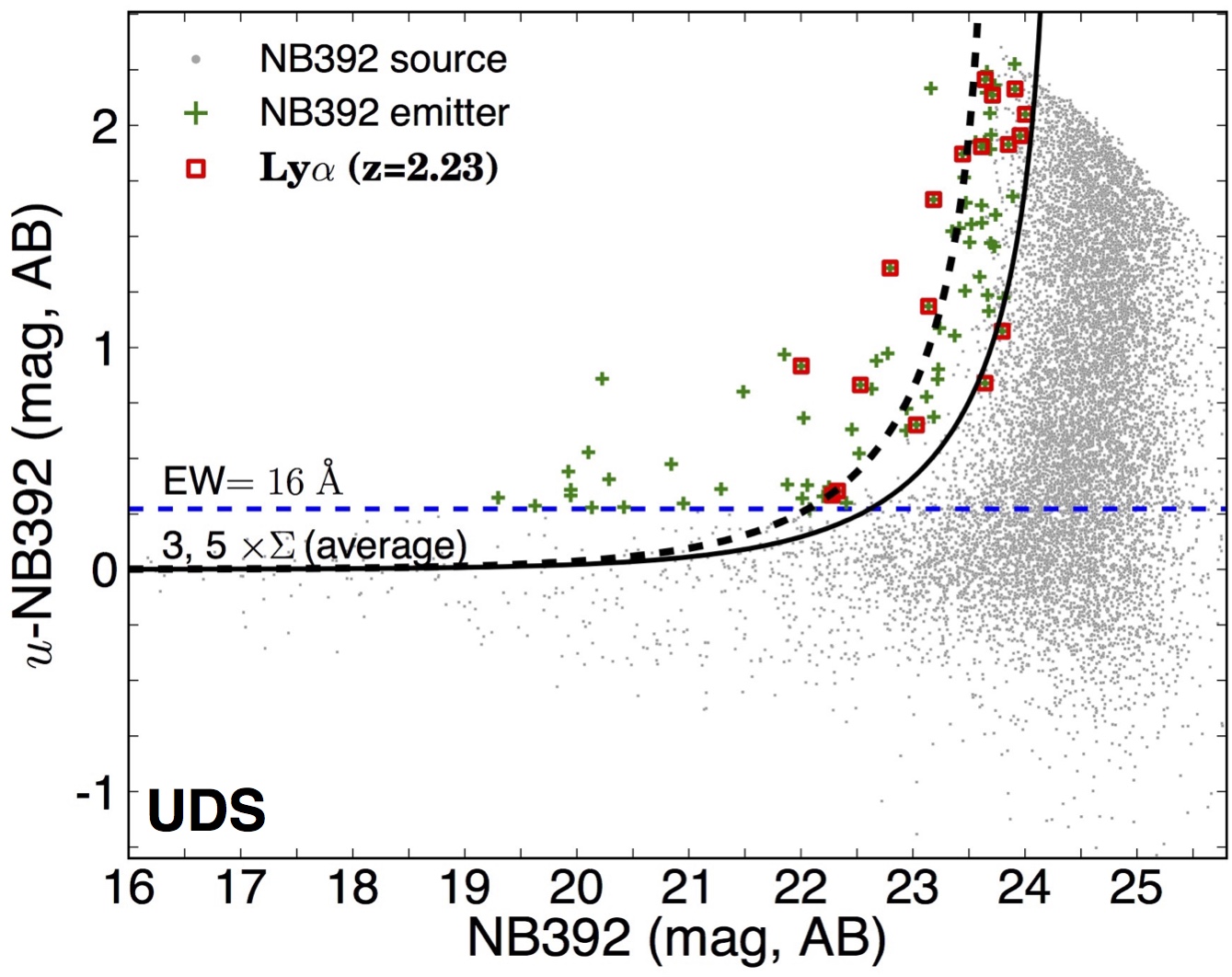}\\
\end{tabular}
\caption{{\it Left:} Selection of potential line emitters in the full COSMOS field (corresponding to about 6 INT/WFC pointings; see Figure \ref{RADEC}). We select these as sources with a significant colour excess ($\Sigma>3$) and with an observed EW$>16$\,\AA. After excluding spurious sources we find 360 potential line emitters. Note that the COSMOS field coverage contains sub-fields which are significantly deeper than others, and thus our $\Sigma$ cut in the figure is indicative only of the average depth: some regions will be deeper, while others are shallower. Our actual selection is done on a chip by chip basis. Also, note that at bright magnitudes, the prevalence of stars, with CaHK absorption features, makes many bright sources have a negative $u-$\,NB392 colour, as a result of this absorption. {\it Right:} The similar selection diagram for the UDS field, targeted with a single WFC/INT pointing (see Figure \ref{RADEC}). We apply the same selection criteria to COSMOS ($\Sigma>3$ and EW$>16$\,\AA). We find 80 candidate line emitters.}
\label{EXCESS_diag}
\end{figure*}

After calibration, we investigate the final stacked images to study their depths. We do this by placing 100,000 random 3$''$ apertures in each of the frames (resulting from combining different independent cameras per pointing). We check that the distribution peaks at 0, consistent with a very good sky subtraction. We then measure the standard deviation which we transform into a magnitude limit (1\,$\sigma$). We find that the deepest images are found in COSMOS P4, reaching M$_{392}=25.0$\,(3\,$\sigma$). The average depth over our entire COSMOS coverage is M$_{392}=24.2\pm0.4$ (3\,$\sigma$). In UDS, the average depth is similar to COSMOS, but with a lower dispersion as only one WFC pointing was obtained: M$_{392}=24.4\pm0.2$ (3\,$\sigma$). The depth of $u$ and $B$ data (PSF matched to our NB data) are 26.6 and 26.8 in COSMOS \citep[27.2 and 27.4 in their original PSF; e.g.][]{Capak2007,Muzzin2013,Santos2016} and 26.4 and 26.7 in UDS \citep[][]{Lawrence2007,Santos2016}.

By using our masks, which avoid noisy regions and pixels which are significantly contaminated by bright stars/haloes, we produce a NB392 selected catalogue. We use {\sc SExtractor} in dual mode to produce our catalogues, and thus obtain PSF matched photometry in all other bands, including $u$ and $B$, which we will use to estimate and remove the continuum and find candidate line emitters. In total, we detect 55,112 sources in COSMOS and 16,242 in UDS in our narrow-band images. All NB392 detected sources are shown in Figure \ref{RADEC}.

\subsection{Multi-wavelength catalogues and photometry}

By using the NB392 image as a detection image, we obtain $uBVgrizJHK$ photometry in COSMOS \citep[][]{Capak2007,McCracken2012} and UDS \citep[][]{Lawrence2007}. We use these excellent data for colour-colour selection in this paper, assuring we measure the photometry from all NB392 sources, even if they result in non-detections/upper limits. Furthermore, we also use publicly available catalogues of the COSMOS field \citep[][]{Ilbert2009} and the UDS field \citep[][]{Cirasuolo2010}, including a large amount of spectroscopic and photometric redshifts \citep[see also][]{Sobral2013}.

\section{NB392 and Ly$\alpha$ emitters selection}\label{Selection_emitters_LAEs}

\subsection{Excess selection: $\Sigma$ and EW cuts}\label{fluxes_EWs}

We correct for any potential dependence of excess on $u-B$ colours (see Figure \ref{Profiles_u_B}) by selecting spectroscopically confirmed galaxies which have no features at the observed 3920\,\AA. In practice, we empirically correct the NB magnitude using:
\begin{equation}
{\rm NB}392 = {\rm NB}392_{\rm uncorrected} + 0.19\times(u-B)-0.09.
\end{equation}
This correction ensures that a zero NB excess translates into a zero line-flux in NB392. For sources which are undetected in $u$ or $B$ we assign the median correction of the sources that are detected in $u$ and $B$: $+0.02$. We note that our corrections empirically tackle potential effects from IGM absorption without any uncertain model assumptions \citep[see e.g.][]{Vasei2016}; but see other studies that correct for IGM effects differently \citep[e.g.][]{Ouchi2008, Konno2016}. This is because, in general, a source with significant IGM absorption (blue-ward of Ly$\alpha$) will end up with a redder $u-B$ colour than a source with e.g. little to no IGM absorption at all. If only $u$ band was used, and significant IGM absorption happens, the total continuum flux we would measure (spread over the full $u$ filter) would be an average over the filter, and thus would be an underestimate of the real continuum flux at Ly$\alpha$. Our correction is able to correct for that.

In order to robustly select sources that have a likely emission line in the NB392 filter, including Ly$\alpha$ emitters at $z=2.23$, we need to find sources which show a real colour excess of the narrow-band (NB392) over the broad-band (in the following, we refer to the broadband $u$ as BB). This is to avoid selecting sources that may mimic such excess due to random scatter or uncertainty in the measurements. In practice, this is assured by using two different selection criteria: 

\begin{itemize}
\item a significance cut ($\Sigma>3$).
\item an equivalent width cut (EW$>16$\,\AA; $u$-NB392\,$>0.3$).
\end{itemize}

The parameter $\Sigma$ \citep[e.g.][]{Bunker1995} is used to quantify the real excess compared to an excess due to random scatter. This means that the difference between counts in the narrow-band and the broad-band must be higher than the total error times $\Sigma$. It can be computed using \citep[][]{Sobral2013}:
\begin{equation}
\Sigma=\frac{1-10^{-0.4(\rm BB-NB)}}{10^{-0.4(\rm ZP-NB)}\sqrt{(\sigma^2_{\rm NB}+\sigma^2_{\rm BB})}}.
\end{equation}

Here ZP is the zeropoint of the narrow-band (NB), NB392, which is the same as the PSF matched $u$ band data (BB); both are scaled to $ZP = 30$ in our analysis. We classify as potential emitters the sources that have $\Sigma>3$ (see Figure \ref{EXCESS_diag}), following \cite{Sobral2013}.

The second criterion for an excess source to be an emitter is that the emission line must have an observed-frame equivalent width (EW, the ratio of the line flux and the continuum flux densities) higher than the scatter at bright magnitudes. This step avoids selecting sources with highly non-uniform continua (with e.g. strong continuum features). We compute EWs by using:
\begin{equation}
EW = \Delta\lambda_{\rm NB}\frac{f_{\rm NB}-f_{\rm BB}}{f_{\rm BB}-f_{\rm NB}(\Delta\lambda_{\rm NB}/\Delta\lambda_{\rm BB})},
\end{equation}
where $\Delta\lambda_{\rm NB}=52$\,\AA \ and $\Delta\lambda_{\rm BB}=720$\,\AA \ are the widths of the filters and $f_{\rm NB}$ and $f_{\rm BB}$ are the flux densities for the narrow (NB392) and broad band ($u$), respectively. In order to identify a source as a potential line emitter we require it to have EW (observed) higher than 16\,\AA, corresponding to an excess of $u-$\,NB392\,$>0.3$ ($>3\times$ the scatter at bright magnitudes). Note that this will correspond to different rest-frame equivalent widths depending on the line/redshift being looked at. We note that specifically to select Ly$\alpha$ emitters at $z=2.23$, our EW cut corresponds to EW$_{0}>5$\,\AA, which is well below the traditional cut of EW$_{0}>25$\,\AA \ \citep[e.g.][]{Ouchi2008} for ``Ly$\alpha$ emitters". This is usually enforced by the typical narrow-band filter widths that do not allow studies to go down to lower EWs. However, this is not the case for our study as we use a narrower filter, and we thus take advantage of that to explore lower EWs.

Fluxes of all emission lines are calculated as follows:
\begin{equation} 
\label{eq:lineflux}
F_{\rm line} = \Delta\lambda_{\rm NB}\frac{f_{\rm NB}-f_{\rm BB}}{1-(\Delta\lambda_{\rm NB}/\Delta\lambda_{\rm BB})},
\end{equation}
with each parameter having been previously defined.

Using our selection criteria, out of the 55,112 NB392 sources individually detected in COSMOS, 394 emitters were selected as potential line emitters (0.7\,\%). For UDS, out of the 16,242 NB392 detections, we identify 83 candidate line emitters (0.5\,\%). However, some of these may still be artefacts and/or sources in very noisy regions. We therefore clean our list of potential emitters by visually inspecting all candidates before flagging them as final emitters and produce a final mask. This leads to a sample of 360 and 80 potential emitters in COSMOS and UDS, respectively, yielding a total of 440 candidate line emitters (see Figure \ref{EXCESS_diag}, Table \ref{catalog} and Table \ref{NB_CAT}), covering an effective area of 1.43\,deg$^2$ after our conservative masking (see Figure \ref{RADEC}).

%%%%%%%%%%%%%%%%%%%%%%%%%%%%%%%%%%
%
% Figure 5 - Double panel, combined COSMOS+UDS
%
% Distribution of Photo-zs and Spect-zs for candidate line emitters
%
%%%%%%%%%%%%%%%%%%%%%%%%%%%%%%%%%%
\begin{figure*}
\begin{tabular}{cc}
\includegraphics[width=8.3cm]{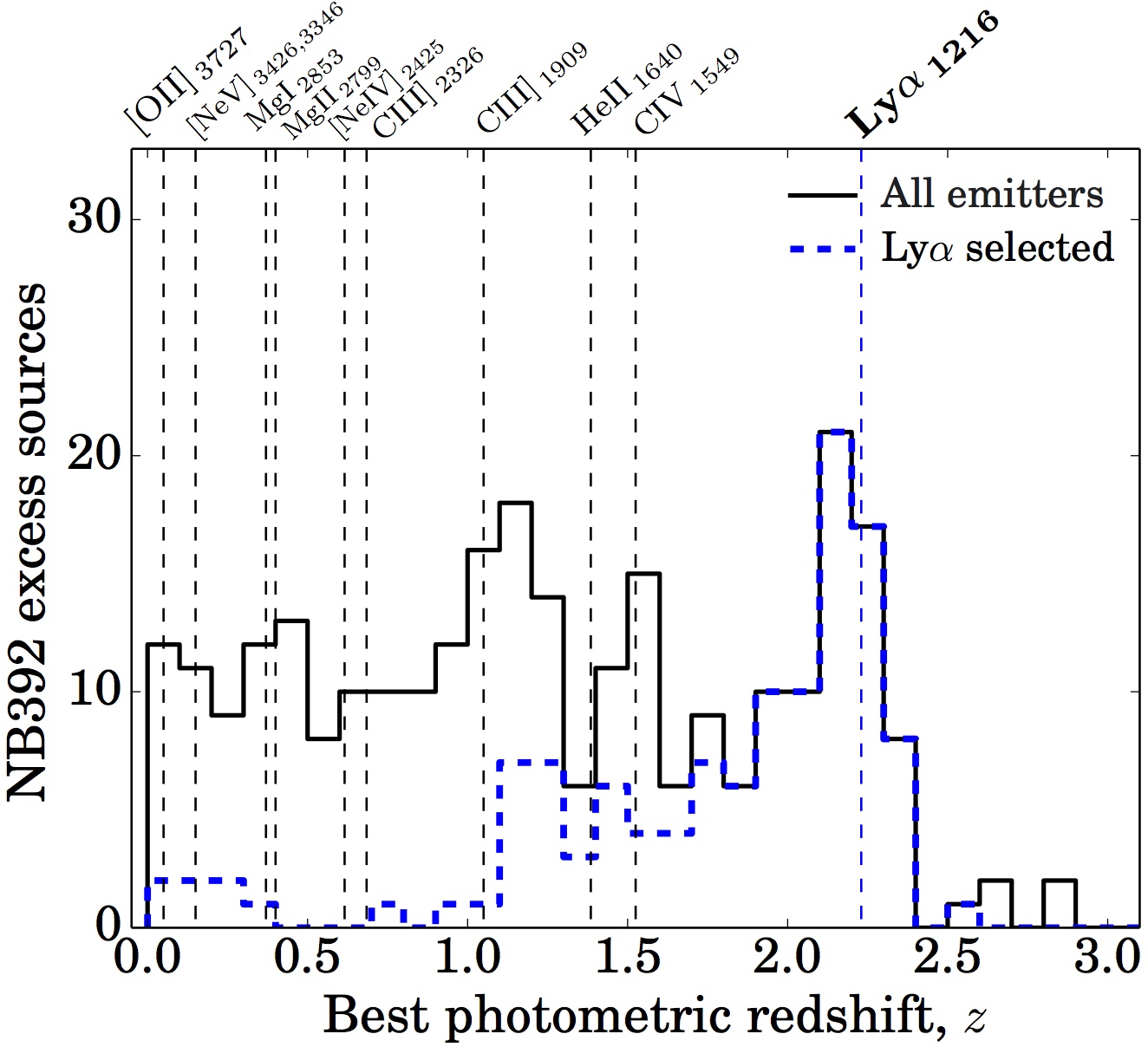}
\includegraphics[width=8.3cm]{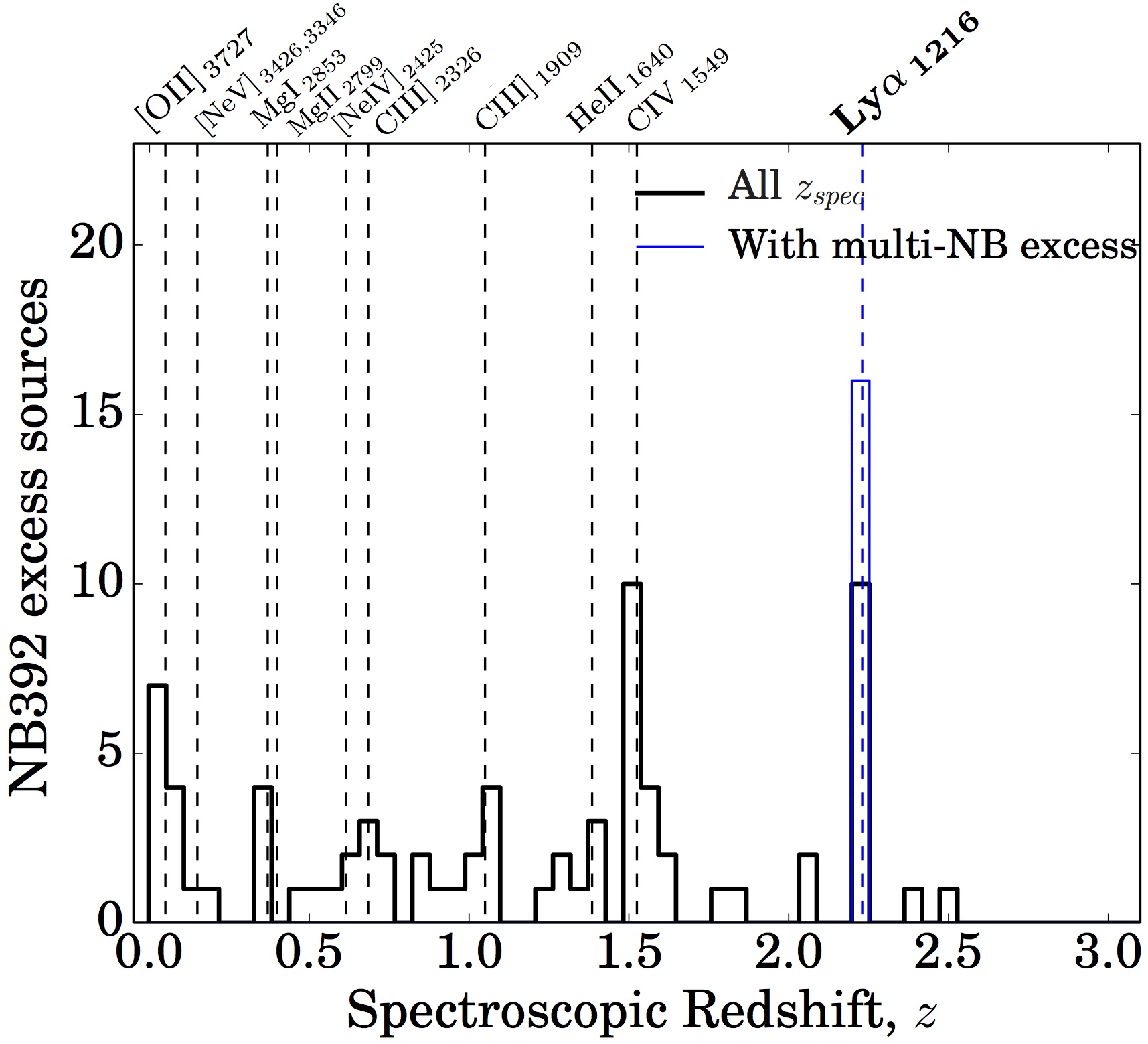}
\end{tabular} 
\caption{{\it Left:} The distribution of photometric redshifts for our candidate NB392 line emitters, indicating the redshift of major emission lines -- see Table \ref{lines_expected}. We find tentative photometric redshift peaks at the redshifts expected from major emission lines. Note that a fraction of the sources ($\sim30$\%) is too faint in the continuum to derive a photometric redshift, and thus is not shown here. For those with a photometric redshift, there is evidence that while Ly$\alpha$ emitters dominate, there is a significant population of C{\sc iv} and C{\sc iii}] emitters, followed by Mg emitters, and Ne{\sc v} + [O{\sc ii}] emitters. {\it Right:} The distribution of spectroscopic redshifts for our sample, from heterogeneous compilations and mostly $i$-band selected spectroscopic surveys. Even though the spectroscopic redshifts available from the literature are not representative of the full sample, and are highly biased towards AGN, the results agree fairly well with the photometric redshift distribution, revealing 5-10 spectroscopic confirmations of all major lines. We also show the NB392 emitters which are emitters in either NB$_J$, NB$_H$ and/or NB$_K$ \citep[from][see \S\ref{selecting_lymanalphaf}]{Sobral2013}, which can be considered as spectroscopically confirmed.}
\label{Photoz_specz_dist}
\end{figure*}

Table \ref{lines_expected} indicates the major emission lines expected to be found with our narrow-band filter. In the following sections we explore the wealth of multi-wavelength data, photometric and spectroscopic redshifts and colour-colour selections, in order to select Ly$\alpha$ emitters at $z=2.23$ (see Figure \ref{RADEC} and Table \ref{catalog}), but also to identify other emission lines. We present a catalogue with all line emitters, and those which we class as likely Ly$\alpha$ emitters in Appendix \ref{catalogue}.

%%%%%%%%%%%%%%%%%%%%%%%%%%%%%
%
% Table 2 - with major lines expected in the NB392 filter
%
%%%%%%%%%%%%%%%%%%%%%%%%%%%%%
\begin{table}
\caption{Our NB392 filter ($\lambda_c = 3918${\AA}, $\Delta\lambda = 52${\AA}) is sensitive to a range of emission lines. Here we list the most prominent (see Figure \ref{Photoz_specz_dist}, which shows these lines in comparison with photometric and spectroscopic redshifts). The redshift ($z$) range shown corresponds to the FWHM of the filter profile. We note that broad emission lines will be picked up over a larger redshift range, and that there may be other, rarer, emission lines, which may also be picked up by our survey. Also, we note that the current spectroscopy is particularly biased towards the UV bright and AGN sources. Fractions given are out of the total number of sources with a robust spectroscopic redshift.}
\begin{center}
\begin{tabular}{ccc}
\hline
Feature/line & Redshift & \# (\%) in sample \\
(rest-frame, \AA) & $z$ & (from $z_{spec}$)  \\ \hline  
[OII]\,$_{3727}$ & 0.044-0.058 & 8 (14\%)  \\ 

[NeV]\,$_{3426,3346}$ & 0.136-0.179 & 2 (4\%)  \\  
MgI\,$_{2853}$ & 0.364-0.382 & 3 (6\%) \\ 
MgII\,$_{2799}$ & 0.390-0.409 & 0 (0\%)  \\ 

[NeIV]\,$_{2425}$ & 0.605-0.626 & 2 (4\%) \\ 
CIII]\,$_{2326}$ & 0.673-0.696 & 3 (6\%) \\ 

CIII]\,$_{1909}$ & 1.039-1.066 & 6 (11\%) \\  
HeII\,$_{1640}$ & 1.373-1.405 & 4 (7\%) \\ 
CIV\,$_{1549}$ & 1.513-1.546 & 14 (25\%)  \\ 
NV\,$_{1239}$ & 2.141-2.183 & 2 (4\%)  \\ 
\hline
\bf Ly$\alpha$\,$_{\bf 1216}$ & \bf 2.201-2.243 & 10 (19\%), 17 (NB)  \\ 
\hline
\end{tabular}
\end{center}
\label{lines_expected}
\end{table}

\subsection{Photometric and spectroscopic redshifts of candidate NB392 line emitters}\label{photoz_spec_dist}

We show the photometric redshift \citep[][]{Ilbert2009,Cirasuolo2010} distributions of the candidate NB392 line emitters in Figure \ref{Photoz_specz_dist}. We have photometric redshifts for 287 out of our 440 NB392 candidate line emitters (65\%). The remaining are typically very faint in the continuum ($i>26$). We note that the photometric redshifts have been derived with a large range of models, including emission lines, AGN and also stars.

The photometric redshift distribution for the sources which we have a reliable photometric redshifts shows tentative peaks associated with strong lines expected to be detected, as detailed in Table \ref{lines_expected}, including Ly$\alpha$ at $z=2.23$, but also [O{\sc ii}]$_{3727}$, MgI\,$_{2853}$, CIII]\,$_{1909}$, HeII\,$_{1640}$ and CIV\,$_{1549}$ (see Figure \ref{Photoz_specz_dist}). The photometric redshifts hint that while the sample of emitters is dominated by Ly$\alpha$ emitters, high excitation Carbon line emitters seem to be an important population.

Spectroscopic redshifts are also available for $\sim16$\% of the selected line-emitters \citep[e.g.][]{Yamada2005, Bart_Simpson2006, Geach008,van_breu2007, Ouchi2008, Smail2008, Lilly2009,Ono2009,Civano2012,Khostovan2016,Civano2016,Sobral2016}, and we show the distribution of those redshifts, for our sample of NB392 line emitters, in Figure \ref{Photoz_specz_dist}. We note that these heterogeneous compilations of redshifts do not allow us to derive robust quantitative conclusions. This is because different spectroscopic surveys have very different selections, and in general they are biased towards the optically brighter sources and/or they result from the follow-up of AGN sources. Also, most surveys do not have the blue sensitivity to detect Ly$\alpha$ at $z\sim2$, and thus the spectroscopically confirmed Ly$\alpha$ emitters are mostly obtained through other AGN lines. Regardless, one can clearly identify the major emission lines one would expect. We find results which are consistent with the distribution of photometric redshifts.

\subsection{Selecting Ly$\alpha$ emitters at $z=2.23$}\label{selecting_lymanalphaf}

The selection of Ly$\alpha$ emitters at $z=2.23$ follows \cite{Sobral2013}, using a combination of photometric redshifts (and spectroscopic redshifts, when available) and colour-colour selections optimised for star-forming galaxies at the redshift of interest ($z\sim2.2$). We note that such selection criteria are optimised for $z\sim2.2$ independently of galaxy colour. In fact, as shown in \cite{Oteo2015}, H$\alpha$ emitters as selected in \cite{Sobral2013} span the full range of galaxy colours expected at $z=2.23$, from the bluest to the reddest galaxies. 

As can be seen in Figure \ref{Photoz_specz_dist}, the photometric redshift distribution can provide a very useful tool to select $z=2.23$ Ly$\alpha$ emitters, for relatively bright optical sources. However, photometric redshifts can be highly uncertain, and have significant systematics, particularly at $z\sim2$ and for blue sources. This is important as many Ly$\alpha$ emitters are expected to be very blue. Furthermore, photometric redshifts are not available for a significant fraction ($\sim30$\,\%) of the typically fainter NB392 emitters. Thus, relying solely on photometric redshifts would not result in a clean, high completeness sample of $z=2.23$ Ly$\alpha$ emitters. We mitigate this by following \cite{Sobral2013}, i.e., by applying colour-colour selections for the fainter NB392 emitters (see \S\ref{selection_pz_colcol}). We also discuss the selection of the faintest sources, which are undetected in the continuum in \S\ref{selection_pz_colcol}.

While spectroscopy is extremely limited for $z=2.23$ sources, double, triple and quadruple narrow-band line detections between NB392 and NB$_{\rm K}$ (H$\alpha$), NB$_{\rm H}$ ([O{\sc iii}]) and/or NB$_{\rm J}$ ([O{\sc ii}]) can be very useful if these lines are bright enough in the observed NIR \citep[][]{Sobral2013}. Those allow the identification of further 7 secure Ly$\alpha$ emitters, while they also recover 6 out of the 11 spectroscopically confirmed ones, including one source that is an emitter in all narrow-bands \citep[see][]{Matthee2016}. Overall, 13 Ly$\alpha$ emitters have information for at least another line from multi-narrow-band imaging (see Figure \ref{Photoz_specz_dist}). Note that \cite{Matthee2016} presents a larger number of Ly$\alpha$+H$\alpha$ emitters, as the study goes down to lower significance in the NB392 filter, by focusing on the H$\alpha$ emitters from \cite{Sobral2013}.

\subsubsection{Selecting continuum-undetected Ly$\alpha$ emitters} \label{undetected_cont}

We note that out of all 440 line emitters, 387 are ``selectable" ($\approx88$\,\%), i.e., we either have a photometric redshift (65\%) or $B-z$ and $z-K$ colours (88\%) that will allow us to test whether they are Ly$\alpha$ emitters in \S\ref{selection_pz_colcol}. For the remaining 53 sources (12\%) this is not possible. We investigate these 53 sources, finding that they present the lowest emission line fluxes in the sample, but, having faint or non-detectable continuum in redder bands than $u$, they have typically very high EWs (median observed EWs $\approx300$\,\AA), consistent with the majority being Ly$\alpha$ emitters at $z=2.23$ (simultaneously the only line able to produce such high EWs and the higher redshift line). For these sources we apply the canonical EW$_0>25$\,\AA \ ($z=2.23$), which selects 46 out of the 53 sources, and flag these as candidate Ly$\alpha$ emitters, including them in our sample \cite[see also][]{Rauch2008}. We note that they all have Ly$\alpha$ luminosities in the range $10^{42.5\pm0.2}$\,erg\,s$^{-1}$, and contribute to the very faintest bin in the Ly$\alpha$ luminosity function. The remaining/excluded 7 sources have lower EWs, likely explained by very low mass lower redshift emitters, such as C{\sc iii}] emitters, although we note that they can still be Ly$\alpha$ emitters (adding these 7 sources does not change any of our results).

In summary, we identify 46 sources as Ly$\alpha$ emitters out of the 53 which are not detected in broad bands.

%%%%%%%%%%%%%%%%%%%%%%%%%%%%%%%%%%%%%%
%
% Figure 6 - Colour-colour separation of emitters - BzK of NB392 emitters
%
%%%%%%%%%%%%%%%%%%%%%%%%%%%%%%%%%%%%%%
\begin{figure}
\includegraphics[width=8.3cm]{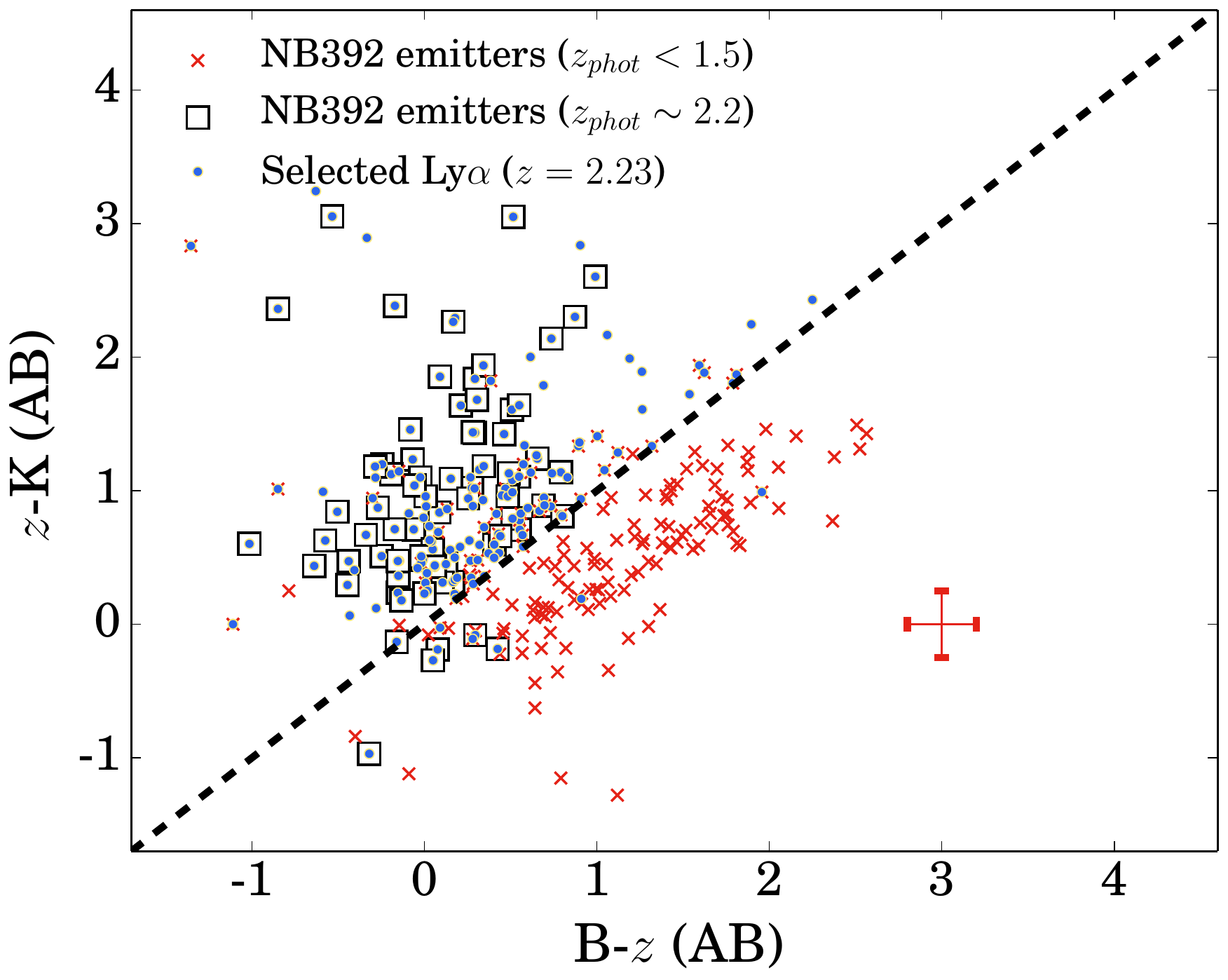}
\caption{In addition to using photometric and spectroscopic redshifts, and in order to increase our completeness, we also use the $BzK$ colour-colour selection to select Ly$\alpha$ emitters, following \citet{Sobral2013}. This allows us to select fainter line emitters for which photometric and spectroscopic redshifts are not available. Note that some real Ly$\alpha$ emitters are slightly outside the selection region, but are recovered by either spectroscopic redshifts or by dual/triple line detections; these are typically AGNs.}
\label{BzK}
\end{figure}

\subsubsection{Selecting continuum-detected Ly$\alpha$ emitters}\label{selection_pz_colcol}

The selection of Ly$\alpha$ emitters is identical for our COSMOS and UDS fields and we follow the selection criteria of \cite{Sobral2013}. An initial sample of $z=2.23$ Ly$\alpha$ emitters is obtained by selecting sources for which $1.7<z_{\rm phot}<2.8$. This selects 77 sources, of which 3 are spectroscopically confirmed to be contaminants, 4 are spectroscopically confirmed $z=2.23$ and 11 are double/triple narrow-band excess sources and thus robust $z=2.23$ Ly$\alpha$ emitters. Because some sources lack reliable photometric redshifts, the colour selection $(z-K)>(B-z)$ is used to recover additional $z\sim2$ continuum-faint emitters. This colour-colour selection is a slightly modified version of the standard $BzK$ \citep{Daddi2004} colour-colour separation \citep[see][]{Sobral2013}. It selects 70 additional Ly$\alpha$ candidates (and re-selects 73\% of those selected through photometric redshifts; four sources are contaminants, two are $z=2.23$ Ly$\alpha$ emitters), and guarantees a high completeness of the Ly$\alpha$ sample (see Figure \ref{BzK}). Finally, two spectroscopically confirmed Ly$\alpha$ sources (AGN, from C-COSMOS) are also selected, which are missed by the photometric redshift and colour-colour selection due to the unusual colours (these are also double/triple narrow-band excess sources). $BzK$ also selects much higher redshift sources, which can be a source of contamination for the H$\alpha$ selection at $z=2.23$ with the NB$_{K}$ filter \citep[e.g. Oxygen lines, see][]{Sobral2013}. This is not a problem for NB392, as no strong emission lines make it into the filter at wavelengths blue-wards of Ly$\alpha$.

Overall, we identify 142 Ly$\alpha$ emitters (see Table \ref{catalog}) which are directly selected, along with the other 46 candidate Ly$\alpha$ emitters that are very faint and/or undetected in the continuum. Our final sample is thus made of 188 Ly$\alpha$ emitters.

With the limited spectroscopy available, it is difficult to accurately determine the completeness and contamination of the sample. However, based on the double/triple narrow-band excess detections and spectroscopically confirmed Ly$\alpha$ emitters (15 are selected out of a total of 17) we infer a likely completeness of $\approx90$\,\%. Of all of the sources initially selected as Ly$\alpha$ emitters ($\sim60$\% of NB392 excess sources are not selected as Ly$\alpha$ emitters). Amongst these, 7 were contaminants (now removed), dominated by C{\sc iv} and C{\sc iii}] emitters. As discussed above, there are reasons to suspect that a larger fraction of the contaminants will have available redshifts (e.g. AGN), and thus we estimate a contamination of between about 5 and 10\%.

%%%%%%%%%%%%%%%%%%%%%%%%%%%%%%%%%%
%
% Table 3 - Number of sub-samples and confirmed emitters
%
%%%%%%%%%%%%%%%%%%%%%%%%%%%%%%%%%%
\begin{table}
\caption{Summary of the information in our CALYMHA catalog. When available, the number of spectroscopic redshifts are shown as well. The number of Ly$\alpha$ emitters within brackets are those with high S/N continuum detections, allowing to be robustly selected using either photometric redshifts or colour-colour selections. We provide the catalogue of all 440 line emitters in Appendix \ref{catalogue}. Out of our 188 Ly$\alpha$ emitters, 13 have a robust detection of either [O{\sc ii}], [O{\sc iii}] or H$\alpha$. See also \citet{Matthee2016} for discussion of Ly$\alpha$ properties of 17 H$\alpha$ emitters recovered down to lower $\Sigma$ in our NB392 data.}
\begin{center}
\begin{tabular}{ccc}
\hline
Sample & No. of sources & $z$-spec\\ \hline
NB392 detections: COSMOS & 55,112 & 5683 \\ 
NB392 detections: UDS & 16,242 & 801 \\  \hline
Emitters (before visual check) & 477 & 70  \\  
Emitters (after visual check) & 440 & 70  \\  
\hline
Ly$\alpha$ emitters $z = 2.23$ & 188 (142) & 17 \\
\hline
\end{tabular}
\end{center}
\label{catalog}
\end{table}

\section{Methods and corrections} \label{methods}

\subsection{Ly$\alpha$ luminosity function calculation} 

\subsubsection{Completeness corrections} \label{compl}

Faint sources and those with weak emission lines and/or low EW might be missed in our selection and thus not included in the sample and/or in a particular sub-volume within our survey. The combination of such effects will result in the underestimation of the number of Ly$\alpha$ emitters, especially at lower luminosities. In order to account for that we follow the method described in \citet{Sobral2013} to estimate completeness corrections per sub-field per emission line.

Very briefly, we use sources which have not been selected as line emitters ($\Sigma<3$ or EW$<16$\, {\AA}), but that satisfy the selection criteria used to select Ly$\alpha$ (photometric and colour-colour selection). We then add emission-line flux to all those sources, and study the recovery fraction as a function of input flux. We do these simulations in a sub-field by sub-field basis. We then apply those corrections in order to obtain our completeness-corrected luminosity functions. We note that in order to deal with the significant differences in depth across our survey areas, and in order to produce robust results, when evaluating the Ly$\alpha$ luminosity function, we only take into account sub-volumes (per chip) if, for that bin, they are complete at a $>50$\% level.

\subsubsection{NB392 filter profile corrections} \label{profil_filt}

The NB392 filter transmission function is not a perfect top-hat (see Figure \ref{Profiles}). Therefore, the real volume surveyed is a weak function of intrinsic luminosity. This is a much stronger effect for filters which are much more gaussian, such as the NB$_K$ filter (see Figure \ref{Profiles}). For example, luminous line emitters will be detectable over a larger volume (even though they will seem fainter) than the fainter ones, as they can be detected in the wings of the filter. Conversely, genuine low luminosity sources will only be detectable in the central regions of the filter, leading to a smaller effective volume. In order to correct for this when deriving luminosity functions, we follow the method described in \citet{Sobral2012}. Briefly, we compute the luminosity function assuming a top-hat narrow-band filter. We then generate a set of $10^{6}$ line emitters with a flux distribution given by the measured luminosity function, but spread evenly over the redshift range being studied (assuming no cosmic structure variation or evolution of the luminosity function over this narrow redshift range). We fold the fake line emitters through the top-hat filter model to confirm that we recover the input luminosity function perfectly. Next, we fold the fake line emitters through the real narrow-band profiles -- their measured flux is not only a function of their real flux, but also of the transmission of the narrow-band filter for their redshift. The simulations show that the number of brighter sources is underestimated relative to the fainter sources. A mean correction factor between the input luminosity function and the one recovered (as a function of luminosity) was then used to correct each bin. In practice, the corrections range from a factor of 0.97 in the faintest bin to 1.3 in the brightest bin.

\subsection{NB392/NB$_K$ filter profile ratios: corrections in measuring Ly$\alpha$/H$\alpha$ ratios}\label{biases_profiles}

As we will compare Ly$\alpha$ and H$\alpha$ directly to obtain line ratios, we derive corrections due to the use of the specific filter profiles. By design, our sample of Ly$\alpha$ emitters have their H$\alpha$ emission in the HiZELS NB$_K$ filter (see Figure \ref{Profiles}). Therefore, it is possible to measure Ly$\alpha$/H$\alpha$ ratios directly. However, the slightly different filter transmission and velocity offsets  between H$\alpha$ and Ly$\alpha$ can introduce biases (see Fig. \ref{Profiles} and discussion in \citealt{Matthee2016}).

We obtain the average relative transmission between Ly$\alpha$ and H$\alpha$ for Ly$\alpha$ selected sources similarly as described in \cite{Matthee2016} (see also e.g. \citealt{Nakajima2012}). We simulate 100,000 Ly$\alpha$ emitters with a redshift probability distribution given by the NB392 filter transmission, as our sample is NB392 (Ly$\alpha$) selected. Note that in \cite{Matthee2016} the sample is NB$_K$ (H$\alpha$) selected, and thus the redshift probability distribution is given by the NB$_K$ filter, leading to different filter corrections. Assuming a dispersion of velocity offsets with a median of 200 km\,s$^{-1}$ \citep[e.g.][]{Steidel2010,Stark2013,Hashimoto2013,Erb2014,Song2014,Sobral2015CR7}, we measure the transmission for the redshifted H$\alpha$ line in the NB$_K$ filter and thus obtain the relative transmission between Ly$\alpha$ and H$\alpha$. 
We find that the Ly$\alpha$ transmission is on average $\approx1.7$ times higher than H$\alpha$ (see Figure \ref{Profiles}), due to the more top-hat-like shape of the NB392 filter as compared to the NB$_K$ filter; i.e. many Ly$\alpha$ emitters (Ly$\alpha$ selected) are observed in the wings of the NB$_K$ filter. We correct for this relative transmission in all our measurements of the Ly$\alpha$ escape fraction, f$_{\rm esc}$. This is a robust correction as long as our Ly$\alpha$ sample has a redshift distribution given by the NB392 filter profile.

We show how the measured line ratio changes as a function of redshift in Figure \ref{Profiles}. We note that the overestimation of the Ly$\alpha$/H$\alpha$ ratio, for a Ly$\alpha$-selected sample, is particularly high towards the wings of the filter and is very uncertain on a source by source basis. Therefore, for the remainder of this paper, we only use Ly$\alpha$/H$\alpha$ ratios obtained by stacking either the full sample of Ly$\alpha$ emitters, or sub-samples, and apply the statistical correction we derive, by dividing observed Ly$\alpha$/H$\alpha$ ratios by 1.7.

\subsection{Stacking and Ly$\alpha$ escape fraction from Ly$\alpha$/H$\alpha$}\label{corrections_caseB}

The observed fraction of Ly$\alpha$ to H$\alpha$ flux encodes information on the fraction of Ly$\alpha$ photons that escape a galaxy, f$_{\rm esc}$. Under the assumption of Case B recombination, a temperature of $T\approx10^4$K and electron density of $n_e \approx 350$ cm$^{-3}$, the intrinsic ratio of Ly$\alpha$ to H$\alpha$ photons is expected to be 8.7 (see e.g. \citealt{Hayes2015} for a recent review and for a discussion on how sensitive this number is to a range of physical conditions). The departure of this ratio is defined as the Ly$\alpha$ escape fraction, f$_{\rm esc} ={\rm L}_{\rm Ly\alpha}/(8.7 {\rm L_{\rm H\alpha}})$, where L$_{\rm H\alpha}$ is corrected for dust attenuation. 

We measure the median f$_{\rm esc}$ of our sample of Ly$\alpha$ emitters by stacking the PSF-matched $U$, $B$, NB392, NB$_K$ and $K$ images on the positions of Ly$\alpha$ emitters, following the same methodology as in \cite{Matthee2016}. Photometry is measured in 3\,$''$ diameter apertures and line fluxes are computed as described in \S\ref{fluxes_EWs}. We correct for dust extinction/dust affecting the H$\alpha$ line by using the median extinction A$_{H\alpha} = 0.9$ \citep[see e.g.][]{Sobral2012,Ibar2013,Sobral2013,Matthee2016} and correct the observed Ly$\alpha$/H$\alpha$ ratio for the relative filter transmission, as described in \S\ref{biases_profiles}.

%%%%%%%%%%%%%%%%%%%%%%%%%%%%%%%%%%%%%%%%%%
%
% Figure 7
%
% Lyman-alpha LF at z=2.23 - with comparison with other z=2.23 studies and scaled H-alpha
%
%%%%%%%%%%%%%%%%%%%%%%%%%%%%%%%%%%%%%%%%%%
\begin{figure*}
\includegraphics[width=12cm]{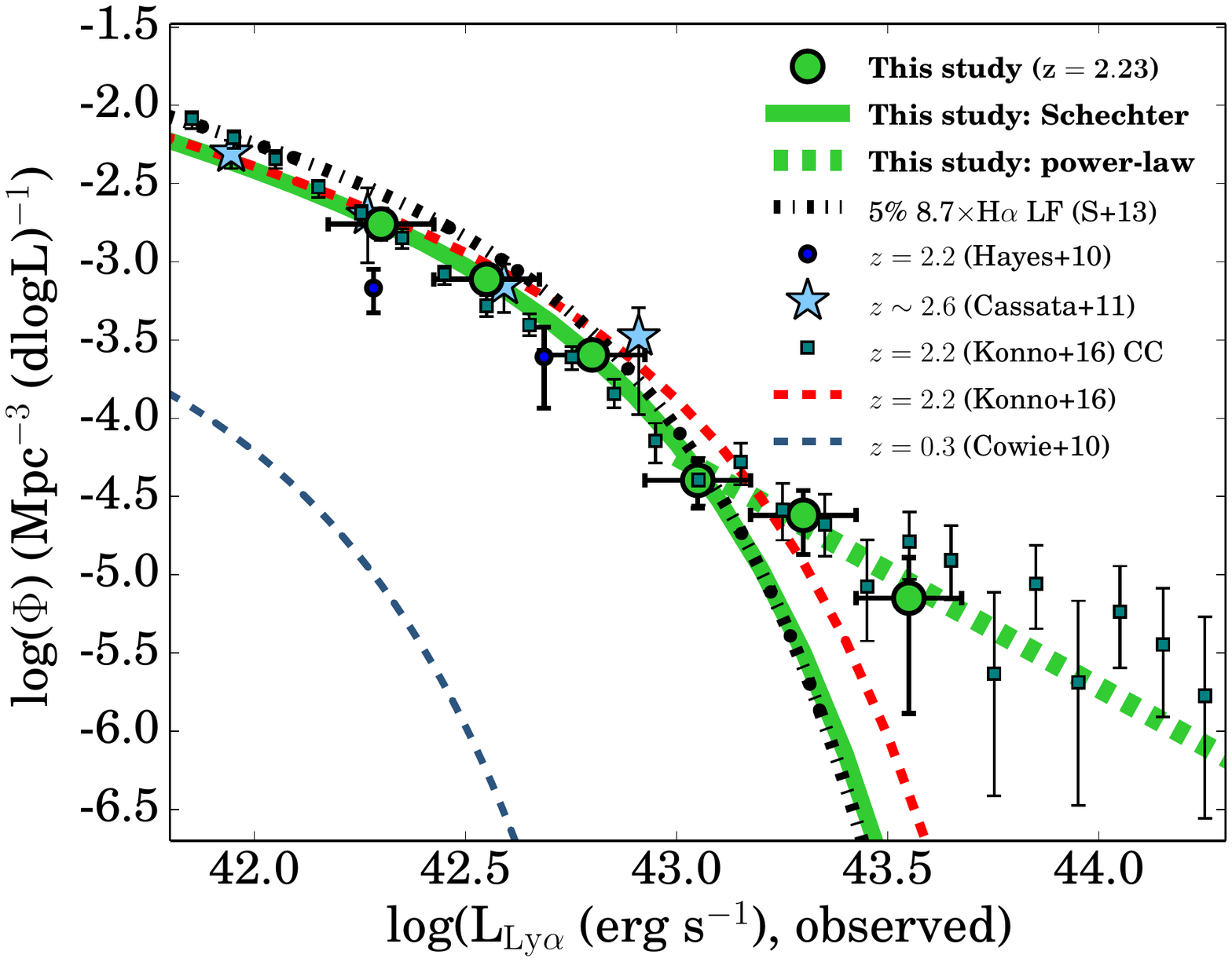}
\caption{The Ly$\alpha$ luminosity function for our combined COSMOS and UDS coverage and down to a Ly$\alpha$ EW$_0>5$\,\AA. We find that the LF is well fitted by a Schechter function up to $\sim10^{43}$\,erg\,s$^{-1}$, but seems to become a power-law for higher luminosities. We also show the Ly$\alpha$ luminosity function presented by \citet{Hayes2010} at $z=2.2$, \citet{Cassata2011} at $z\sim2.6$ and the recent determination at $z=2.2$ by \citet{Konno2016} before (dashed red line and Figure \ref{LF_Lya_KONNO_comparison}) and after our contamination correction (CC; see \S\ref{contam_LF}). We find good agreement with the wide and deep survey of \citet{Konno2016}, including the departure from the Schechter function. We are also in very good agreement with \citet{Cassata2011}. While it may seem that we are in disagreement with \citet{Hayes2010}, we note that their data-points, due to probing a very deep, but very small single volume, only overlap with the faintest of our two bins, and there is likely to be strong cosmic variance in their survey. We also show the extinction-corrected H$\alpha$ luminosity function from \citet{Sobral2013}, transformed into Ly$\alpha$ with a 5\% escape fraction.}
\label{LF_Lya}
\end{figure*}

\section{Results}\label{results}

\subsection{Ly$\alpha$ luminosity function at $\bf z=2.23$: comparison to other surveys and evolution}

We estimate source densities in a luminosity bin of width $\Delta(\log L)$ centred on $\log L_c$ by obtaining the sum of the inverse volumes of all the sources in that bin, after correcting for completeness. The volume probed is calculated taking into account the survey area and the narrow-band filter width, followed by applying the appropriate real filter profile corrections obtained in \S\ref{profil_filt}.

The luminosity functions presented here are fitted with Schechter functions defined by three parameters: $\alpha$ (the faint-end slope), $L^*$ (the transition between a power law at lower luminosities and an exponential decline at higher luminosities) and $\phi ^*$ (the number density/normalisation at $L^*$). We can still get a reasonable constraint on $\alpha$, but we also fit the luminosity function by fixing $\alpha$ to common values found in the literature \citep[$\alpha=-1.5,-1.7$, e.g.][]{Ouchi2008,Hayes2010,Konno2016}, particularly so we can make a direct comparison. Finally, we also explore power law fits with the form: $\log_{10}\phi=A\times \log_{10}(\rm L)+B$.

We present our final $z=2.23$ Ly$\alpha$ luminosity function in Figure \ref{LF_Lya} and in Table \ref{LF_funccc}. We find it to be well fit by a Schechter function up to 10$^{43.0}$\,erg\,s$^{-1}$. Our best-fit parameters for L\,$<10^{43.0}$\, erg\,s$^{-1}$ are:

\medskip
\noindent $\log L^*_{\rm Ly\alpha}=42.59^{+0.16}_{-0.08}$\,erg\,s$^{-1}$

\smallskip
\noindent $\log\phi^*_{\rm Ly\alpha}=-3.09^{+0.14}_{-0.34}$\,Mpc$^{-3}$

\smallskip
\noindent $\alpha_{\rm Ly\alpha}=-1.75\pm0.25$.
\smallskip

Our results favour a steep $\alpha$ for the Ly$\alpha$ luminosity function at $z=2.23$ ($\alpha\approx-1.8$), in very good agreement with \cite{Konno2016}. Beyond 10$^{43.0}$\,erg\,s$^{-1}$ we find evidence of a significant deviation from a Schechter function, similarly to what was found by \cite{Ouchi2008} and \cite{Konno2016}. We thus fit a power law ($\log_{10}\phi=A\times \log_{10}(\rm L)+B$), with parameters $\rm A=-1.48$ and $\rm B=59.4$. We show our results and the best fits in Figure \ref{LF_Lya}. We also attempt to fit a single power-law to our full Ly$\alpha$ luminosity function. The best fit yields a reduced $\chi^2=1.4$ with $\rm A=-1.9\pm0.2$ and $\rm B=79\pm6$.

We compare our results with other studies at $z=2.23$ \citep[e.g.][]{Hayes2010,Konno2016}. We correct the \cite{Konno2016} data-points for potential contamination (particularly important at the bright end; see \S\ref{contam_LF}), but we also show the Schechter fit derived without such corrections; see Figure \ref{LF_Lya}. We find very good agreement with \cite{Konno2016} across most luminosities regardless of the contamination correction, but after such correction our results agree at all luminosities. We find a higher number density of Ly$\alpha$ emitters at comparable luminosities than \cite{Hayes2010}, but we note that we probe a significantly larger volume ($\approx150$ times larger), and thus cosmic variance is likely able to explain the apparent discrepancies \citep[$\phi^*$ expected to vary by more than a factor of 2 for surveys of the size of theirs; see][]{Sobral2015}. 

We also compare our results with other previous determinations presented in the literature at slightly different redshifts \citep[e.g.][]{Cassata2011,Blanc2011,Ciardullo2012,Ciardullo2014}, finding good agreement. Other studies have made contributions towards unveiling the Ly$\alpha$ luminosity function at $z<2$ \citep[see e.g.][]{Cowie2010,Barger2012}. Comparing to these, we find a very strong evolution in the Ly$\alpha$ luminosity function from $z=0.3$ to $z=2.23$. For $\alpha=-1.6$, the characteristic luminosity evolves by almost 1\,dex from $z=0.3$ to $z=2.23$, a very similar behaviour to the evolution of $L^*$ of the H$\alpha$ luminosity function \citep[][]{Sobral2013}. $\phi^*$ evolves by about 0.8\,dex, thus much more than the mild $\sim0.2-0.3$\,dex evolution seen for the H$\alpha$ luminosity function \citep[][]{Sobral2013}.

Comparing our results with higher redshift \citep[e.g.][]{Ouchi2008,Ouchi2010,Matthee2015,Santos2016,Drake2016}, we find that the Ly$\alpha$ luminosity function continues to evolve at least up to $z=3.1$. We note that issues with contamination and/or completeness, due to the simple EW cut usually used may play an important role at $z\sim3$ and at higher redshift, although it is expected to be less important than at $z\sim2$.

We note that the bright-end power-law component of the Ly$\alpha$ luminosity function is consistent with being dominated by luminous X-ray AGN. We can conclude this because 10 out of the 12 ($83\pm36$\%) Ly$\alpha$ emitters with $L>10^{43}$\,erg\,s$^{-1}$ are detected in {\it Chandra}/X-rays with luminosities in excess of $\approx10^{43.5}$\,erg\,s$^{-1}$ \citep[][]{Civano2016}. We note that while these sources have significant Lyman-breaks, and all are X-ray sources, two of our Ly$\alpha$ emitters are also candidates for being strong Lyman continuum (LyC) leakers \citep[][]{Matthee2016b}. This is consistent with the potential connection between the escape of Ly$\alpha$ and LyC photons \citep[see e.g.][]{Verhamme2015,Verhamme2016,DijkstraLyaLyC2016,Vanzella2016}.

\subsection{Ly$\alpha$ luminosity function: how important is it to remove contaminants?}\label{contam_LF}

\begin{figure}
\begin{tabular}{cccc}
\includegraphics[width=8.3cm]{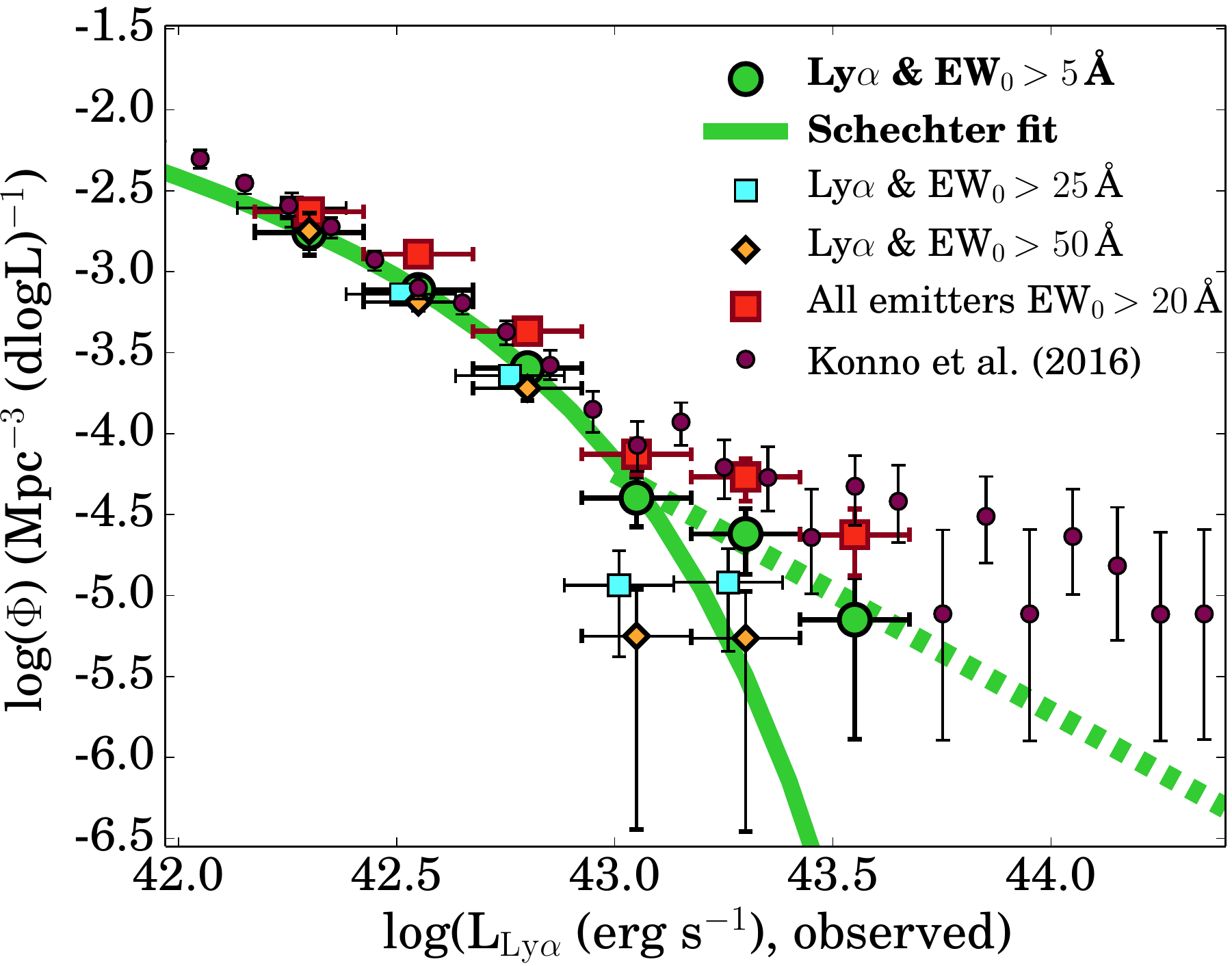}
\end{tabular} 
\caption{The Ly$\alpha$ luminosity function for our combined COSMOS and UDS coverages down to a Ly$\alpha$ EW$_0>5$\,\AA. We compare with what we would obtain by not removing contaminants, but instead applying only a higher EW cut (EW$_0>20$\,\AA), to directly compare with \citet{Konno2016}. We find that we can fully recover the results of \citealt{Konno2016}, including a much higher number density of very bright sources. However, as our spectroscopic (we have spectroscopic redshifts for 50\% of all $>10^{43}$\,erg\,s$^{-1}$ line emitters) and photometric redshift analysis shows, this is driven by the presence of C{\sc iii}] and C{\sc iv} emitters. We also investigate and show the effect of varying the Ly$\alpha$ EW$_0$ cut in addition to our robust Ly$\alpha$ selection (redshifts and colour-colour selection). For different EW cuts, we re-compute all our completeness corrections per field to take into account that our selection changes (a higher EW cut means a lower completeness, so our completeness corrections increase). We find that completeness corrections can compensate for incompleteness at the faint end, but the bright end becomes significantly incomplete for higher EW cuts.}
\label{LF_Lya_KONNO_comparison}
\end{figure}

We have presented the Ly$\alpha$ luminosity function at $z=2.23$ with our robust Ly$\alpha$ selected sample (see Figures \ref{LF_Lya_KONNO_comparison} and \ref{LF_Lya}), which goes down to EW$_0\approx5$\,\AA. We stress that for the highest Ly$\alpha$ luminosities ($>10^{43}$\,erg\,s$^{-1}$), we have spectroscopic redshifts for 50\% of all line emitters. We now investigate the role of selecting Ly$\alpha$ among all narrow-band emitters (see Figure \ref{LF_Lya_KONNO_comparison}). This is particularly relevant as most studies until now have made the assumption that contamination from other lines should be negligible. We have already showed how important it actually is in practice when we presented the distribution of photometric and spectroscopic redshifts in \S\ref{photoz_spec_dist}, but here we place that into the context of deriving Ly$\alpha$ luminosity functions. This may be particularly relevant to understand and discuss significant differences in results with other studies.

% Fig 8
%
%%%%%%%%%%%%%%%%%%%%%%%%%%%%%%%%%%%%%%%%%%%%%%%
%
% Lyman-alpha LF at z=2.23 - importance of selection
%
%%%%%%%%%%%%%%%%%%%%%%%%%%%%%%%%%%%%%%%%%%%%%%%

In order to address this issue, we compare our most robust results, after carefully selecting Ly$\alpha$ emitters (and using the wealth of spectroscopic redshifts available), with those we would have derived if we assumed that the sample was dominated by Ly$\alpha$ emitters (as long as we apply a particular EW cut). We show the results in Figure \ref{LF_Lya_KONNO_comparison}. It is particularly interesting to compare the results from a recent study, that also targeted COSMOS and UDS, with a slightly different filter \citep[][]{Konno2016}. The crucial difference between our study and \cite{Konno2016} is that we use spectroscopic and photometric redshifts, colour-colour selections and take advantage of dual/triple and quadruple narrow-band detections for other emission lines. We thus obtain a very robust sample of Ly$\alpha$ emitters, and exclude confirmed and very likely contaminants. As presented in \S\ref{photoz_spec_dist}, down to the flux limit of our study, around $\approx50$\% of the emitters are likely not Ly$\alpha$, with the bulk of them being C{\sc iii}] and C{\sc iv}, not [O{\sc ii}]. However, \cite{Konno2016} assume that all narrow-band excess sources above a certain EW correspond to Ly$\alpha$. While such assumption may work relatively well for very low fluxes, it breaks down at the highest fluxes, as our spectroscopic results show.

In order to compare our results, we apply the EW$_0$ cut (EW$_0>20$\,\AA) of \cite{Konno2016}, and no other selection criteria. Based on our spectroscopic redshifts (dominated by sources with fluxes corresponding to L$_{\rm Ly\alpha}>10^{43}$\,erg\,s$^{-1}$), this results in a highly contaminated sample at the bright end (16 confirmed contaminants out of 21 sources with spectroscopy; 76\% contamination), whilst being relatively incomplete for bright Ly$\alpha$ emitters: only 5 spectroscopically confirmed Ly$\alpha$ emitters are recovered out of the 11 (completeness $\sim45$\%).

We can now derive a new luminosity function, fully comparable with \cite{Konno2016}, which we show in Figure \ref{LF_Lya_KONNO_comparison}. Our results show a remarkable agreement at all luminosities, and we recover the much higher number density of very luminous sources. We also confirm that those additional sources are all X-ray sources, but we check that the vast majority are spectroscopically confirmed C{\sc iii}] and C{\sc iv} emitters. We note that since {\it GALEX} data are also available it is relatively easy to identify C{\sc iii}] and C{\sc iv} emitters, as they will have Lyman-breaks at shorter wavelength than Ly$\alpha$ emitters, even if spectroscopic redshifts are not available.

Only spectroscopic follow-up can completely establish the exact shape of the bright end of the Ly$\alpha$ luminosity function (for the remaining 50\% of the sources spectroscopic redshifts are not currently available). We have already followed-up further two of the bright line-emitters with XSHOOTER on the VLT in October 2016 without any Ly$\alpha$ pre-selection, confirming a N{\sc v}\,$_{1239}$ emitter (with broad Ly$\alpha$) at $z=2.15$, and one Ly$\alpha$ emitter at $z=2.2088$, in line with our expectations of relatively high contamination. These source will be presented in a future paper, together with the rest of the on-going follow-up on the VLT. Nevertheless, we can already conclude that it is crucial to remove contaminants, even for surveys in the bluest optical bands like ours. Our ``Ly$\alpha$" luminosity function obtained by using all NB392 emitters can also be seen as a strong upper limit for the real Ly$\alpha$ luminosity function, as it already contains a significant number of confirmed contaminants, which become more and more significant at the highest luminosities. As our data allow us to derive contamination fractions per bin, we compute them and apply them to \cite{Konno2016}, to derive a Ly$\alpha$ luminosity function which is fully comparable to ours. We show the results in Figure \ref{LF_Lya}. The contamination corrections ($CC$) to $\log(\Phi)$ we derive are well described as a function of Ly$\alpha$ luminosity: $CC=-0.28{\rm L_{Ly\alpha}}+11.732$ for L$_{\rm Ly\alpha}\approx10^{42-44.5}$\,erg\,s$^{-1}$. We note that if one fits the Ly$\alpha$ luminosity function with a Schechter function up to L$_{\rm Ly\alpha}\sim10^{43}$\,erg\,s$^{-1}$ the contamination effect is still relatively small with $\log L^*_{\rm Ly\alpha}$ being overestimated by $\approx0.15$\,dex and $\log\phi^*_{\rm Ly\alpha}$ being underestimated (as a consequence of the change in L$^*$) by $\approx0.1$\,dex. However, contamination plays a major role for the highest luminosities and for determining the apparent power-law component of the Ly$\alpha$ luminosity function.

%%%%%%%%%%%%%%%%%%%%%%%
%
% Figure 9
%
% Ew distribution of sources at z=2.23
%
%%%%%%%%%%%%%%%%%%%%%%%
\begin{figure}
\includegraphics[width=8.3cm]{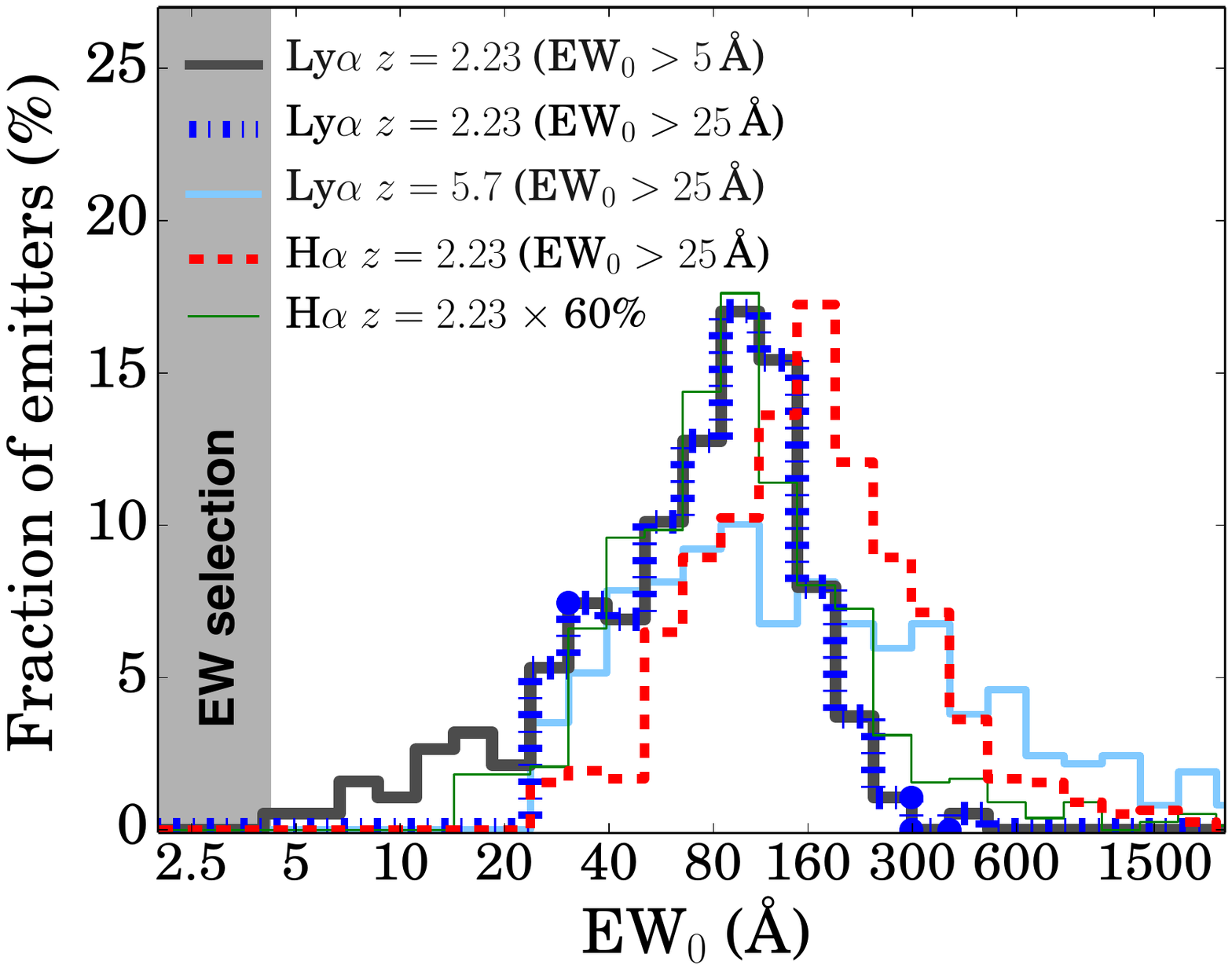}
\caption{The rest-frame EW distribution of Ly$\alpha$ selected emitters at $z=2.23$. We find an average EW$_0=85\pm57$\,\AA, with a median of $\approx100$\,\AA. We find that 11\% of all Ly$\alpha$ emitters have $5<$\,EW$_0<25$\,\AA, but that the lower EW$_0$ Ly$\alpha$ emitters are preferentially the brightest in Ly$\alpha$ luminosity, and are particularly important for the bright end of the Ly$\alpha$ luminosity function at $z=2.23$ (X-ray AGN). We also show the H$\alpha$ EW$_0$ distribution of H$\alpha$ emitters from \citet{Sobral2014}, from roughly the same volume surveyed with Ly$\alpha$. This clearly shows that the rest-frame EW distribution of H$\alpha$ is shifted to higher values, but scaling them by 60\% recovers a similar distribution. For comparison at higher redshift, but avoiding potential re-ionisation effects, we also show the EW$_0$ of Ly$\alpha$ emitters at $z=5.7$ from \citet{Santos2016}, clearly showing evolution not only in the average, but even more so on the spread, revealing very high EWs that simply are not seen at the peak of star-formation history.}
\label{EW_dist}
\end{figure}

\subsection{The EW distribution of Ly$\alpha$ emitters at $\bf z=2.23$ and implications for the Ly$\alpha$ luminosity function}

As discussed in \S\ref{contam_LF}, the choice of Ly$\alpha$ rest-frame EW cut may have important effects in conclusions regarding the nature of Ly$\alpha$ emitters. Traditionally, due to the FWHM of typical narrow-band filters, and particularly due to the early difficulty in applying colour-colour and/or photometric redshift selections to differentiate between Ly$\alpha$ and other line emitters\footnote{This becomes more problematic for higher redshift Ly$\alpha$ surveys, as Ly$\alpha$ emitters become a progressively lower fraction of the full sample of emitters; see e.g. \cite{Matthee2014} or \cite{Matthee2015}}, a relatively high EW cut was used. This assured that lower redshift emitters would be excluded. The typical value for this cut has been EW$_0\sim25$\,\AA.

As we are able to probe down to a Ly$\alpha$ rest-frame EW of 5\,\AA, we have the opportunity to investigate how complete samples with higher rest-frame EW cuts may be and what is the effect on e.g. the Ly$\alpha$ luminosity function. Figure \ref{EW_dist} shows the distribution of Ly$\alpha$ rest-frame EWs at $z=2.23$. We find that the median EW$_0$ at $z=2.23$ is $\approx100$\,\AA, with a tail at both higher rest-frame EWs (highest: 390\,\AA) and lower (lowest: 5.1\,\AA). If we were to apply a cut at EW$_0>25$\,\AA, we would still recover 89\% of our full sample of Ly$\alpha$ emitters. By imposing a cut of EW$_0>50$\,\AA, we would only recover 69\% of all Ly$\alpha$ emitters. 

In Figure \ref{EW_dist} we also compare the rest-frame EW distribution of our Ly$\alpha$ emitters with H$\alpha$ emitters at the same redshift \citep[][]{Sobral2014} and the EW distribution of Ly$\alpha$ emitters at higher redshift \citep[$z=5.7$;][]{Santos2016}. We find that H$\alpha$ emitters at $z=2.23$ show much higher EWs than Ly$\alpha$ selected sources at the same redshift. Interestingly, if one reduces the H$\alpha$ EWs by $\approx60$\%, the distribution becomes relatively similar to the one observed in Ly$\alpha$, i.e., Ly$\alpha$ and H$\alpha$ have a similar dispersion of EWs. This is not at all the case for the distribution of EWs for higher redshift Ly$\alpha$ emitters, selected over a similar range in luminosities from \cite{Santos2016}. Ly$\alpha$ emitters at $z\sim6$ present a much broader EW distribution, with a tail at very high EWs. These high EW Ly$\alpha$ emitters become much rarer at lower redshift.

By applying different EW cuts, we also study the effect of those on the Ly$\alpha$ luminosity function at $z=2.23$. For all EW$_0$ cuts, we repeat our Ly$\alpha$ selection, in order to eliminate interlopers, as described in \S\ref{selection_pz_colcol}. Also, for each new selection, as our EW cut changes, our completeness also changes, and thus we re-compute it and apply the appropriate corrections for each cut. This means that while a higher EW$_0$ cut results in a lower completeness, our corrections can account for at least part of that. We show our results in Figure \ref{LF_Lya_KONNO_comparison}, which shows the effect of varying the Ly$\alpha$ EW$_0$.

We find that for Ly$\alpha$ selected samples a higher EW cut preferentially lowers the number densities at the bright end, eliminating the power-law component, and making the LF look steeper. On the other hand, a simple EW cut, without filtering out the non Ly$\alpha$ emitters from the sample, still leads to significant contamination at all luminosities, particularly at the bright end. We find that in order to eliminate such contaminants effectively one requires a relatively high EW$_0$ of at least $>50$\,\AA, but that is far from ideal, as it will also eliminate a significant fraction of real luminous Ly$\alpha$ emitters, which we know are spectroscopically confirmed to be at $z=2.23$.

\section{The Ly$\alpha$ escape fraction at $\bf z=2.23$} \label{global_escape_section}

\subsection{Ly$\alpha$ emitters at $\bf z=2.23$: the H$\alpha$ view}\label{Lyaselected_Hastack}

The H$\alpha$ stack of our Ly$\alpha$ emitters allows us to compute the typical star formation rate of our Ly$\alpha$ emitters. We use \cite{Kennicutt98} with a Chabrier initial mass function \citep[][]{Chabrier2003}, and correct H$\alpha$ for extinction using \cite{GarnBest2010}, following e.g. \cite{Sobral2014}. Our results show that our sample of Ly$\alpha$ selected sources has a median dust corrected star formation rate (SFR) of $7.7\pm0.6$\,M$_{\odot}$\,yr$^{-1}$. Such median SFR implies that our Ly$\alpha$ emitters are $\sim0.1$\,SFR$^*$ star-forming galaxies at $z=2.2$ \citep[][]{Sobral2014}.

\begin{figure}
\begin{tabular}{cccc}
\includegraphics[width=8.3cm]{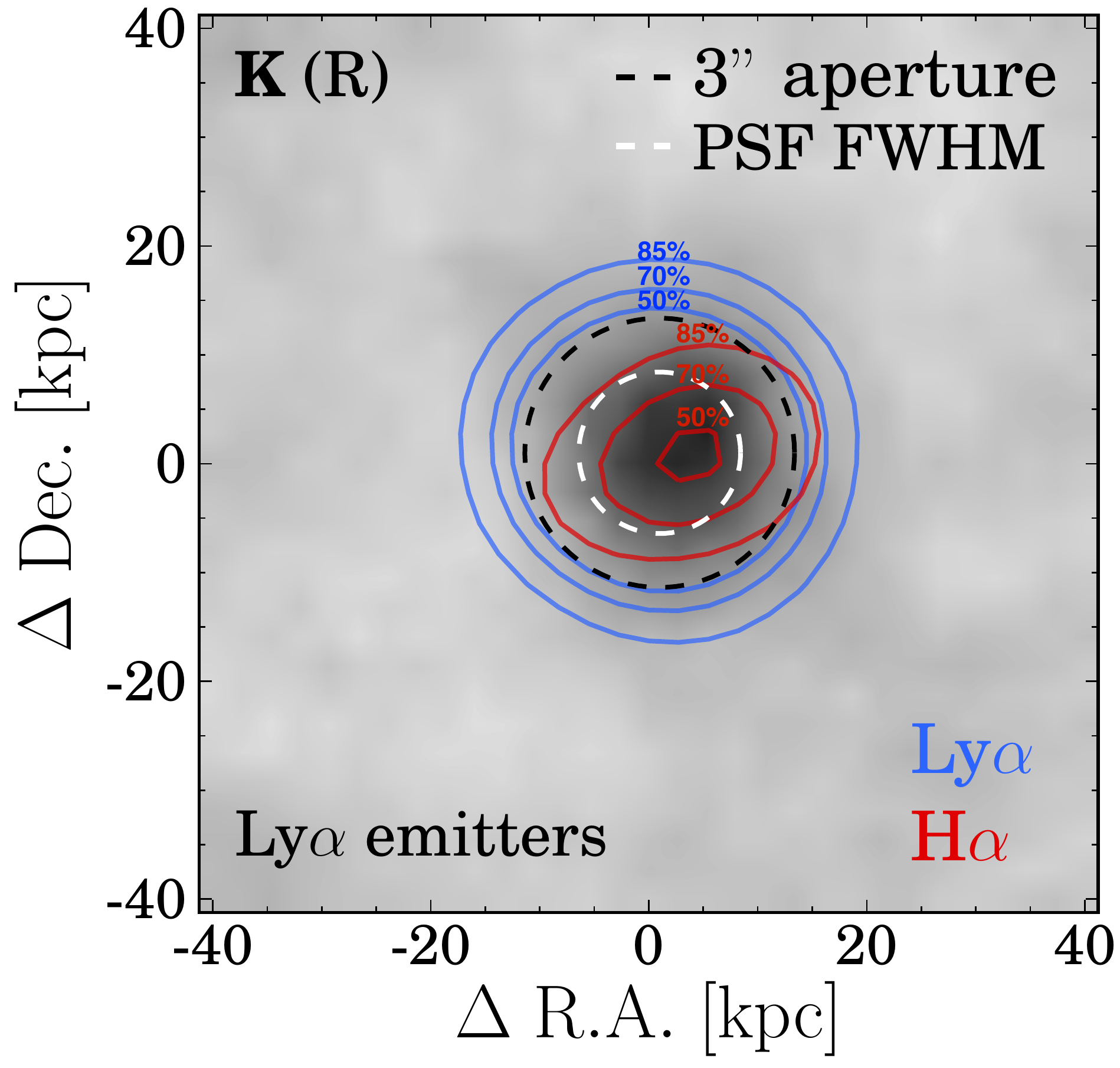}
\end{tabular} 
\caption{Stacked H$\alpha$ and Ly$\alpha$ images of all our Ly$\alpha$ emitters. All bands are PSF matched and we show the common PSF in the image. We compare the stacked rest-frame R band (observed $K$ band, H$\alpha$ subtracted), tracing the older stellar population/stellar mass, with both H$\alpha$ and Ly$\alpha$ emission from our sample of Ly$\alpha$ selected emitters. Ly$\alpha$ and H$\alpha$ contours show the 50\%, 70\% and 85\% contours of the total flux. For a 3$''$ diameter aperture we recover 82\% of the total H$\alpha$ flux ($0.4\times10^{-16}$\,erg\,s$^{-1}$) but only 50\% of the total Ly$\alpha$ flux ($2.2\times10^{-16}$\,erg\,s$^{-1}$). We thus find that while H$\alpha$ is slightly more extended than the continuum emission, the Ly$\alpha$ emission extends to much larger radii. This is consistent with the results from \citet{Matthee2016}.}
\label{Stack_Ha_Lya}
\end{figure}

In Figure \ref{Stack_Ha_Lya} we show the H$\alpha$ stack, a comparison to the rest-frame (H$\alpha$ subtracted) $R$ band stack, and the Ly$\alpha$ stack of all our Ly$\alpha$ emitters. We find that Ly$\alpha$ is significantly more extended (diameter of about $\sim40$\,kpc) than H$\alpha$ by about a factor of 2. Our results are consistent with those presented in \cite{Matthee2016} for a sub-set of Ly$\alpha$-H$\alpha$ emitters at $z=2.23$, and reveal that Ly$\alpha$ emitters have ubiquitous extended Ly$\alpha$ emission \citep[see also e.g.][]{Momose2014,Matthee2016,Wisotzki2016}. When compared to \cite{Momose2014}, we seem to find slightly larger Ly$\alpha$ extents, although our sample is dominated by brighter Ly$\alpha$ emitters than those in \cite{Momose2014}, while our PSF is also larger than \cite{Momose2014}. The combined effects (more luminous Ly$\alpha$ emitters in our sample and larger PSF) can likely explain the larger extents that we measure.

%%%%%%%%%%%%%%%%%%%%%%%%%%%%%%%%%%%%%%%%%%%%%%%
%
% Figure 10
%
% Figure showing the full stack as example: Lya, Ha, R-restframe
%
%
%%%%%%%%%%%%%%%%%%%%%%%%%%%%%%%%%%%%%%%%%%%%%

%%%%%%%%%%%%%%%%%%%%%%%%%%%%%
%
% Figure 11 - double panel
%
% Lyman-alpha escape fraction for the full population and for bins of Lyman-alpha luminosity. 
%
%%%%%%%%%%%%%%%%%%%%%%%%%%%%%%%%%%%%%%%%%%%%%%%%%%%
\begin{figure*}
\begin{tabular}{cc}
\includegraphics[width=16.3cm]{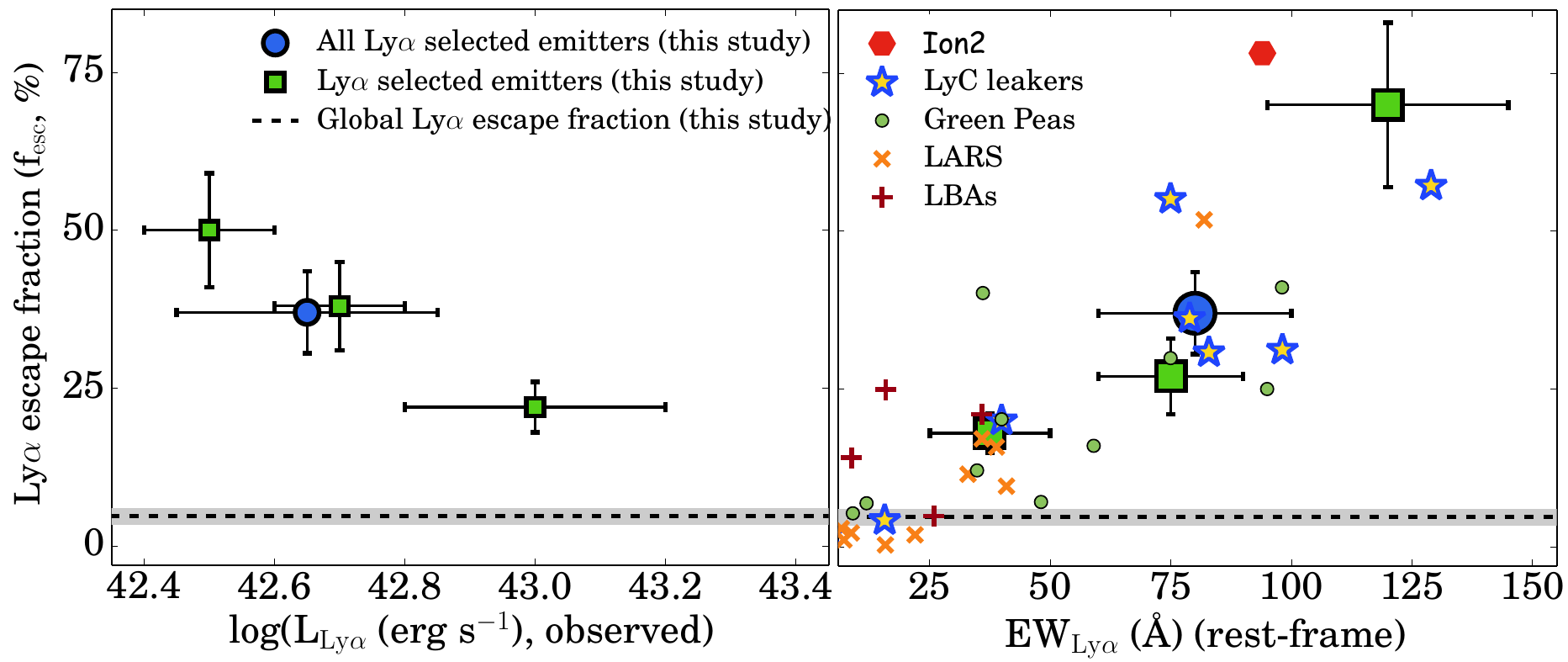}
\end{tabular} 
\caption{{\it Left:} The escape fraction (within $\approx13$\,kpc) of Ly$\alpha$ emitters as a function of Ly$\alpha$ luminosity. The results show that more luminous Ly$\alpha$ emitters have lower escape fractions than those which are less luminous, but consistent with still having very high escape fractions even at the highest Ly$\alpha$ luminosities. {\it Right:} The escape fraction of Ly$\alpha$ photons increases as function of rest-frame Ly$\alpha$ equivalent width. We also compare our results with similar measurements done for a range of Ly$\alpha$ emitting galaxies mostly in the local Universe, and find that they all follow a similar relation to our $z=2.2$ Ly$\alpha$ emitters. These include the recently discovered LyC leakers \citep[$z\sim0.3$;][]{Izotov2016bI,Izotov2016,Verhamme2016}, a Ly$\alpha$-LyC leaker at $z=3.2$, ``Ion2" \citep[][]{Vanzella2016,deBarros2016}, green peas \citep[$z\sim0.3$, e.g.][]{Cardamone,Henry2015,Yang2016}, Lyman-break Analogues \citep[LBAs, $z\sim0.2$, e.g.][]{Heckman2005,Overzier2009} and the Lyman Alpha Reference Survey \citep[LARS, $z\sim0.1$, e.g.][]{Ostlin2014,Hayes2014}. The fact that our Ly$\alpha$ emitters follow the relation of confirmed LyC emitters could suggest that at least part of our Ly$\alpha$ emitters may be LyC leakers.}
\label{Escape_vs_L}
\end{figure*}

\subsection{Ly$\alpha$ escape fraction and dependence on Ly$\alpha$ luminosity and EW$_0$} \label{esc_vs_LyaL}

Assuming Case B recombination, we use the H$\alpha$ stack (after applying all corrections; see \S\ref{corrections_caseB}) to measure an escape fraction of $37\pm7$\,\% for a 3$''$ aperture. We also use larger apertures for both Ly$\alpha$ and H$\alpha$ and find that the Ly$\alpha$ escape fraction increases with increasing aperture. We find this to be the case up to an aperture of 8$''$, when the Ly$\alpha$ escape fraction reaches an apparent plateau of $65\pm20$\,\%, consistent with \cite{Matthee2016}. Regardless of the aperture used, the values are significantly above the global average or the escape fraction for H$\alpha$ selected/more typical star-forming galaxies, which is only a few percent \citep[see e.g.][]{Hayes2010,Matthee2016,Konno2016}, as we will also show in \S\ref{global_escape}. However, this is not surprising, as, by definition, Ly$\alpha$ emitters will have to have relatively high escape fractions, otherwise they would not be selected as such.

We then split our Ly$\alpha$ emitters according to their Ly$\alpha$ luminosity and EW$_{0}$ (see Figure \ref{Escape_vs_L}). We find that the Ly$\alpha$ escape fraction increases with increasing EW$_0$, with f$_{\rm esc}$ increasing from $\approx 18$\% for EW$_0$$\approx 40$ {\AA} to f$_{\rm esc} \approx 70$\% for EW$_0$$\approx 120$ {\AA}. This is consistent with the younger/more star-bursting sources having higher Ly$\alpha$ escape fractions \citep[see also][]{Verhamme2016}. We compare our results with measurements from the literature, including a sample of recently discovered LyC leakers at $z\sim0.3$ \citep[][]{Izotov2016bI,Izotov2016,Verhamme2016}, a Ly$\alpha$-LyC leaker at $z=3.2$, ``Ion2" \citep[][]{Vanzella2016,deBarros2016}, and $z\sim0.1-0.3$ sources such as: green peas \citep[e.g.][]{Cardamone,Henry2015,Yang2016}, Lyman-break Analogues \citep[LBAs, e.g.][]{Heckman2005,Overzier2009} and the Lyman Alpha Reference Survey \citep[LARS, e.g.][]{Ostlin2014,Hayes2014}. Interestingly, we find that all these Ly$\alpha$ emitters follow a similar relation, with the Ly$\alpha$ escape fraction rising with increasing Ly$\alpha$ rest-frame equivalent width, even though they are found at very different redshifts and have been pre-selected through different methods. It is particularly interesting that green peas and the confirmed LyC emitters at low redshift, and ``Ion2", a confirmed LyC leaker at $z=3.2$ populate the highest escape fraction and highest EWs that we find in our sample. The results suggest that our high EW Ly$\alpha$ selected sources at $z\sim2$ may be very good LyC leaker candidates.

Naively f$_{\rm esc}$ could be expected to increase with Ly$\alpha$ luminosity. However, as Figure \ref{Escape_vs_L} shows, we find that f$_{\rm esc}$ decreases with increasing Ly$\alpha$ luminosity. For our faintest luminosity bin,  $L_{\rm Ly\alpha}\approx10^{42.5}$ erg\,s$^{-1}$, we measure f$_{\rm esc} = 50\pm9$\%, while we only measure f$_{\rm esc} = 22\pm4$\% for the most luminous bin, $L_{\rm Ly\alpha}\approx10^{43.0}$\,erg\,s$^{-1}$. When interpreted together with results from \S\ref{global_escape}, it is expected that f$_{\rm esc}$ will drop again for very low Ly$\alpha$ luminosities. We discuss this in more detail in \S\ref{conclusions}.

\subsection{The global Ly$\bf \alpha$ escape fraction at $\bf z=2.23$} \label{global_escape}

Here we investigate the global escape fraction of Ly$\alpha$ photons (with a fixed 3$''$ diameter aperture) at the peak epoch of the star formation history of the Universe. We focus on the global escape fraction (from the integral of the Ly$\alpha$ and H$\alpha$ luminosity functions) and use the extinction corrected H$\alpha$ luminosity function presented by \cite{Sobral2013}.

The Schechter component of our fit to the Ly$\alpha$ luminosity function yields an integrated luminosity density (full integral) of $1.1\times10^{40}$\,erg\,s$^{-1}$\,Mpc$^{-3}$. The additional power-law component adds a further $1.1\times10^{39}$\,erg\,s$^{-1}$\,Mpc$^{-3}$, or $\sim10$\% of the Schechter contribution. However, it should be noted that if one integrates down to e.g. $L_{\rm Ly\alpha}>10^{41.6}$\,erg\,s$^{-1}$, the Schechter component becomes only $0.4\times10^{40}$\,erg\,s$^{-1}$\,Mpc$^{-3}$, and thus the power-law component becomes more important for shallower Ly$\alpha$ surveys.

By integrating our Ly$\alpha$ luminosity function at $z=2.23$, assuming Case B recombination, and directly comparing with the equivalent integral of the extinction-corrected H$\alpha$ luminosity function, we find that, on average, within the same apertures used for H$\alpha$ and Ly$\alpha$ (corresponding to roughly to a 13\,kpc radius), only $5.1\pm0.2$\,\% of Ly$\alpha$ photons escape. This is in very good agreement with the measurement from \cite{Hayes2010} of $5\pm4$\,\%, but our result greatly reduces the errors due to a much larger volume and significantly larger samples. More recently, \cite{Matthee2016} studied an H$\alpha$-selected sample, finding that, down to the detection limit of the sample, the Ly$\alpha$ escape fraction is $1.6\pm0.5$\,\%. However, those authors show that the escape fraction strongly anti-correlates with H$\alpha$ flux/star formation rate, with the low H$\alpha$ flux and low star-formation rate galaxies having the highest Ly$\alpha$ escape fractions. Thus, the results in \cite{Matthee2016} are in very good agreement with our global escape fraction of $5.1\pm0.2$\,\%, particularly due to the contribution of much lower SFR sources to the global measurement. 

The results presented by \cite{Matthee2016} already hint that Ly$\alpha$ escape fractions will strongly depend on e.g. the H$\alpha$ luminosity limit of a survey (and also depend strongly on the aperture used). Thus, while we find a typical escape fraction of $5.1\pm0.2$\,\% by integrating both the H$\alpha$ and the Ly$\alpha$ luminosity functions, we also study the effect of integrating down to different luminosity limits. Our full results are presented in Table \ref{integral_results}.

Furthermore, as shown in Figure \ref{Stack_Ha_Lya}, we find that at a fixed 3$''$ we recover a much larger fraction of the total H$\alpha$ flux (82\%; consistent with e.g. \citealt{Sobral2014}) than the total Ly$\alpha$ flux (50\%). If we apply these results to correct the integral of the H$\alpha$ and Ly$\alpha$ luminosity functions, we find that the total (aperture corrected) average Ly$\alpha$ escape fraction would be 1.64 times larger, or $8.4\pm0.3$\,\%. This means that, potentially, a further 3.3\% of Ly$\alpha$ photons still escape, but at larger radii than those that our 3$''$ diameter apertures capture.

%%%%%%%%%%%%%%%%%%%%%%%%%%%%%%%%%%%%%%%%%%%%%%%
%
% Table 4 - Results for Lya and Ha integrations and implied escape fractions
%
%%%%%%%%%%%%%%%%%%%%%%%%%%%%%%%%%%%%%%%%%%%%%%%
\begin{table}
\caption{Ly$\alpha$ (observed) and H$\alpha$ (extinction corrected) luminosity densities for different integration limits and different assumptions, including integrating down to the same limits, but also integrating down to limits scaled by a factor 8.7 (labeled ``B"). For each set of luminosity densities, we assume case B recombination and compute the escape fraction of Ly$\alpha$ photons (factor 8.7). All Ly$\alpha$ luminosity densities have been computed taking into account the power-law component. To obtain the Schechter luminosity density only one simply has to remove the fixed contribution of $1.1\times10^{39}$\,erg\,s$^{-1}$\,Mpc$^{-3}$. Note that all these measurements are based on 3$''$ apertures; for Ly$\alpha$ emitters we find that those may only recover about 50\% of the Ly$\alpha$ flux, but recover around 80\% of the H$\alpha$, implying a potential aperture correction of 1.64.}
\begin{center}
\begin{tabular}{cccc}
\hline
Integration & log$\rho_{\rm L_{Ly\alpha}}$ & log$\rho_{\rm L_{H\alpha}}$  & Ly$\alpha$ f$_{\rm esc}$ \\ 
limit  & erg\,s$^{-1}$\,Mpc$^{-3}$ & erg\,s$^{-1}$\,Mpc$^{-3}$  & (\%) \\ 
\hline
10$^{41.0}$ & $39.87\pm0.02$ & $40.34\pm0.02$ & $3.9\pm0.3$  \\  
10$^{41.6}$& $39.74\pm0.02$ & $40.25\pm0.02$ & $3.5\pm0.3$   \\ 
10$^{42.0}$ & $39.59\pm0.02$ & $40.15\pm0.02$ & $3.2\pm0.3$ \\ 
 \hline
10$^{41.0}$ (B) & $39.97\pm0.02$ & $40.34\pm0.02$ & $4.9\pm0.4$  \\  
10$^{41.6}$ (B) & $39.91\pm0.02$ & $40.25\pm0.02$ & $5.3\pm0.4$ \\ 
10$^{42.0}$ (B) & $39.86\pm0.02$ & $40.15\pm0.02$ & $5.9\pm0.5$  \\
 \hline
0.01\,$L^*$ & $39.92\pm0.02$ & $40.35\pm0.02$ & $4.3\pm0.4$  \\ 
0.1\,$L^*$  & $39.74\pm0.01$ & $40.18\pm0.02$ & $4.1\pm0.3$ \\
0.2\,$L^*$  & $39.64\pm0.01$& $40.08\pm0.02$ & $4.2\pm0.3$ \\ 
$L^*$  & $39.27\pm0.02$ & $39.51\pm0.02$ & $6.5\pm0.4$ \\ 
\hline
\bf Full & $40.08\pm0.02$ & $40.43\pm0.02$ & $\bf 5.1\pm0.2$ \\  \hline
\end{tabular}
\end{center}
\label{integral_results}
\end{table}

Of particular importance is the fact that if one does not integrate both e.g. H$\alpha$ (or UV) and Ly$\alpha$ fully, one needs to be careful about the limit one integrates down to. It is very clear from all observational work that Ly$\alpha$ luminosities show significant scatter and a non-linear behaviour as a function of either UV or H$\alpha$ \citep[e.g.][]{Matthee2016}, due to significant changes in escape fraction as function of various properties. This is further highlighted by our results on the escape fraction by stacking in H$\alpha$ (only possible with our data-set). This shows that the escape fraction changes significantly with Ly$\alpha$ luminosity, and that there certainly is not a 1:1 correlation between H$\alpha$ and Ly$\alpha$. Thus, the results of integrating down to a specific luminosity are not easily interpreted. For example, a 0.1\,$L^*$ H$\alpha$ emitter is not necessarily a 0.1\,$L^*$ Ly$\alpha$ emitter and vice-versa. This means that integrating down to a different $L^*$ will lead to a different escape fraction. We illustrate this by obtaining escape fractions based on integrations of the H$\alpha$ and Ly$\alpha$ luminosity functions down to different limits in Table \ref{integral_results}.

\section{Conclusions} \label{conclusions}

We presented the first results from the CALYMHA pilot survey conducted at the Isaac Newton Telescope over the COSMOS and UDS fields. We used a custom-built Ly$\alpha$ narrow-band filter, NB392 ($\lambda_c = 3918${\AA}, $\Delta\lambda = 52${\AA}), on the Wide Field Camera, to survey large extragalactic fields at $z=2.23$. Our NB392 filter ($\lambda_c = 3918${\AA}, $\Delta\lambda = 52${\AA}) has been designed to provide a matched volume coverage to the $z=2.23$ HiZELS survey conducted with UKIRT \citep[][]{Sobral2013}. CALYMHA currently reaches a line flux limit of $\sim4\times10^{-17}$\,erg\,s$^{-1}$\,cm$^{-2}$, and a Ly$\alpha$ luminosity limit of $\sim10^{42.3}$\,erg\,s$^{-1}$ (3$\sigma$). Our main results are:

\begin{itemize}

\item We obtained a sample of 440 line emitters in COSMOS and UDS. Among them, and apart from Ly$\alpha$ emitters, we find a significant population of spectroscopically confirmed [O{\sc ii}], He{\sc ii}, C{\sc iii}] and C{\sc iv} line emitters. C{\sc iv} emitters at $z\sim1.5$ represent $\sim25$\% of line emitters with an available spectroscopic redshift. We show how important it is for Ly$\alpha$ surveys to remove contaminants, especially C{\sc iii}] and C{\sc iv} (which many have incorrectly assumed to be unimportant). Removing those contaminants is essential to robustly determine the bright end of the Ly$\alpha$ luminosity function.

\item We use spectroscopic and photometric redshifts, together with colour-colour selections, to select a clean and complete sample of 188 Ly$\alpha$ emitters over a volume of $7.3\times10^5$\,Mpc$^{3}$.

\item We show that the Ly$\alpha$ luminosity function is significantly overestimated if all line emitters are used, with a simple equivalent cut. Such simple selections (single EW cut) are particularly problematic at higher fluxes, where contaminants become more and more important, particularly spectroscopically confirmed C{\sc iii}] and C{\sc iv} emitters (AGN) which can have EW$_0>25$\,\AA.

\item The Ly$\alpha$ luminosity function at $z=2.23$ is very well described by a Schechter function up to $L_{\rm Ly\alpha}\approx10^{43}$\,erg\,s$^{-1}$ with $L^*=10^{42.59^{+0.16}_{-0.08}}$\,erg\,s$^{-1}$, $\phi^*=10^{-3.09^{+0.14}_{-0.34}}$\,Mpc$^{-3}$ and $\alpha=-1.75\pm0.25$.

\item Beyond $L_{\rm Ly\alpha}\approx10^{43}$\,erg\,s$^{-1}$ the Ly$\alpha$ luminosity function becomes power-law like, similarly to what has been found in \cite{Konno2016} at $z=2.2$, due to the prevalence of bright X-ray AGN with similar X-ray and Ly$\alpha$ luminosities. However our normalisation of the power-law component is significantly below that of \cite{Konno2016}. We note that our results are based on a sample which is $\sim50$\,\% spectroscopically complete and cleaned of contaminants.

\item We show that the bright end of the Ly$\alpha$ luminosity function depends strongly on the choice of EW cut applied, as the sample of bright Ly$\alpha$ emitters becomes increasingly incomplete as a function of EW cut. Selections with a very high EW cut (usually motivated to eliminate contaminants) lose the power-law component and fail to select real, spectroscopically confirmed Ly$\alpha$ emitters.

\item By stacking the H$\alpha$ narrow-band images of our Ly$\alpha$ emitters in H$\alpha$, we find they have a median dust corrected H$\alpha$ star formation rate of $7.7\pm0.6$\,M$_{\odot}$\,yr$^{-1}$ ($\sim0.1$\,SFR$^*$ at $z=2.2$), and have an escape fraction (Ly$\alpha$ photons) of $37\pm7$\,\%. Ly$\alpha$ emission from our stack of Ly$\alpha$ emitters extends ($\approx40$\,kpc) by about 2 times that of the H$\alpha$ emission, in very good agreement with \cite{Matthee2016}.

\item We find that the Ly$\alpha$ escape fraction of Ly$\alpha$ emitters at $z=2.23$ drops with increasing Ly$\alpha$ luminosity, and increases with increasing Ly$\alpha$ rest-frame equivalent width. This may be due to sources with high equivalent widths being generally younger, less dusty and less massive, favouring high escape fractions. Sources with the highest Ly$\alpha$ luminosities are dominated by X-ray detected AGN.

\item By directly comparing our Ly$\alpha$ and H$\alpha$ luminosity functions, which are not affected by cosmic variance and are obtained over the same multiple large volumes, we find that the global escape fraction of Ly$\alpha$ from star-forming galaxies at $z=2.23$ is $5.1\pm0.2$\%. We also show how important the choice of integration limits is, given that the Ly$\alpha$ escape fraction varies significantly both with Ly$\alpha$ luminosity, as shown in this paper, but also as a function of H$\alpha$ luminosity \citep[][]{Matthee2016} in a non-linear way.

\end{itemize}

Our results imply that $94.9\pm0.2$\% of the total Ly$\alpha$ luminosity density produced at the peak of the star formation history ($z\sim2$) does not escape the host galaxies within a radius of $\sim13$\,kpc (3$''$ diameter aperture). Integrating the luminosity functions down to observed values yields a lower escape fraction, in agreement with e.g. \cite{Matthee2016} and \cite{Konno2016}. Also, we show that for Ly$\alpha$ selected samples, the escape fraction is, not surprisingly, significantly above the cosmic average we measure ($5.1\pm0.2$\%), and around 37\% for a 3$''$ aperture. Interestingly, and even though this already corresponds to a quite high escape fraction, it is only a lower limit as far as the total escape fraction (at any radii) is concerned, as we clearly see that Ly$\alpha$ extends beyond the H$\alpha$ emission by a factor of $\sim2$ and the radius usually used to measure emission line properties \citep[see also][]{Wisotzki2016,Drake2016}. This means that an extra $3.3\pm0.3$\% of Ly$\alpha$ photons likely still escape, but at larger radii, potentially adding up to a total escape fraction at any radii of $\sim8$\%, although this is highly uncertain. Nevertheless, significant progress can be further achieved with new instruments such as MUSE \citep[e.g.][]{Wisotzki2016,Borisova2016}

Our results provide important empirical measurements that are useful to interpret observations at higher redshift. Significantly deeper Ly$\alpha$-H$\alpha$ observations, and observations spread over more fields will allow for further significant progress. Moreover, once {\it JWST} is launched it will be possible to directly measure H$\alpha$ from both UV and Ly$\alpha$ selected sources, and thus some of the results from our survey can then be tested at higher redshift.

\section*{Acknowledgements}

We thank the reviewer for their helpful comments and suggestions which have greatly improved this work. D.S. and J.M. acknowledge financial support from the Netherlands Organisation for Scientific research (NWO) through a Veni fellowship. D.S. also acknowledges funding from FCT through a FCT Investigator Starting Grant and Start-up Grant (IF/01154/2012/CP0189/CT0010). PNB is grateful for support from the UK STFC via grant ST/M001229/1. I.R.S. acknowledges support from STFC (ST/L00075X/1), the ERC Advanced Investigator programme DUSTYGAL 321334 and a Royal Society/Wolfson merit award. We thank Matthew Hayes, Ryan Trainor, Kimihiko Nakajima and Anne Verhamme for many helpful discussions and Ana Sobral, Carolina Duarte and Miguel Domingos for taking part in observations with the NB392 filter. We also thank Sergio Santos for helpful comments. The authors acknowledge the award of time from programmes: I13AN002, I14AN002, 088-INT7/14A, I14BN006, 118-INT13/14B, I15AN008 on the Isaac Newton Telescope (INT). INT is operated on the island of La Palma by the Isaac Newton Group in the Spanish Observatorio del Roque de los Muchachos of the Instituto de Astrofisica de Canarias. Based on observations made with ESO Telescopes at the La Silla Paranal Observatory under programme ID 098.A-0819. We have benefited greatly from the public available programming language {\sc Python}, including the {\sc numpy, matplotlib, pyfits, scipy} and {\sc astropy} packages, the astronomical imaging tools {\sc SExtractor, Swarp} \citep[][]{Bertin1996,Bertin2010} and {\sc Scamp} \citep[][]{SCAMP} and {\sc Topcat} \citep{Topcat}. Dedicated to the memory of M. L. Nicolau and M. C. Serrano.

%%%%%%%%%%%%%%%%%%%% REFERENCES %%%%%%%%%%%%%%%%%%

\bibliographystyle{mnras}
\bibliography{myBib.bib}

\begin{thebibliography}{}
\makeatletter
\relax
\def\mn@urlcharsother{\let\do\@makeother \do\$\do\&\do\#\do\^\do\_\do\%\do\~}
\def\mn@doi{\begingroup\mn@urlcharsother \@ifnextchar [ {\mn@doi@}
  {\mn@doi@[]}}
\def\mn@doi@[#1]#2{\def\@tempa{#1}\ifx\@tempa\@empty \href
  {http://dx.doi.org/#2} {doi:#2}\else \href {http://dx.doi.org/#2} {#1}\fi
  \endgroup}
\def\mn@eprint#1#2{\mn@eprint@#1:#2::\@nil}
\def\mn@eprint@arXiv#1{\href {http://arxiv.org/abs/#1} {{\tt arXiv:#1}}}
\def\mn@eprint@dblp#1{\href {http://dblp.uni-trier.de/rec/bibtex/#1.xml}
  {dblp:#1}}
\def\mn@eprint@#1:#2:#3:#4\@nil{\def\@tempa {#1}\def\@tempb {#2}\def\@tempc
  {#3}\ifx \@tempc \@empty \let \@tempc \@tempb \let \@tempb \@tempa \fi \ifx
  \@tempb \@empty \def\@tempb {arXiv}\fi \@ifundefined
  {mn@eprint@\@tempb}{\@tempb:\@tempc}{\expandafter \expandafter \csname
  mn@eprint@\@tempb\endcsname \expandafter{\@tempc}}}

\bibitem[\protect\citeauthoryear{{Atek}, {Kunth}, {Hayes}, {{\"O}stlin}  \&
  {Mas-Hesse}}{{Atek} et~al.}{2008}]{Atek2008}
{Atek} H.,  {Kunth} D.,  {Hayes} M.,  {{\"O}stlin} G.,   {Mas-Hesse} J.~M.,
  2008, \mn@doi [AAP] {10.1051/0004-6361:200809527}, \href
  {http://adsabs.harvard.edu/abs/2008A%26A...488..491A} {488, 491}

\bibitem[\protect\citeauthoryear{{Atek}, {Kunth}, {Schaerer}, {Hayes},
  {Deharveng}, {{\"O}stlin}  \& {Mas-Hesse}}{{Atek} et~al.}{2009}]{Atek09}
{Atek} H.,  {Kunth} D.,  {Schaerer} D.,  {Hayes} M.,  {Deharveng} J.~M.,
  {{\"O}stlin} G.,   {Mas-Hesse} J.~M.,  2009, \mn@doi [A\&A]
  {10.1051/0004-6361/200912787}, \href
  {http://adsabs.harvard.edu/abs/2009A%26A...506L...1A} {506, L1}

\bibitem[\protect\citeauthoryear{{Atek}, {Kunth}, {Schaerer}, {Mas-Hesse},
  {Hayes}, {{\"O}stlin}  \& {Kneib}}{{Atek} et~al.}{2014}]{Atek2014}
{Atek} H.,  {Kunth} D.,  {Schaerer} D.,  {Mas-Hesse} J.~M.,  {Hayes} M.,
  {{\"O}stlin} G.,   {Kneib} J.-P.,  2014, \mn@doi [AAP]
  {10.1051/0004-6361/201321519}, \href
  {http://adsabs.harvard.edu/abs/2014A%26A...561A..89A} {561, A89}

\bibitem[\protect\citeauthoryear{{Bacon}, {Brinchmann}, {Richard}
  et~al.}{{Bacon} et~al.}{2015}]{Bacon15}
{Bacon} R.,  {Brinchmann} J.,  {Richard} J.,   et~al., 2015, \mn@doi [A\&A]
  {10.1051/0004-6361/201425419}, \href
  {http://adsabs.harvard.edu/abs/2015A%26A...575A..75B} {575, A75}

\bibitem[\protect\citeauthoryear{{Barger}, {Cowie}  \& {Wold}}{{Barger}
  et~al.}{2012}]{Barger2012}
{Barger} A.~J.,  {Cowie} L.~L.,   {Wold} I.~G.~B.,  2012, \mn@doi [ApJ]
  {10.1088/0004-637X/749/2/106}, \href
  {http://adsabs.harvard.edu/abs/2012ApJ...749..106B} {749, 106}

\bibitem[\protect\citeauthoryear{{Bertin}}{{Bertin}}{2006}]{SCAMP}
{Bertin} E.,  2006, in {Gabriel} C.,  {Arviset} C.,  {Ponz} D.,   {Enrique} S.,
   eds,  Astronomical Society of the Pacific Conference Series Vol. 351,
  Astronomical Data Analysis Software and Systems XV. p.~112

\bibitem[\protect\citeauthoryear{{Bertin}}{{Bertin}}{2010}]{Bertin2010}
{Bertin} E.,  2010, {SWarp: Resampling and Co-adding FITS Images Together},
  Astrophysics Source Code Library (\mn@eprint {ascl} {1010.068})

\bibitem[\protect\citeauthoryear{{Bertin} \& {Arnouts}}{{Bertin} \&
  {Arnouts}}{1996}]{Bertin1996}
{Bertin} E.,  {Arnouts} S.,  1996, AAPS, \href
  {http://adsabs.harvard.edu/abs/1996A%26AS..117..393B} {117, 393}

\bibitem[\protect\citeauthoryear{{Blanc} et~al.,}{{Blanc}
  et~al.}{2011}]{Blanc2011}
{Blanc} G.~A.,  et~al., 2011, \mn@doi [ApJ] {10.1088/0004-637X/736/1/31}, \href
  {http://adsabs.harvard.edu/abs/2011ApJ...736...31B} {736, 31}

\bibitem[\protect\citeauthoryear{{Bongiovanni}, {Oteo}, {Cepa}
  et~al.}{{Bongiovanni} et~al.}{2010}]{Bongiovanni2010}
{Bongiovanni} A.,  {Oteo} I.,  {Cepa} J.,   et~al., 2010, \mn@doi [A\&A]
  {10.1051/0004-6361/201014719}, \href
  {http://adsabs.harvard.edu/abs/2010A%26A...519L...4B} {519, L4}

\bibitem[\protect\citeauthoryear{{Borisova} et~al.}{{Borisova}
  et~al.}{2016}]{Borisova2016}
{Borisova} E.,  et~al., 2016, preprint, \href
  {http://adsabs.harvard.edu/abs/2016arXiv160501422B} {} (\mn@eprint {arXiv}
  {1605.01422})

\bibitem[\protect\citeauthoryear{{Bouwens}, {Illingworth}, {Oesch}
  et~al.}{{Bouwens} et~al.}{2015}]{Bouwens15}
{Bouwens} R.~J.,  {Illingworth} G.~D.,  {Oesch} P.~A.,   et~al., 2015, \mn@doi
  [ApJ] {10.1088/0004-637X/803/1/34}, \href
  {http://adsabs.harvard.edu/abs/2015ApJ...803...34B} {803, 34}

\bibitem[\protect\citeauthoryear{{Bunker}, {Warren}, {Hewett}  \&
  {Clements}}{{Bunker} et~al.}{1995}]{Bunker1995}
{Bunker} A.~J.,  {Warren} S.~J.,  {Hewett} P.~C.,   {Clements} D.~L.,  1995,
  MNRAS, \href {http://adsabs.harvard.edu/abs/1995MNRAS.273..513B} {273, 513}

\bibitem[\protect\citeauthoryear{{Capak} et~al.}{{Capak}
  et~al.}{2007}]{Capak2007}
{Capak} P.,  et~al., 2007, \mn@doi [ApJS] {10.1086/519081}, \href
  {http://adsabs.harvard.edu/abs/2007ApJS..172...99C} {172, 99}

\bibitem[\protect\citeauthoryear{{Cardamone} et~al.}{{Cardamone}
  et~al.}{2009}]{Cardamone}
{Cardamone} C.,  et~al., 2009, \mn@doi [MNRAS]
  {10.1111/j.1365-2966.2009.15383.x}, \href
  {http://adsabs.harvard.edu/abs/2009MNRAS.399.1191C} {399, 1191}

\bibitem[\protect\citeauthoryear{{Cassata}, {Le F{\`e}vre}, {Garilli}  \&
  others.}{{Cassata} et~al.}{2011}]{Cassata2011}
{Cassata} P.,  {Le F{\`e}vre} O.,  {Garilli} B.,   others. 2011, \mn@doi [A\&A]
  {10.1051/0004-6361/201014410}, \href
  {http://adsabs.harvard.edu/abs/2011A%26A...525A.143C} {525, A143}

\bibitem[\protect\citeauthoryear{{Chabrier}}{{Chabrier}}{2003}]{Chabrier2003}
{Chabrier} G.,  2003, \mn@doi [PASP] {10.1086/376392}, \href
  {http://adsabs.harvard.edu/abs/2003PASP..115..763C} {115, 763}

\bibitem[\protect\citeauthoryear{{Chapman}, {Blain}, {Smail}  \&
  {Ivison}}{{Chapman} et~al.}{2005}]{Chapman2005}
{Chapman} S.~C.,  {Blain} A.~W.,  {Smail} I.,   {Ivison} R.~J.,  2005, \mn@doi
  [ApJ] {10.1086/428082}, \href
  {http://adsabs.harvard.edu/abs/2005ApJ...622..772C} {622, 772}

\bibitem[\protect\citeauthoryear{{Ciardullo} et~al.,}{{Ciardullo}
  et~al.}{2012}]{Ciardullo2012}
{Ciardullo} R.,  et~al., 2012, \mn@doi [ApJ] {10.1088/0004-637X/744/2/110},
  \href {http://adsabs.harvard.edu/abs/2012ApJ...744..110C} {744, 110}

\bibitem[\protect\citeauthoryear{{Ciardullo} et~al.,}{{Ciardullo}
  et~al.}{2014}]{Ciardullo2014}
{Ciardullo} R.,  et~al., 2014, \mn@doi [ApJ] {10.1088/0004-637X/796/1/64},
  \href {http://adsabs.harvard.edu/abs/2014ApJ...796...64C} {796, 64}

\bibitem[\protect\citeauthoryear{{Cirasuolo}, {McLure}, {Dunlop}, {Almaini},
  {Foucaud}  \& {Simpson}}{{Cirasuolo} et~al.}{2010}]{Cirasuolo2010}
{Cirasuolo} M.,  {McLure} R.~J.,  {Dunlop} J.~S.,  {Almaini} O.,  {Foucaud} S.,
    {Simpson} C.,  2010, \mn@doi [MNRAS] {10.1111/j.1365-2966.2009.15710.x},
  \href {http://adsabs.harvard.edu/abs/2010MNRAS.401.1166C} {401, 1166}

\bibitem[\protect\citeauthoryear{{Civano} et~al.}{{Civano}
  et~al.}{2012}]{Civano2012}
{Civano} F.,  et~al., 2012, \mn@doi [ApJS] {10.1088/0067-0049/201/2/30}, \href
  {http://adsabs.harvard.edu/abs/2012ApJS..201...30C} {201, 30}

\bibitem[\protect\citeauthoryear{{Civano}, {Marchesi}, {Comastri}
  et~al.}{{Civano} et~al.}{2016}]{Civano2016}
{Civano} F.,  {Marchesi} S.,  {Comastri} A.,   et~al., 2016, \mn@doi [ApJ]
  {10.3847/0004-637X/819/1/62}, \href
  {http://adsabs.harvard.edu/abs/2016ApJ...819...62C} {819, 62}

\bibitem[\protect\citeauthoryear{{Cowie}, {Barger}  \& {Hu}}{{Cowie}
  et~al.}{2010}]{Cowie2010}
{Cowie} L.~L.,  {Barger} A.~J.,   {Hu} E.~M.,  2010, \mn@doi [ApJ]
  {10.1088/0004-637X/711/2/928}, \href
  {http://adsabs.harvard.edu/abs/2010ApJ...711..928C} {711, 928}

\bibitem[\protect\citeauthoryear{{Daddi}, {Cimatti}, {Renzini}, {Fontana},
  {Mignoli}, {Pozzetti}, {Tozzi}  \& {Zamorani}}{{Daddi}
  et~al.}{2004}]{Daddi2004}
{Daddi} E.,  {Cimatti} A.,  {Renzini} A.,  {Fontana} A.,  {Mignoli} M.,
  {Pozzetti} L.,  {Tozzi} P.,   {Zamorani} G.,  2004, \mn@doi [ApJ]
  {10.1086/425569}, \href {http://adsabs.harvard.edu/abs/2004ApJ...617..746D}
  {617, 746}

\bibitem[\protect\citeauthoryear{{Dijkstra}}{{Dijkstra}}{2014}]{DijkstraReview}
{Dijkstra} M.,  2014, \mn@doi [PASA] {10.1017/pasa.2014.33}, \href
  {http://adsabs.harvard.edu/abs/2014PASA...31...40D} {31, 40}

\bibitem[\protect\citeauthoryear{{Dijkstra}, {Lidz}  \& {Wyithe}}{{Dijkstra}
  et~al.}{2007}]{Dijkstra2007}
{Dijkstra} M.,  {Lidz} A.,   {Wyithe} J.~S.~B.,  2007, \mn@doi [MNRAS]
  {10.1111/j.1365-2966.2007.11666.x}, \href
  {http://adsabs.harvard.edu/abs/2007MNRAS.377.1175D} {377, 1175}

\bibitem[\protect\citeauthoryear{{Dijkstra}, {Gronke}  \&
  {Venkatesan}}{{Dijkstra} et~al.}{2016}]{DijkstraLyaLyC2016}
{Dijkstra} M.,  {Gronke} M.,   {Venkatesan} A.,  2016, \mn@doi [ApJ]
  {10.3847/0004-637X/828/2/71}, \href
  {http://adsabs.harvard.edu/abs/2016ApJ...828...71D} {828, 71}

\bibitem[\protect\citeauthoryear{{Drake} et~al.}{{Drake}
  et~al.}{2016}]{Drake2016}
{Drake} A.~B.,  et~al., 2016, preprint, \href
  {http://adsabs.harvard.edu/abs/2016arXiv160902920D} {} (\mn@eprint {arXiv}
  {1609.02920})

\bibitem[\protect\citeauthoryear{{Erb} et~al.,}{{Erb} et~al.}{2014}]{Erb2014}
{Erb} D.~K.,  et~al., 2014, \mn@doi [ApJ] {10.1088/0004-637X/795/1/33}, \href
  {http://adsabs.harvard.edu/abs/2014ApJ...795...33E} {795, 33}

\bibitem[\protect\citeauthoryear{{Erb}, {Pettini}, {Steidel}, {Strom}, {Rudie},
  {Trainor}, {Shapley}  \& {Reddy}}{{Erb} et~al.}{2016}]{Erb2016}
{Erb} D.~K.,  {Pettini} M.,  {Steidel} C.~C.,  {Strom} A.~L.,  {Rudie} G.~C.,
  {Trainor} R.~F.,  {Shapley} A.~E.,   {Reddy} N.~A.,  2016, preprint, \href
  {http://adsabs.harvard.edu/abs/2016arXiv160504919E} {} (\mn@eprint {arXiv}
  {1605.04919})

\bibitem[\protect\citeauthoryear{{Garel}, {Blaizot}, {Guiderdoni},
  {Michel-Dansac}, {Hayes}  \& {Verhamme}}{{Garel} et~al.}{2015}]{Garel2015}
{Garel} T.,  {Blaizot} J.,  {Guiderdoni} B.,  {Michel-Dansac} L.,  {Hayes} M.,
   {Verhamme} A.,  2015, \mn@doi [MNRAS] {10.1093/mnras/stv374}, \href
  {http://adsabs.harvard.edu/abs/2015MNRAS.450.1279G} {450, 1279}

\bibitem[\protect\citeauthoryear{{Garn} \& {Best}}{{Garn} \&
  {Best}}{2010}]{GarnBest2010}
{Garn} T.,  {Best} P.~N.,  2010, \mn@doi [MNRAS]
  {10.1111/j.1365-2966.2010.17321.x}, \href
  {http://adsabs.harvard.edu/abs/2010MNRAS.409..421G} {409, 421}

\bibitem[\protect\citeauthoryear{{Gawiser} et~al.}{{Gawiser}
  et~al.}{2007}]{Gawiser2007}
{Gawiser} E.,  et~al., 2007, \mn@doi [ApJ] {10.1086/522955}, \href
  {http://adsabs.harvard.edu/abs/2007ApJ...671..278G} {671, 278}

\bibitem[\protect\citeauthoryear{{Geach}, {Simpson}, {Rawlings}, {Read}  \&
  {Watson}}{{Geach} et~al.}{2008}]{Geach008}
{Geach} J.~E.,  {Simpson} C.,  {Rawlings} S.,  {Read} A.~M.,   {Watson} M.,
  2008, VizieR Online Data Catalog, \href
  {http://adsabs.harvard.edu/abs/2008yCat..83811369G} {838, 11369}

\bibitem[\protect\citeauthoryear{{Giavalisco}}{{Giavalisco}}{2002}]{Giavalisco2002}
{Giavalisco} M.,  2002, \mn@doi [ARAA]
  {10.1146/annurev.astro.40.121301.111837}, \href
  {http://adsabs.harvard.edu/abs/2002ARA%26A..40..579G} {40, 579}

\bibitem[\protect\citeauthoryear{{Gronke} \& {Dijkstra}}{{Gronke} \&
  {Dijkstra}}{2016}]{Gronke2016}
{Gronke} M.,  {Dijkstra} M.,  2016, \mn@doi [ApJ] {10.3847/0004-637X/826/1/14},
  \href {http://adsabs.harvard.edu/abs/2016ApJ...826...14G} {826, 14}

\bibitem[\protect\citeauthoryear{{Gronke}, {Bull}  \& {Dijkstra}}{{Gronke}
  et~al.}{2015}]{Gronke2015b}
{Gronke} M.,  {Bull} P.,   {Dijkstra} M.,  2015, \mn@doi [ApJ]
  {10.1088/0004-637X/812/2/123}, \href
  {http://adsabs.harvard.edu/abs/2015ApJ...812..123G} {812, 123}

\bibitem[\protect\citeauthoryear{{Gronwall} et~al.}{{Gronwall}
  et~al.}{2007}]{Gronwall2007}
{Gronwall} C.,  et~al., 2007, \mn@doi [ApJ] {10.1086/520324}, \href
  {http://adsabs.harvard.edu/abs/2007ApJ...667...79G} {667, 79}

\bibitem[\protect\citeauthoryear{{Hashimoto}, {Ouchi}, {Shimasaku}, {Ono},
  {Nakajima}, {Rauch}, {Lee}  \& {Okamura}}{{Hashimoto}
  et~al.}{2013}]{Hashimoto2013}
{Hashimoto} T.,  {Ouchi} M.,  {Shimasaku} K.,  {Ono} Y.,  {Nakajima} K.,
  {Rauch} M.,  {Lee} J.,   {Okamura} S.,  2013, \mn@doi [ApJ]
  {10.1088/0004-637X/765/1/70}, \href
  {http://adsabs.harvard.edu/abs/2013ApJ...765...70H} {765, 70}

\bibitem[\protect\citeauthoryear{{Hathi} et~al.}{{Hathi}
  et~al.}{2016}]{Hathi2016}
{Hathi} N.~P.,  et~al., 2016, \mn@doi [A\&A] {10.1051/0004-6361/201526012},
  \href {http://adsabs.harvard.edu/abs/2016A%26A...588A..26H} {588, A26}

\bibitem[\protect\citeauthoryear{{Hayes}}{{Hayes}}{2015}]{Hayes2015}
{Hayes} M.,  2015, \mn@doi [PASA] {10.1017/pasa.2015.25}, \href
  {http://adsabs.harvard.edu/abs/2015PASA...32...27H} {32, e027}

\bibitem[\protect\citeauthoryear{{Hayes} et~al.}{{Hayes}
  et~al.}{2010}]{Hayes2010}
{Hayes} M.,  et~al., 2010, \mn@doi [Nature] {10.1038/nature08881}, \href
  {http://adsabs.harvard.edu/abs/2010Natur.464..562H} {464, 562}

\bibitem[\protect\citeauthoryear{{Hayes}, {Schaerer}, {{\"O}stlin},
  {Mas-Hesse}, {Atek}  \& {Kunth}}{{Hayes} et~al.}{2011}]{Hayes2011}
{Hayes} M.,  {Schaerer} D.,  {{\"O}stlin} G.,  {Mas-Hesse} J.~M.,  {Atek} H.,
  {Kunth} D.,  2011, \mn@doi [ApJ] {10.1088/0004-637X/730/1/8}, \href
  {http://adsabs.harvard.edu/abs/2011ApJ...730....8H} {730, 8}

\bibitem[\protect\citeauthoryear{{Hayes} et~al.}{{Hayes}
  et~al.}{2014}]{Hayes2014}
{Hayes} M.,  et~al., 2014, \mn@doi [ApJ] {10.1088/0004-637X/782/1/6}, \href
  {http://adsabs.harvard.edu/abs/2014ApJ...782....6H} {782, 6}

\bibitem[\protect\citeauthoryear{{Heckman} et~al.}{{Heckman}
  et~al.}{2005}]{Heckman2005}
{Heckman} T.~M.,  et~al., 2005, \mn@doi [ApJLl] {10.1086/425979}, \href
  {http://adsabs.harvard.edu/abs/2005ApJ...619L..35H} {619, L35}

\bibitem[\protect\citeauthoryear{{Henry}, {Scarlata}, {Martin}  \&
  {Erb}}{{Henry} et~al.}{2015}]{Henry2015}
{Henry} A.,  {Scarlata} C.,  {Martin} C.~L.,   {Erb} D.,  2015, \mn@doi [ApJ]
  {10.1088/0004-637X/809/1/19}, \href
  {http://adsabs.harvard.edu/abs/2015ApJ...809...19H} {809, 19}

\bibitem[\protect\citeauthoryear{{Hernquist}, {Katz}, {Weinberg}  \&
  {Miralda-Escud{\'e}}}{{Hernquist} et~al.}{1996}]{Hernquist1996}
{Hernquist} L.,  {Katz} N.,  {Weinberg} D.~H.,   {Miralda-Escud{\'e}} J.,
  1996, \mn@doi [ApJL] {10.1086/309899}, \href
  {http://adsabs.harvard.edu/abs/1996ApJ...457L..51H} {457, L51}

\bibitem[\protect\citeauthoryear{{Ibar} et~al.}{{Ibar} et~al.}{2013}]{Ibar2013}
{Ibar} E.,  et~al., 2013, \mn@doi [MNRAS] {10.1093/mnras/stt1258}, \href
  {http://adsabs.harvard.edu/abs/2013MNRAS.434.3218I} {434, 3218}

\bibitem[\protect\citeauthoryear{{Ilbert} et~al.}{{Ilbert}
  et~al.}{2009}]{Ilbert2009}
{Ilbert} O.,  et~al., 2009, \mn@doi [ApJ] {10.1088/0004-637X/690/2/1236}, \href
  {http://adsabs.harvard.edu/abs/2009ApJ...690.1236I} {690, 1236}

\bibitem[\protect\citeauthoryear{{Izotov}, {Schaerer}, {Thuan}, {Worseck},
  {Guseva}, {Orlitov{\'a}}  \& {Verhamme}}{{Izotov} et~al.}{2016a}]{Izotov2016}
{Izotov} Y.~I.,  {Schaerer} D.,  {Thuan} T.~X.,  {Worseck} G.,  {Guseva} N.~G.,
   {Orlitov{\'a}} I.,   {Verhamme} A.,  2016a, \mn@doi [MNRAS]
  {10.1093/mnras/stw1205}, \href
  {http://adsabs.harvard.edu/abs/2016MNRAS.461.3683I} {461, 3683}

\bibitem[\protect\citeauthoryear{{Izotov}, {Orlitov{\'a}}, {Schaerer}, {Thuan},
  {Verhamme}, {Guseva}  \& {Worseck}}{{Izotov} et~al.}{2016b}]{Izotov2016bI}
{Izotov} Y.~I.,  {Orlitov{\'a}} I.,  {Schaerer} D.,  {Thuan} T.~X.,  {Verhamme}
  A.,  {Guseva} N.~G.,   {Worseck} G.,  2016b, \mn@doi [Nature]
  {10.1038/nature16456}, \href
  {http://adsabs.harvard.edu/abs/2016Natur.529..178I} {529, 178}

\bibitem[\protect\citeauthoryear{{Karim} et~al.}{{Karim}
  et~al.}{2011}]{Karim11}
{Karim} A.,  et~al., 2011, \mn@doi [ApJ] {10.1088/0004-637X/730/2/61}, \href
  {http://adsabs.harvard.edu/abs/2011ApJ...730...61K} {730, 61}

\bibitem[\protect\citeauthoryear{{Karman}, {Caputi}, {Grillo}  et~al.}{{Karman}
  et~al.}{2015}]{Karman15}
{Karman} W.,  {Caputi} K.~I.,  {Grillo} C.,   et~al., 2015, \mn@doi [A\&A]
  {10.1051/0004-6361/201424962}, \href
  {http://adsabs.harvard.edu/abs/2015A%26A...574A..11K} {574, A11}

\bibitem[\protect\citeauthoryear{{Kashikawa} et~al.}{{Kashikawa}
  et~al.}{2006}]{Kashikawa2006}
{Kashikawa} N.,  et~al., 2006, \mn@doi [ApJ] {10.1086/504966}, \href
  {http://adsabs.harvard.edu/abs/2006ApJ...648....7K} {648, 7}

\bibitem[\protect\citeauthoryear{{Kashikawa} et~al.,}{{Kashikawa}
  et~al.}{2012}]{Kashikawa2012}
{Kashikawa} N.,  et~al., 2012, \mn@doi [ApJ] {10.1088/0004-637X/761/2/85},
  \href {http://adsabs.harvard.edu/abs/2012ApJ...761...85K} {761, 85}

\bibitem[\protect\citeauthoryear{{Kennicutt}}{{Kennicutt}}{1998}]{Kennicutt98}
{Kennicutt} Jr. R.~C.,  1998, \mn@doi [ARAA] {10.1146/annurev.astro.36.1.189},
  \href {http://adsabs.harvard.edu/abs/1998ARA%26A..36..189K} {36, 189}

\bibitem[\protect\citeauthoryear{{Khostovan}, {Sobral}, {Mobasher}, {Best},
  {Smail}, {Stott}, {Hemmati}  \& {Nayyeri}}{{Khostovan}
  et~al.}{2015}]{Khostovan2015}
{Khostovan} A.~A.,  {Sobral} D.,  {Mobasher} B.,  {Best} P.~N.,  {Smail} I.,
  {Stott} J.~P.,  {Hemmati} S.,   {Nayyeri} H.,  2015, \mn@doi [MNRAS]
  {10.1093/mnras/stv1474}, \href
  {http://adsabs.harvard.edu/abs/2015MNRAS.452.3948K} {452, 3948}

\bibitem[\protect\citeauthoryear{{Khostovan}, {Sobral}, {Mobasher}, {Smail},
  {Darvish}, {Nayyeri}, {Hemmati}  \& {Stott}}{{Khostovan}
  et~al.}{2016}]{Khostovan2016}
{Khostovan} A.~A.,  {Sobral} D.,  {Mobasher} B.,  {Smail} I.,  {Darvish} B.,
  {Nayyeri} H.,  {Hemmati} S.,   {Stott} J.~P.,  2016, preprint, \href
  {http://adsabs.harvard.edu/abs/2016MNRAS.tmp.1290K} {} (\mn@eprint {arXiv}
  {1604.02456})

\bibitem[\protect\citeauthoryear{{Konno}, {Ouchi}, {Nakajima}, {Duval},
  {Kusakabe}, {Ono}  \& {Shimasaku}}{{Konno} et~al.}{2016}]{Konno2016}
{Konno} A.,  {Ouchi} M.,  {Nakajima} K.,  {Duval} F.,  {Kusakabe} H.,  {Ono}
  Y.,   {Shimasaku} K.,  2016, \mn@doi [ApJ] {10.3847/0004-637X/823/1/20},
  \href {http://adsabs.harvard.edu/abs/2016ApJ...823...20K} {823, 20}

\bibitem[\protect\citeauthoryear{{Lawrence} et~al.}{{Lawrence}
  et~al.}{2007}]{Lawrence2007}
{Lawrence} A.,  et~al., 2007, \mn@doi [MNRAS]
  {10.1111/j.1365-2966.2007.12040.x}, \href
  {http://adsabs.harvard.edu/abs/2007MNRAS.379.1599L} {379, 1599}

\bibitem[\protect\citeauthoryear{{Le Delliou}, {Lacey}, {Baugh}  \&
  {Morris}}{{Le Delliou} et~al.}{2006}]{LeDelliou06}
{Le Delliou} M.,  {Lacey} C.~G.,  {Baugh} C.~M.,   {Morris} S.~L.,  2006,
  \mn@doi [MNRAS] {10.1111/j.1365-2966.2005.09797.x}, \href
  {http://adsabs.harvard.edu/abs/2006MNRAS.365..712L} {365, 712}

\bibitem[\protect\citeauthoryear{{Lilly}, {Le Fevre}, {Hammer}  \&
  {Crampton}}{{Lilly} et~al.}{1996}]{Lilly96}
{Lilly} S.~J.,  {Le Fevre} O.,  {Hammer} F.,   {Crampton} D.,  1996, \mn@doi
  [ApJL] {10.1086/309975}, \href
  {http://adsabs.harvard.edu/abs/1996ApJ...460L...1L} {460, L1}

\bibitem[\protect\citeauthoryear{{Lilly} et~al.}{{Lilly}
  et~al.}{2009}]{Lilly2009}
{Lilly} S.~J.,  et~al., 2009, \mn@doi [ApJS] {10.1088/0067-0049/184/2/218},
  \href {http://adsabs.harvard.edu/abs/2009ApJS..184..218L} {184, 218}

\bibitem[\protect\citeauthoryear{{Malhotra} \& {Rhoads}}{{Malhotra} \&
  {Rhoads}}{2004}]{MalhotraRhoads2004}
{Malhotra} S.,  {Rhoads} J.~E.,  2004, \mn@doi [ApJL] {10.1086/427182}, \href
  {http://adsabs.harvard.edu/abs/2004ApJ...617L...5M} {617, L5}

\bibitem[\protect\citeauthoryear{{Matthee} et~al.}{{Matthee}
  et~al.}{2014}]{Matthee2014}
{Matthee} J.~J.~A.,  et~al., 2014, \mn@doi [MNRAS] {10.1093/mnras/stu392},
  \href {http://adsabs.harvard.edu/abs/2014MNRAS.440.2375M} {440, 2375}

\bibitem[\protect\citeauthoryear{{Matthee}, {Sobral}, {Santos},
  {R{\"o}ttgering}, {Darvish}  \& {Mobasher}}{{Matthee}
  et~al.}{2015}]{Matthee2015}
{Matthee} J.,  {Sobral} D.,  {Santos} S.,  {R{\"o}ttgering} H.,  {Darvish} B.,
   {Mobasher} B.,  2015, \mn@doi [MNRAS] {10.1093/mnras/stv947}, \href
  {http://adsabs.harvard.edu/abs/2015MNRAS.451..400M} {451, 400}

\bibitem[\protect\citeauthoryear{{Matthee}, {Sobral}, {Best}, {Khostovan},
  {Oteo}, {Bouwens}  \& {R{\"o}ttgering}}{{Matthee}
  et~al.}{2016a}]{Matthee2016b}
{Matthee} J.,  {Sobral} D.,  {Best} P.,  {Khostovan} A.~A.,  {Oteo} I.,
  {Bouwens} R.,   {R{\"o}ttgering} H.,  2016a, preprint, \href
  {http://adsabs.harvard.edu/abs/2016arXiv160508782M} {} (\mn@eprint {arXiv}
  {1605.08782})

\bibitem[\protect\citeauthoryear{{Matthee}, {Sobral}, {Oteo}, {Best}, {Smail},
  {R{\"o}ttgering}  \& {Paulino-Afonso}}{{Matthee} et~al.}{2016b}]{Matthee2016}
{Matthee} J.,  {Sobral} D.,  {Oteo} I.,  {Best} P.,  {Smail} I.,
  {R{\"o}ttgering} H.,   {Paulino-Afonso} A.,  2016b, \mn@doi [MNRAS]
  {10.1093/mnras/stw322}, \href
  {http://adsabs.harvard.edu/abs/2016MNRAS.458..449M} {458, 449}

\bibitem[\protect\citeauthoryear{{McCracken} et~al.}{{McCracken}
  et~al.}{2012}]{McCracken2012}
{McCracken} H.~J.,  et~al., 2012, \mn@doi [AAP] {10.1051/0004-6361/201219507},
  \href {http://adsabs.harvard.edu/abs/2012A%26A...544A.156M} {544, A156}

\bibitem[\protect\citeauthoryear{{Momose} et~al.}{{Momose}
  et~al.}{2014}]{Momose2014}
{Momose} R.,  et~al., 2014, \mn@doi [MNRAS] {10.1093/mnras/stu825}, \href
  {http://adsabs.harvard.edu/abs/2014MNRAS.442..110M} {442, 110}

\bibitem[\protect\citeauthoryear{{Murayama} et~al.}{{Murayama}
  et~al.}{2007}]{Murayama2007}
{Murayama} T.,  et~al., 2007, \mn@doi [ApJS] {10.1086/516597}, \href
  {http://adsabs.harvard.edu/abs/2007ApJS..172..523M} {172, 523}

\bibitem[\protect\citeauthoryear{{Muzzin} et~al.}{{Muzzin}
  et~al.}{2013}]{Muzzin2013}
{Muzzin} A.,  et~al., 2013, \mn@doi [ApJs] {10.1088/0067-0049/206/1/8}, \href
  {http://adsabs.harvard.edu/abs/2013ApJS..206....8M} {206, 8}

\bibitem[\protect\citeauthoryear{{Nagamine}, {Ouchi}, {Springel}  \&
  {Hernquist}}{{Nagamine} et~al.}{2010}]{Nagamine08}
{Nagamine} K.,  {Ouchi} M.,  {Springel} V.,   {Hernquist} L.,  2010, \mn@doi
  [PASJ] {10.1093/pasj/62.6.1455}, \href
  {http://adsabs.harvard.edu/abs/2010PASJ...62.1455N} {62, 1455}

\bibitem[\protect\citeauthoryear{{Nakajima} et~al.}{{Nakajima}
  et~al.}{2012}]{Nakajima2012}
{Nakajima} K.,  et~al., 2012, \mn@doi [ApJ] {10.1088/0004-637X/745/1/12}, \href
  {http://adsabs.harvard.edu/abs/2012ApJ...745...12N} {745, 12}

\bibitem[\protect\citeauthoryear{{Nakajima}, {Ellis}, {Iwata}, {Inoue},
  {Kusakabe}, {Ouchi}  \& {Robertson}}{{Nakajima} et~al.}{2016}]{Nakajima2016}
{Nakajima} K.,  {Ellis} R.~S.,  {Iwata} I.,  {Inoue} A.,  {Kusakabe} H.,
  {Ouchi} M.,   {Robertson} B.,  2016, preprint, \href
  {http://adsabs.harvard.edu/abs/2016arXiv160808222N} {} (\mn@eprint {arXiv}
  {1608.08222})

\bibitem[\protect\citeauthoryear{{Nilsson}, {Tapken}, {M{\o}ller}, {Freudling},
  {Fynbo}, {Meisenheimer}, {Laursen}  \& {{\"O}stlin}}{{Nilsson}
  et~al.}{2009}]{Nilsson2009}
{Nilsson} K.~K.,  {Tapken} C.,  {M{\o}ller} P.,  {Freudling} W.,  {Fynbo}
  J.~P.~U.,  {Meisenheimer} K.,  {Laursen} P.,   {{\"O}stlin} G.,  2009,
  \mn@doi [AAP] {10.1051/0004-6361/200810881}, \href
  {http://adsabs.harvard.edu/abs/2009A%26A...498...13N} {498, 13}

\bibitem[\protect\citeauthoryear{{Oesch} et~al.}{{Oesch}
  et~al.}{2015}]{Oesch2015}
{Oesch} P.~A.,  et~al., 2015, \mn@doi [ApJL] {10.1088/2041-8205/804/2/L30},
  \href {http://adsabs.harvard.edu/abs/2015ApJ...804L..30O} {804, L30}

\bibitem[\protect\citeauthoryear{{Ono}, {Ouchi}, {Shimasaku}, {Akiyama},
  {Dunlop}, {Farrah}  \& {et al.}.}{{Ono} et~al.}{2010a}]{Ono2009}
{Ono} Y.,  {Ouchi} M.,  {Shimasaku} K.,  {Akiyama} M.,  {Dunlop} J.,  {Farrah}
  D.,   {et al.}. 2010a, \mn@doi [MNRAS] {10.1111/j.1365-2966.2009.16034.x},
  \href {http://adsabs.harvard.edu/abs/2010MNRAS.402.1580O} {402, 1580}

\bibitem[\protect\citeauthoryear{{Ono}, {Ouchi}, {Shimasaku}, {Dunlop},
  {Farrah}, {McLure}  \& {Okamura}}{{Ono} et~al.}{2010b}]{Ono2010}
{Ono} Y.,  {Ouchi} M.,  {Shimasaku} K.,  {Dunlop} J.,  {Farrah} D.,  {McLure}
  R.,   {Okamura} S.,  2010b, \mn@doi [ApJ] {10.1088/0004-637X/724/2/1524},
  \href {http://adsabs.harvard.edu/abs/2010ApJ...724.1524O} {724, 1524}

\bibitem[\protect\citeauthoryear{{Ono} et~al.}{{Ono} et~al.}{2012}]{Ono2012}
{Ono} Y.,  et~al., 2012, \mn@doi [ApJ] {10.1088/0004-637X/744/2/83}, \href
  {http://adsabs.harvard.edu/abs/2012ApJ...744...83O} {744, 83}

\bibitem[\protect\citeauthoryear{{{\"O}stlin} et~al.}{{{\"O}stlin}
  et~al.}{2014}]{Ostlin2014}
{{\"O}stlin} G.,  et~al., 2014, \mn@doi [ApJ] {10.1088/0004-637X/797/1/11},
  \href {http://adsabs.harvard.edu/abs/2014ApJ...797...11O} {797, 11}

\bibitem[\protect\citeauthoryear{{Oteo}, {Bongiovanni}, {P{\'e}rez
  Garc{\'{\i}}a}  et~al.}{{Oteo} et~al.}{2012a}]{Oteo2012}
{Oteo} I.,  {Bongiovanni} A.,  {P{\'e}rez Garc{\'{\i}}a} A.~M.,   et~al.,
  2012a, \mn@doi [A\&A] {10.1051/0004-6361/201016261}, \href
  {http://adsabs.harvard.edu/abs/2012A%26A...541A..65O} {541, A65}

\bibitem[\protect\citeauthoryear{{Oteo}, {Bongiovanni}, {P{\'e}rez
  Garc{\'{\i}}a}  et~al.}{{Oteo} et~al.}{2012b}]{Oteo2012b}
{Oteo} I.,  {Bongiovanni} A.,  {P{\'e}rez Garc{\'{\i}}a} A.~M.,   et~al.,
  2012b, \mn@doi [ApJ] {10.1088/0004-637X/751/2/139}, \href
  {http://adsabs.harvard.edu/abs/2012ApJ...751..139O} {751, 139}

\bibitem[\protect\citeauthoryear{{Oteo}, {Sobral}, {Ivison}, {Smail}, {Best},
  {Cepa}  \& {P{\'e}rez-Garc{\'{\i}}a}}{{Oteo} et~al.}{2015}]{Oteo2015}
{Oteo} I.,  {Sobral} D.,  {Ivison} R.~J.,  {Smail} I.,  {Best} P.~N.,  {Cepa}
  J.,   {P{\'e}rez-Garc{\'{\i}}a} A.~M.,  2015, \mn@doi [MNRAS]
  {10.1093/mnras/stv1284}, \href
  {http://adsabs.harvard.edu/abs/2015MNRAS.452.2018O} {452, 2018}

\bibitem[\protect\citeauthoryear{{Ouchi} et~al.}{{Ouchi}
  et~al.}{2008}]{Ouchi2008}
{Ouchi} M.,  et~al., 2008, \mn@doi [ApJs] {10.1086/527673}, \href
  {http://adsabs.harvard.edu/abs/2008ApJS..176..301O} {176, 301}

\bibitem[\protect\citeauthoryear{{Ouchi} et~al.}{{Ouchi}
  et~al.}{2010}]{Ouchi2010}
{Ouchi} M.,  et~al., 2010, \mn@doi [ApJ] {10.1088/0004-637X/723/1/869}, \href
  {http://adsabs.harvard.edu/abs/2010ApJ...723..869O} {723, 869}

\bibitem[\protect\citeauthoryear{{Overzier} et~al.}{{Overzier}
  et~al.}{2009}]{Overzier2009}
{Overzier} R.~A.,  et~al., 2009, \mn@doi [ApJ] {10.1088/0004-637X/706/1/203},
  \href {http://adsabs.harvard.edu/abs/2009ApJ...706..203O} {706, 203}

\bibitem[\protect\citeauthoryear{{Partridge} \& {Peebles}}{{Partridge} \&
  {Peebles}}{1967}]{PartridgePeebles1967}
{Partridge} R.~B.,  {Peebles} P.~J.~E.,  1967, \mn@doi [ApJ] {10.1086/149079},
  \href {http://adsabs.harvard.edu/abs/1967ApJ...147..868P} {147, 868}

\bibitem[\protect\citeauthoryear{{Pritchet}}{{Pritchet}}{1994}]{Pritchet1994}
{Pritchet} C.~J.,  1994, \mn@doi [PASP] {10.1086/133479}, \href
  {http://adsabs.harvard.edu/abs/1994PASP..106.1052P} {106, 1052}

\bibitem[\protect\citeauthoryear{{Rauch}, {Haehnelt}, {Bunker}  et~al.}{{Rauch}
  et~al.}{2008}]{Rauch2008}
{Rauch} M.,  {Haehnelt} M.,  {Bunker} A.,   et~al., 2008, \mn@doi [ApJ]
  {10.1086/525846}, \href {http://adsabs.harvard.edu/abs/2008ApJ...681..856R}
  {681, 856}

\bibitem[\protect\citeauthoryear{{Saintonge}, {Kauffmann}, {Kramer}
  et~al.}{{Saintonge} et~al.}{2011}]{Saintonge2011}
{Saintonge} A.,  {Kauffmann} G.,  {Kramer} C.,   et~al., 2011, \mn@doi [MNRAS]
  {10.1111/j.1365-2966.2011.18677.x}, \href
  {http://adsabs.harvard.edu/abs/2011MNRAS.415...32S} {415, 32}

\bibitem[\protect\citeauthoryear{{Sandberg}, {Guaita}, {{\"O}stlin}, {Hayes}
  \& {Kiaeerad}}{{Sandberg} et~al.}{2015}]{Sandberg2015}
{Sandberg} A.,  {Guaita} L.,  {{\"O}stlin} G.,  {Hayes} M.,   {Kiaeerad} F.,
  2015, \mn@doi [A\&A] {10.1051/0004-6361/201525728}, \href
  {http://adsabs.harvard.edu/abs/2015A%26A...580A..91S} {580, A91}

\bibitem[\protect\citeauthoryear{{Santos}, {Sobral}  \& {Matthee}}{{Santos}
  et~al.}{2016}]{Santos2016}
{Santos} S.,  {Sobral} D.,   {Matthee} J.,  2016, \mn@doi [\mnras]
  {10.1093/mnras/stw2076}, \href
  {http://adsabs.harvard.edu/abs/2016MNRAS.tmp.1202S} {463, 1678}

\bibitem[\protect\citeauthoryear{{Sargent}, {Young}, {Boksenberg}  \&
  {Tytler}}{{Sargent} et~al.}{1980}]{Sargent1980}
{Sargent} W.~L.~W.,  {Young} P.~J.,  {Boksenberg} A.,   {Tytler} D.,  1980,
  \mn@doi [ApJS] {10.1086/190644}, \href
  {http://adsabs.harvard.edu/abs/1980ApJS...42...41S} {42, 41}

\bibitem[\protect\citeauthoryear{{Simpson} et~al.,}{{Simpson}
  et~al.}{2006}]{Bart_Simpson2006}
{Simpson} C.,  et~al., 2006, \mn@doi [MNRAS]
  {10.1111/j.1365-2966.2006.10907.x}, \href
  {http://adsabs.harvard.edu/abs/2006MNRAS.372..741S} {372, 741}

\bibitem[\protect\citeauthoryear{{Smail}, {Sharp}, {Swinbank}, {Akiyama},
  {Ueda}, {Foucaud}, {Almaini}  \& {Croom}}{{Smail} et~al.}{2008}]{Smail2008}
{Smail} I.,  {Sharp} R.,  {Swinbank} A.~M.,  {Akiyama} M.,  {Ueda} Y.,
  {Foucaud} S.,  {Almaini} O.,   {Croom} S.,  2008, \mn@doi [MNRAS]
  {10.1111/j.1365-2966.2008.13579.x}, \href
  {http://adsabs.harvard.edu/abs/2008MNRAS.389..407S} {389, 407}

\bibitem[\protect\citeauthoryear{{Smit}, {Bouwens}, {Franx}, {Illingworth},
  {Labb{\'e}}, {Oesch}  \& {van Dokkum}}{{Smit} et~al.}{2012}]{Smit2012}
{Smit} R.,  {Bouwens} R.~J.,  {Franx} M.,  {Illingworth} G.~D.,  {Labb{\'e}}
  I.,  {Oesch} P.~A.,   {van Dokkum} P.~G.,  2012, \mn@doi [ApJ]
  {10.1088/0004-637X/756/1/14}, \href
  {http://adsabs.harvard.edu/abs/2012ApJ...756...14S} {756, 14}

\bibitem[\protect\citeauthoryear{{Sobral}, {Best}, {Matsuda}, {Smail}, {Geach}
  \& {Cirasuolo}}{{Sobral} et~al.}{2012}]{Sobral2012}
{Sobral} D.,  {Best} P.~N.,  {Matsuda} Y.,  {Smail} I.,  {Geach} J.~E.,
  {Cirasuolo} M.,  2012, \mn@doi [MNRAS] {10.1111/j.1365-2966.2011.19977.x},
  \href {http://adsabs.harvard.edu/abs/2012MNRAS.420.1926S} {420, 1926}

\bibitem[\protect\citeauthoryear{{Sobral}, {Smail}, {Best}, {Geach}, {Matsuda},
  {Stott}, {Cirasuolo}  \& {Kurk}}{{Sobral} et~al.}{2013}]{Sobral2013}
{Sobral} D.,  {Smail} I.,  {Best} P.~N.,  {Geach} J.~E.,  {Matsuda} Y.,
  {Stott} J.~P.,  {Cirasuolo} M.,   {Kurk} J.,  2013, \mn@doi [MNRAS]
  {10.1093/mnras/sts096}, \href
  {http://adsabs.harvard.edu/abs/2013MNRAS.428.1128S} {428, 1128}

\bibitem[\protect\citeauthoryear{{Sobral}, {Best}, {Smail}, {Mobasher}, {Stott}
   \& {Nisbet}}{{Sobral} et~al.}{2014}]{Sobral2014}
{Sobral} D.,  {Best} P.~N.,  {Smail} I.,  {Mobasher} B.,  {Stott} J.,
  {Nisbet} D.,  2014, \mn@doi [MNRAS] {10.1093/mnras/stt2159}, \href
  {http://adsabs.harvard.edu/abs/2014MNRAS.437.3516S} {437, 3516}

\bibitem[\protect\citeauthoryear{{Sobral} et~al.,}{{Sobral}
  et~al.}{2015a}]{Sobral2015}
{Sobral} D.,  et~al., 2015a, \mn@doi [MNRAS] {10.1093/mnras/stv1076}, \href
  {http://adsabs.harvard.edu/abs/2015MNRAS.451.2303S} {451, 2303}

\bibitem[\protect\citeauthoryear{{Sobral}, {Matthee}, {Darvish}, {Schaerer},
  {Mobasher}, {R{\"o}ttgering}, {Santos}  \& {Hemmati}}{{Sobral}
  et~al.}{2015b}]{Sobral2015CR7}
{Sobral} D.,  {Matthee} J.,  {Darvish} B.,  {Schaerer} D.,  {Mobasher} B.,
  {R{\"o}ttgering} H.~J.~A.,  {Santos} S.,   {Hemmati} S.,  2015b, \mn@doi
  [ApJ] {10.1088/0004-637X/808/2/139}, \href
  {http://adsabs.harvard.edu/abs/2015ApJ...808..139S} {808, 139}

\bibitem[\protect\citeauthoryear{{Sobral}, {Kohn}, {Best}, {Smail}, {Harrison},
  {Stott}, {Calhau}  \& {Matthee}}{{Sobral} et~al.}{2016}]{Sobral2016}
{Sobral} D.,  {Kohn} S.~A.,  {Best} P.~N.,  {Smail} I.,  {Harrison} C.~M.,
  {Stott} J.,  {Calhau} J.,   {Matthee} J.,  2016, \mn@doi [MNRAS]
  {10.1093/mnras/stw022}, \href
  {http://adsabs.harvard.edu/abs/2016MNRAS.457.1739S} {457, 1739}

\bibitem[\protect\citeauthoryear{{Song} et~al.,}{{Song}
  et~al.}{2014}]{Song2014}
{Song} M.,  et~al., 2014, \mn@doi [ApJ] {10.1088/0004-637X/791/1/3}, \href
  {http://adsabs.harvard.edu/abs/2014ApJ...791....3S} {791, 3}

\bibitem[\protect\citeauthoryear{{Stark}, {Schenker}, {Ellis}, {Robertson},
  {McLure}  \& {Dunlop}}{{Stark} et~al.}{2013}]{Stark2013}
{Stark} D.~P.,  {Schenker} M.~A.,  {Ellis} R.,  {Robertson} B.,  {McLure} R.,
  {Dunlop} J.,  2013, \mn@doi [ApJ] {10.1088/0004-637X/763/2/129}, \href
  {http://adsabs.harvard.edu/abs/2013ApJ...763..129S} {763, 129}

\bibitem[\protect\citeauthoryear{{Steidel}, {Giavalisco}, {Pettini},
  {Dickinson}  \& {Adelberger}}{{Steidel} et~al.}{1996}]{Steidel96}
{Steidel} C.~C.,  {Giavalisco} M.,  {Pettini} M.,  {Dickinson} M.,
  {Adelberger} K.~L.,  1996, \mn@doi [ApJL] {10.1086/310029}, \href
  {http://adsabs.harvard.edu/abs/1996ApJ...462L..17S} {462, L17}

\bibitem[\protect\citeauthoryear{{Steidel}, {Erb}, {Shapley}, {Pettini},
  {Reddy}, {Bogosavljevi{\'c}}, {Rudie}  \& {Rakic}}{{Steidel}
  et~al.}{2010}]{Steidel2010}
{Steidel} C.~C.,  {Erb} D.~K.,  {Shapley} A.~E.,  {Pettini} M.,  {Reddy} N.,
  {Bogosavljevi{\'c}} M.,  {Rudie} G.~C.,   {Rakic} O.,  2010, \mn@doi [ApJ]
  {10.1088/0004-637X/717/1/289}, \href
  {http://adsabs.harvard.edu/abs/2010ApJ...717..289S} {717, 289}

\bibitem[\protect\citeauthoryear{{Stiavelli}, {Scarlata}, {Panagia}, {Treu},
  {Bertin}  \& {Bertola}}{{Stiavelli} et~al.}{2001}]{Stiavelli2001}
{Stiavelli} M.,  {Scarlata} C.,  {Panagia} N.,  {Treu} T.,  {Bertin} G.,
  {Bertola} F.,  2001, \mn@doi [ApJL] {10.1086/324513}, \href
  {http://adsabs.harvard.edu/abs/2001ApJ...561L..37S} {561, L37}

\bibitem[\protect\citeauthoryear{{Stott}, {Swinbank}, {Johnson}
  et~al.}{{Stott} et~al.}{2016}]{Stott2016}
{Stott} J.~P.,  {Swinbank} A.~M.,  {Johnson} H.~L.,   et~al., 2016, \mn@doi
  [MNRAS] {10.1093/mnras/stw129}, \href
  {http://adsabs.harvard.edu/abs/2016MNRAS.457.1888S} {457, 1888}

\bibitem[\protect\citeauthoryear{{Stroe} \& {Sobral}}{{Stroe} \&
  {Sobral}}{2015}]{StroeSobral2015}
{Stroe} A.,  {Sobral} D.,  2015, \mn@doi [MNRAS] {10.1093/mnras/stv1555}, \href
  {http://adsabs.harvard.edu/abs/2015MNRAS.453..242S} {453, 242}

\bibitem[\protect\citeauthoryear{{Stroe}, {Sobral}, {R{\"o}ttgering}  \& {van
  Weeren}}{{Stroe} et~al.}{2014}]{Stroe2014}
{Stroe} A.,  {Sobral} D.,  {R{\"o}ttgering} H.~J.~A.,   {van Weeren} R.~J.,
  2014, \mn@doi [MNRAS] {10.1093/mnras/stt2286}, \href
  {http://adsabs.harvard.edu/abs/2014MNRAS.438.1377S} {438, 1377}

\bibitem[\protect\citeauthoryear{{Swinbank} et~al.,}{{Swinbank}
  et~al.}{2015}]{Swinbank2015}
{Swinbank} A.~M.,  et~al., 2015, \mn@doi [MNRAS] {10.1093/mnras/stv366}, \href
  {http://adsabs.harvard.edu/abs/2015MNRAS.449.1298S} {449, 1298}

\bibitem[\protect\citeauthoryear{{Tacconi} et~al.}{{Tacconi}
  et~al.}{2010}]{Tacconi2010}
{Tacconi} L.~J.,  et~al., 2010, \mn@doi [Nature] {10.1038/nature08773}, \href
  {http://adsabs.harvard.edu/abs/2010Natur.463..781T} {463, 781}

\bibitem[\protect\citeauthoryear{{Taylor}}{{Taylor}}{2005}]{Topcat}
{Taylor} M.~B.,  2005, in {Shopbell} P.,  {Britton} M.,   {Ebert} R.,  eds,
  Astronomical Society of the Pacific Conference Series Vol. 347, Astronomical
  Data Analysis Software and Systems XIV. p.~29

\bibitem[\protect\citeauthoryear{{Trainor}, {Steidel}, {Strom}  \&
  {Rudie}}{{Trainor} et~al.}{2015}]{Trainor2015}
{Trainor} R.~F.,  {Steidel} C.~C.,  {Strom} A.~L.,   {Rudie} G.~C.,  2015,
  \mn@doi [ApJ] {10.1088/0004-637X/809/1/89}, \href
  {http://adsabs.harvard.edu/abs/2015ApJ...809...89T} {809, 89}

\bibitem[\protect\citeauthoryear{{Trainor}, {Strom}, {Steidel}  \&
  {Rudie}}{{Trainor} et~al.}{2016}]{Trainor2016}
{Trainor} R.~F.,  {Strom} A.~L.,  {Steidel} C.~C.,   {Rudie} G.~C.,  2016,
  preprint, \href {http://adsabs.harvard.edu/abs/2016arXiv160807280T} {}
  (\mn@eprint {arXiv} {1608.07280})

\bibitem[\protect\citeauthoryear{{Vanzella} et~al.}{{Vanzella}
  et~al.}{2016}]{Vanzella2016}
{Vanzella} E.,  et~al., 2016, \mn@doi [ApJ] {10.3847/0004-637X/825/1/41}, \href
  {http://adsabs.harvard.edu/abs/2016ApJ...825...41V} {825, 41}

\bibitem[\protect\citeauthoryear{{Vasei} et~al.,}{{Vasei}
  et~al.}{2016}]{Vasei2016}
{Vasei} K.,  et~al., 2016, preprint, \href
  {http://adsabs.harvard.edu/abs/2016arXiv160302309V} {} (\mn@eprint {arXiv}
  {1603.02309})

\bibitem[\protect\citeauthoryear{{Verhamme}, {Schaerer}  \&
  {Maselli}}{{Verhamme} et~al.}{2006}]{Verhamme2006}
{Verhamme} A.,  {Schaerer} D.,   {Maselli} A.,  2006, \mn@doi [A\&A]
  {10.1051/0004-6361:20065554}, \href
  {http://adsabs.harvard.edu/abs/2006A%26A...460..397V} {460, 397}

\bibitem[\protect\citeauthoryear{{Verhamme}, {Schaerer}, {Atek}  \&
  {Tapken}}{{Verhamme} et~al.}{2008}]{Verhamme2008}
{Verhamme} A.,  {Schaerer} D.,  {Atek} H.,   {Tapken} C.,  2008, \mn@doi [AAP]
  {10.1051/0004-6361:200809648}, \href
  {http://adsabs.harvard.edu/abs/2008A%26A...491...89V} {491, 89}

\bibitem[\protect\citeauthoryear{{Verhamme}, {Orlitov{\'a}}, {Schaerer}  \&
  {Hayes}}{{Verhamme} et~al.}{2015}]{Verhamme2015}
{Verhamme} A.,  {Orlitov{\'a}} I.,  {Schaerer} D.,   {Hayes} M.,  2015, \mn@doi
  [A\&A] {10.1051/0004-6361/201423978}, \href
  {http://adsabs.harvard.edu/abs/2015A%26A...578A...7V} {578, A7}

\bibitem[\protect\citeauthoryear{{Verhamme}, {Orlitova}, {Schaerer}, {Izotov},
  {Worseck}, {Thuan}  \& {Guseva}}{{Verhamme} et~al.}{2016}]{Verhamme2016}
{Verhamme} A.,  {Orlitova} I.,  {Schaerer} D.,  {Izotov} Y.,  {Worseck} G.,
  {Thuan} T.~X.,   {Guseva} N.,  2016, preprint, \href
  {http://adsabs.harvard.edu/abs/2016arXiv160903477V} {} (\mn@eprint {arXiv}
  {1609.03477})

\bibitem[\protect\citeauthoryear{{Wardlow} et~al.}{{Wardlow}
  et~al.}{2014}]{Wardlow2014}
{Wardlow} J.~L.,  et~al., 2014, \mn@doi [ApJ] {10.1088/0004-637X/787/1/9},
  \href {http://adsabs.harvard.edu/abs/2014ApJ...787....9W} {787, 9}

\bibitem[\protect\citeauthoryear{{Wisotzki} et~al.}{{Wisotzki}
  et~al.}{2016}]{Wisotzki2016}
{Wisotzki} L.,  et~al., 2016, \mn@doi [A\&A] {10.1051/0004-6361/201527384},
  \href {http://adsabs.harvard.edu/abs/2016A%26A...587A..98W} {587, A98}

\bibitem[\protect\citeauthoryear{{Wold}, {Barger}  \& {Cowie}}{{Wold}
  et~al.}{2014}]{Wold2014}
{Wold} I.~G.~B.,  {Barger} A.~J.,   {Cowie} L.~L.,  2014, \mn@doi [ApJ]
  {10.1088/0004-637X/783/2/119}, \href
  {http://adsabs.harvard.edu/abs/2014ApJ...783..119W} {783, 119}

\bibitem[\protect\citeauthoryear{{Yamada} et~al.,}{{Yamada}
  et~al.}{2005}]{Yamada2005}
{Yamada} T.,  et~al., 2005, \mn@doi [ApJ] {10.1086/496954}, \href
  {http://adsabs.harvard.edu/abs/2005ApJ...634..861Y} {634, 861}

\bibitem[\protect\citeauthoryear{{Yang}, {Malhotra}, {Gronke}, {Rhoads},
  {Dijkstra}, {Jaskot}, {Zheng}  \& {Wang}}{{Yang} et~al.}{2016}]{Yang2016}
{Yang} H.,  {Malhotra} S.,  {Gronke} M.,  {Rhoads} J.~E.,  {Dijkstra} M.,
  {Jaskot} A.,  {Zheng} Z.,   {Wang} J.,  2016, \mn@doi [ApJ]
  {10.3847/0004-637X/820/2/130}, \href
  {http://adsabs.harvard.edu/abs/2016ApJ...820..130Y} {820, 130}

\bibitem[\protect\citeauthoryear{{Zitrin} et~al.,}{{Zitrin}
  et~al.}{2015}]{Zitrin2015}
{Zitrin} A.,  et~al., 2015, \mn@doi [ApJL] {10.1088/2041-8205/810/1/L12}, \href
  {http://adsabs.harvard.edu/abs/2015ApJ...810L..12Z} {810, L12}

\bibitem[\protect\citeauthoryear{{de Barros} et~al.}{{de Barros}
  et~al.}{2016}]{deBarros2016}
{de Barros} S.,  et~al., 2016, \mn@doi [A\&A] {10.1051/0004-6361/201527046},
  \href {http://adsabs.harvard.edu/abs/2016A%26A...585A..51D} {585, A51}

\bibitem[\protect\citeauthoryear{{van Breukelen} et~al.,}{{van Breukelen}
  et~al.}{2007}]{van_breu2007}
{van Breukelen} C.,  et~al., 2007, \mn@doi [MNRAS]
  {10.1111/j.1365-2966.2007.12433.x}, \href
  {http://adsabs.harvard.edu/abs/2007MNRAS.382..971V} {382, 971}

\makeatother
\end{thebibliography}

%%%%%%%%%%%%%%%%% APPENDICES %%%%%%%%%%%%%%%%%%%%%

\appendix

\section{Catalogues of line emitters} \label{catalogue}

%
% Table A2 - Luminosity function
%
\begin{table}
 \centering
  \caption{The $z=2.23$ Ly$\alpha$ luminosity function. We present both observed/raw number densities and number densities corrected taken into account incompleteness and the filter profile corrections.}
  \begin{tabular}{@{}cccc@{}}
  \hline
  \bf $\bf \log$\bf L$_{\rm \bf Ly\alpha}$ &$\bf  \phi$ \bf obs & $\bf  \phi$ \bf corr & \bf Volume \\

  & Mpc$^{-3}$ & Mpc$^{-3}$  & 10$^5$\,Mpc$^3$  \\
  \hline

$42.30\pm0.125$  & $-2.74\pm0.10$ & $-2.76\pm0.11$ & 0.539  \\
$42.55\pm0.125$  & $-3.06\pm0.04$ & $-3.11\pm0.05$ & 5.208  \\
$42.80\pm0.125$  & $-3.53\pm0.06$ & $-3.60\pm0.07$ & 7.123  \\
$43.05\pm0.125$  & $-4.32\pm0.18$ & $-4.40\pm0.21$ & 7.330  \\
$43.30\pm0.125$  & $-4.54\pm0.25$ & $-4.62\pm0.30$ & 7.330  \\
$43.55\pm0.125$  & $-5.07\pm0.70$ & $-5.15\pm1.88$ & 7.330  \\
 \hline
\end{tabular}
\label{LF_funccc}
\end{table}

Table \ref{NB_CAT} presents example entries of the full catalogue of 440 line emitters, and a flag for sources selected as Ly$\alpha$ emitters. The full catalogue is available on the on-line version of the paper.

%
% Table A1 - Catalogues
%
\begin{table*}
 \centering
  \caption{The first 9 entries from the catalogue of 440 candidate line emitters selected in COSMOS and UDS. The full catalogue is available on-line.}
  \begin{tabular}{@{}ccccccccccc@{}}
  \hline
   ID & R.A. & Dec. & NB392 & $u$ & log Flux  & EW$_{\rm obs}$  & $\Sigma$ & Ly$\alpha$ selec  \\
         & {(J2000)} &(J2000) & (AB) &(AB) & erg\,s$^{-1}$\,cm$^{-2}$ & \AA &  \\
 \hline
   \noalign{\smallskip}
  CALYMHA-S16-1 & 10\, 02\, 42.916 & +02\, 12\, 52.93 & 23.6 & 25.1 & -15.9 & 248 & 3.1 & no \\
  CALYMHA-S16-2 & 10\, 02\, 42.318 & +02\, 35\, 29.47 & 23.3 & 24.5 & -15.9 & 121 & 3.9 & yes \\
  CALYMHA-S16-3 & 10\, 02\, 39.342 & +02\, 26\, 10.08 & 22.9 & 23.6 & -15.8 & 55 & 3.3 & no \\
  CALYMHA-S16-4 & 10\, 02\, 38.841 & +02\, 35\, 58.76 & 23.7 & 25.2 & -15.9 & 246 & 3.4 & no \\
  CALYMHA-S16-5 & 10\, 02\, 37.020 & +02\, 40\, 29.08 & 23.2 & 24.2 & -15.9 & 95 & 4.0 & no \\
  CALYMHA-S16-6 & 10\, 02\, 36.829 & +02\, 39\, 47.33 & 23.8 & 25.6 & -16.0 & 375 & 3.1 & yes\\
  CALYMHA-S16-7 & 10\, 02\, 36.770 & +02\, 32\, 25.79 & 23.3 & 24.8 & -15.8& 212 & 4.6 & yes\\
  CALYMHA-S16-8 & 10\, 02\, 35.458 & +02\, 17\, 57.62 & 23.2 & 24.3 & -15.8 & 113 & 3.5 & yes\\
  CALYMHA-S16-9 & 10\, 02\, 35.106 & +02\, 32\, 23.91 & 23.6 & 25.5 & -15.9 & 377 & 3.6 & yes\\
 \hline
\end{tabular}
\label{NB_CAT}
\end{table*}

%%%%%%%%%%%%%%%%%%%%%%%%%%%%%%%%%%%%%%%%%%%%%%%%%%
\bsp	% typesetting comment
\label{lastpage}
\end{document}